\documentclass[useAMS,usenatbib]{mn2e}

\usepackage{graphicx}
\usepackage{ amsmath }
\usepackage{booktabs}
\usepackage{rotating}
\usepackage{morefloats}
\usepackage{enumerate}
\usepackage{ amssymb }
\usepackage[hyphens]{url}

\title[Hot bubbles in PNe (II)]{Formation and X-ray Emission from Hot bubbles in 
Planetary Nebulae\\ {\LARGE II. Hot bubble X-ray emission}}

\author[Toal\'{a} \& Arthur]{J.A.\,Toal\'{a}$^{1,2}$\thanks{E-mail:toala@asiaa.sinica.edu.tw} and S.J.\,Arthur$^{3}$\\ 
$^{1}$Institute of Astronomy and Astrophysics, Academia Sinica (ASIAA), 10617 Taipei, Taiwan, R.O.C.\\
$^{2}$Instituto de Astrof\'{\i}sica de Andaluc\'{\i}a, IAA-CSIC, Glorieta de la Astronom\'{\i}a s/n, 18008 Granada, Spain\\ 
$^{3}$Instituto de Radioastronom\'{i}a y Astrof\'{i}sica, UNAM Campus Morelia, Apartado Postal 3-72, 58090, Morelia, Michoac\'{a}n, M\'{e}xico.}
\begin{document}


\maketitle

\label{firstpage}

\begin{abstract}
  We present a study of the X-ray emission from numerical simulations
  of hot bubbles in planetary nebulae (PNe). High-resolution,
  two-dimensional, radiation-hydrodynamical simulations of the
  formation and evolution of hot bubbles in PNe, with and without
  thermal conduction, are used to calculate the X-ray emission and
  study its time-dependence and relationship to the changing stellar
  parameters. Instabilities in the wind-wind interaction zone produce
  clumps and filaments in the swept-up shell of nebular
  material. Turbulent mixing and thermal conduction at the corrugated
  interface can produce quantities of intermediate temperature and
  density gas between the hot, shocked wind bubble and the swept-up
  photoionized nebular material, which can emit in soft, diffuse
  X-rays.  We use the {\sc chianti} software to compute synthetic
  spectra for the models and calculate their luminosities. We find
  that models both with conduction and those without can produce the
  X-ray temperatures and luminosities that are in the ranges reported
  in observations, although the models including thermal conduction
  are an order of magnitude more luminous than those without. Our
  results show that at early times the diffuse X-ray emission should
  be dominated by the contribution from the hot, shocked stellar wind,
  whereas at later times the nebular gas will dominate the
  spectrum. We analyse the effect of sampling on the resultant spectra
  and conclude that a minimum of 200~counts is required to reliably
  reproduce the spectral shape. Likewise, heavily smoothed
  surface-brightness profiles obtained from low-count detections of
  PNe do not provide a reliable description of the spatial
  distribution of the X-ray emitting gas.
\end{abstract}


\section{Introduction}
\label{sec:intro}

The presence of diffuse X-ray emission associated with hot bubbles in
planetary nebulae (PNe) has been confirmed by means of high-quality
X-ray observations performed with the {\it XMM-Newton} and {\it
  Chandra} observatories \citep[e.g.,][and references
  therein]{Ruiz2013}, lending strong support to the interacting
stellar winds model of {hot bubble formation in PNe}
\citep{Kwok1978,Balick1987}. In particular, the {\it Chandra}
Planetary Nebula Survey ({\sc ChanPlaNS}) is producing a
{statistically significant} study of the X-ray emission from PNe
within 1.5\,kpc of the Sun \citep{Kastner2012,Freeman2014,Montez2015}
looking for both point source X-ray emission from the central stars
(CSPNe) and diffuse X-ray emission (from the hot bubbles). {The
  results published to date have reported a diffuse X-ray emission
  detection rate of $\sim 30\%$ \citep{Kastner2012,Freeman2014}. The
  detected objects are generally compact ($R_\mathrm{HB} < 0.2$~pc in
  radius) with closed structures, suggesting that they are young
  objects ($< 5\times10^3$~yrs), with high nebular electron densities
  $n_{e} > 10^3$~cm$^{-3}$ (as determined via their H$\alpha$ line
  luminosities). Detailed spectral analysis of some of the detected
  objects indicates that the temperature of the X-ray-emitting gas
  lies in the narrow range $(1$--$3)\times10^6$~K with electronic
  densities of the order of 1--$10^{2}$~cm$^{-3}$ \citep[e.g.,][and
    references therein]{Ruiz2013,Kastner2012}. Some of the objects
  detected in diffuse emission can be described using the abundances
  of that of the nebular material, while others require stellar
  abundances \citep[e.g.,][]{{Yu2009}}.

The low temperatures derived from the X-ray observations contrast with
expectations from adiabatic shock theory as applied to the fast
stellar winds from CSPN \citep[$v_{\infty}=1000$--4000~km~s$^{-1}$;
see][]{GuerrerodeMarco2013}, which would indicate post-shock
temperatures at least an order of magnitude higher\footnote{For an
  adiabatic shock in a free-flowing stellar wind with
  velocity $v_\infty$, the post-shock temperature is $k_\mathrm{B}
  T_\mathrm{s} = 3 \mu m_\mathrm{H} v_\infty^2/16$, where $\mu
  m_\mathrm{H}$ is the mean mass per particle and $k_\mathrm{B}$ is
  Boltzmann's constant \citep[see, e.g.,][]{Dyson1997}.}. 

There have been several numerical studies that have addressed the
production of diffuse X-ray emission in PNe
\citep[e.g.][]{Mellema1995,Stute2006,Akashi2007,Lou2010}, even for the
case of non-isotropic wind ejections \citep[see][]{Akashi2008}. Given
that diffuse X-ray emission is a relatively common phenomenon in
planetary nebulae, any theoretical model should be able to explain all
the features of the X-ray emission for a wide range of model
parameters. That is, the distribution of the X-ray-emitting gas,
temperature, and abundances of the gas responsible for the diffuse
emission. The main sequence mass of the central star sets the
timescale for the mass loss during the AGB stage and thus determines
the density distribution of the circumstellar medium around the
CSPN. In the post-AGB stage the stellar parameters are determined by
the mass of the central hot object that remains after the hydrogen
envelope has been expelled. Previous 1D numerical work by
\citet{Akashi2007} and \citet{Steffen2008} has shown that it is
important to take into account the time history of the stellar wind as
it increases in velocity and decreases in mass-loss rate over the few
hundred years after the AGB envelope is ejected. In order to explain
the low X-ray temperatures, either the X-rays come from wind shocked
while the velocity was still low ($v_\mathrm{w} < 500$~km~s$^{-1}$),
or some physical process operates to produce large quantities of gas
at about a million degrees even when the stellar wind velocity is
higher than 1000~km~s$^{-1}$. \citet{Soker1994} suggested that thermal
conduction could contribute to the production of the diffuse X-ray
emission, and that the soft X-rays come from a conduction layer of
evaporated nebular gas adjacent to the hot bubble, rather than from
the shocked fast stellar wind.

The one-dimensional (1D) radiation-hydrodynamic simulations presented
in \citet{Steffen2008} provided a coherent view of the formation and
X-ray emission from PNe by considering the detailed time evolution of
the central star parameters and ionizing photon flux from the
asymptotic giant branch (AGB) through to the post-AGB phase. These
authors included heat diffusion in their simulations and considered
both classical and saturated thermal conduction \citep[see,
  e.g.,][]{Spitzer1962}. They used two different formalisms for
limiting the electron mean free path: one that depends essentially on
the numerical resolution and can be interpreted as inefficient
conduction, and the other, which defaults to the saturation heat flux
wherever the classical heat flux exceeds the saturated value, and
represents efficent conduction.  They found that their simulations
reproduce the plasma temperatures and luminosities as long as
conduction is included. However, they favoured the numerical results
from their efficient conduction method, arguing that the
low-efficiency conduction method results in limb-brightened synthetic
X-ray surface brightness profiles, which did not match some
observations at the time. This group has recently extended its
numerical simulations to study the X-ray emission from PNe with
hydrogen-deficient winds \citep{Sandin2016}.

\begin{figure*}
\includegraphics[width=0.25\linewidth]{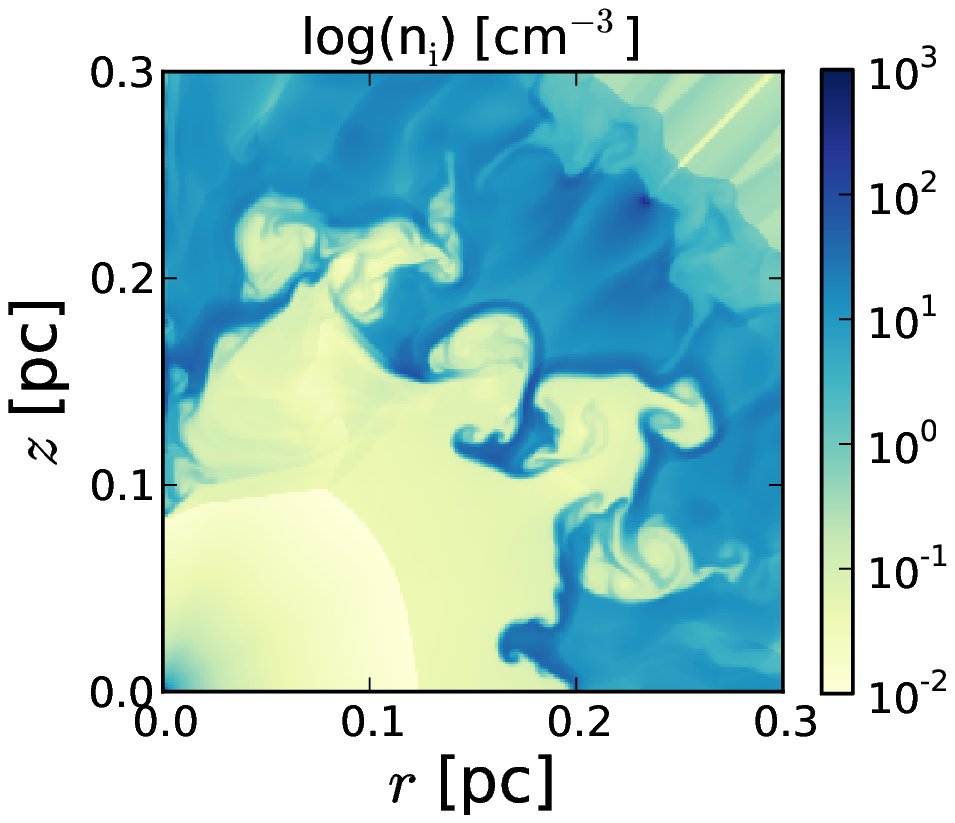}~
\includegraphics[width=0.25\linewidth]{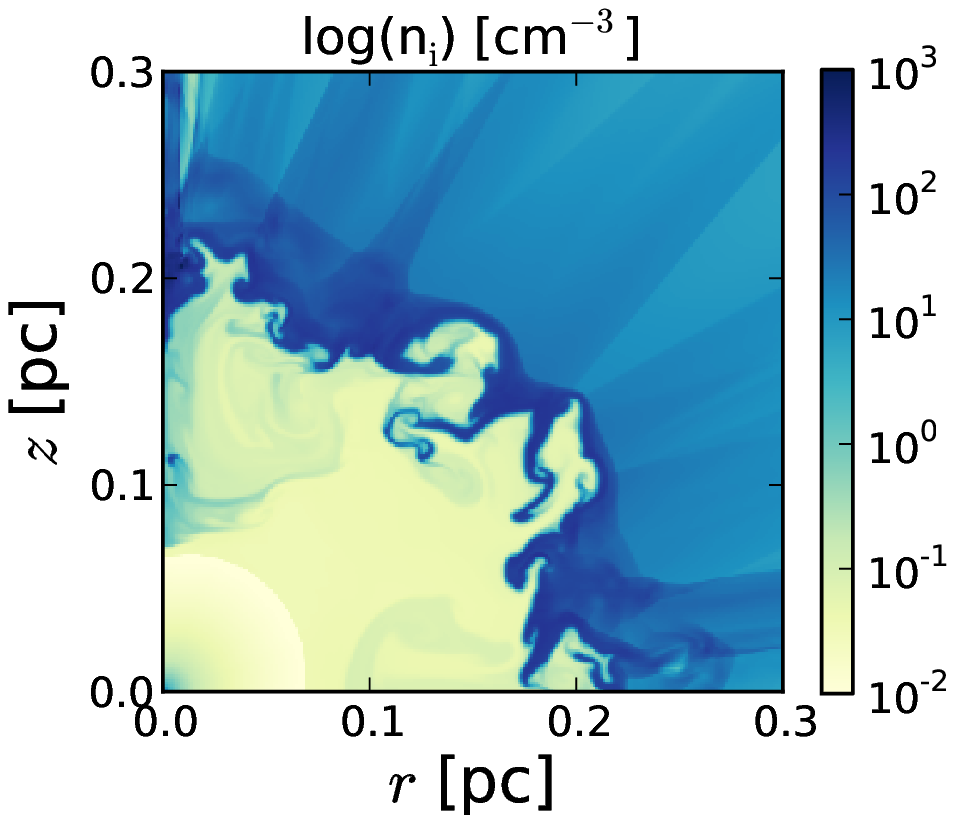}~
\includegraphics[width=0.25\linewidth]{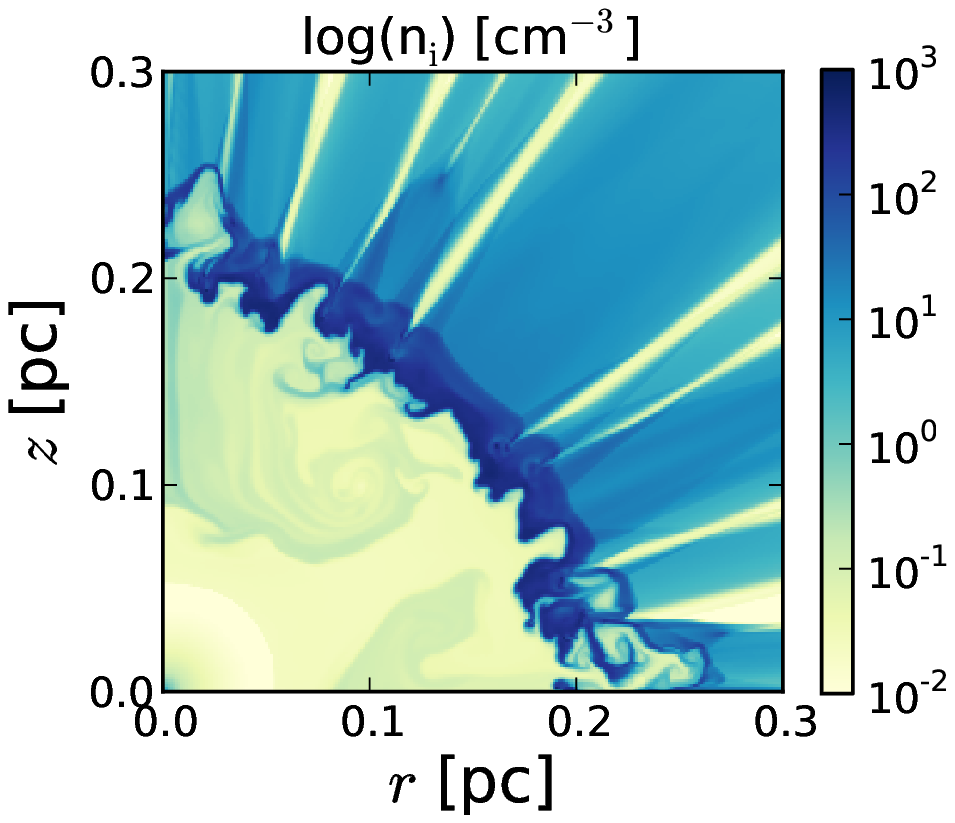}~
\includegraphics[width=0.25\linewidth]{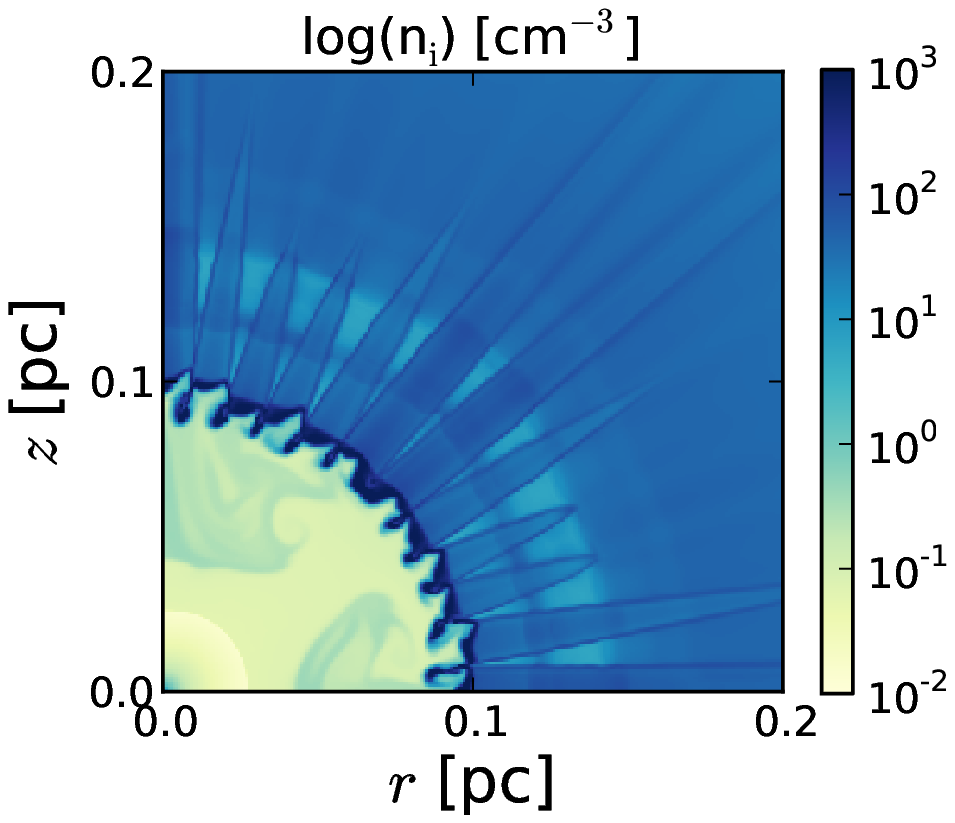}
~%
\includegraphics[width=0.25\linewidth]{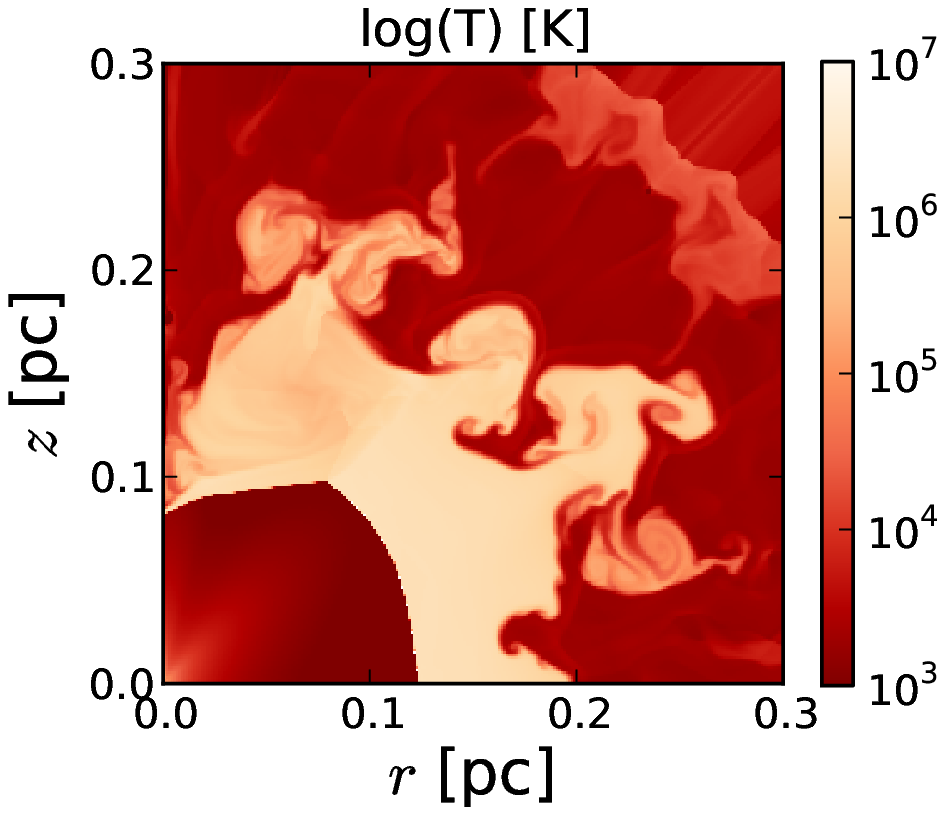}~
\includegraphics[width=0.25\linewidth]{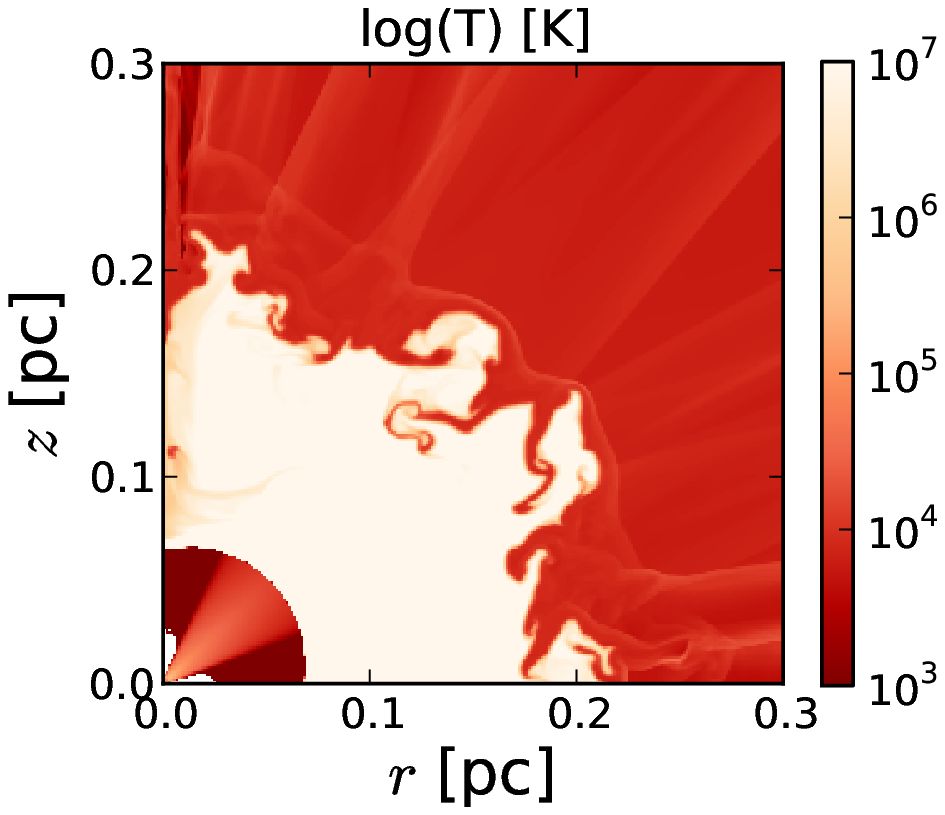}~
\includegraphics[width=0.25\linewidth]{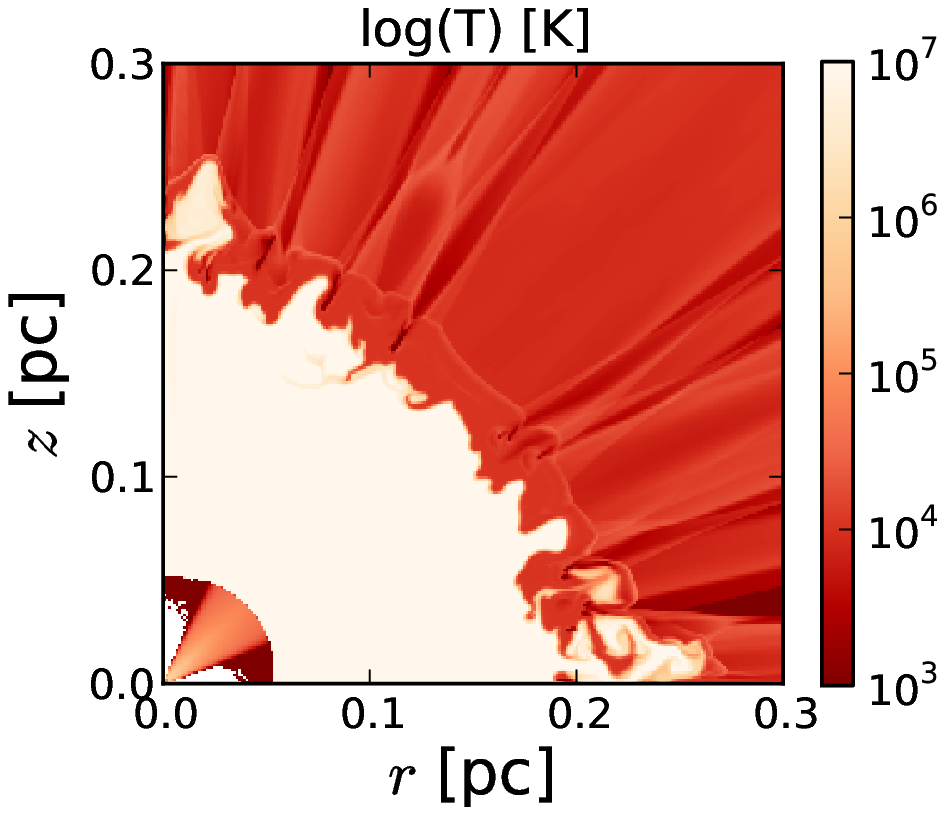}~
\includegraphics[width=0.25\linewidth]{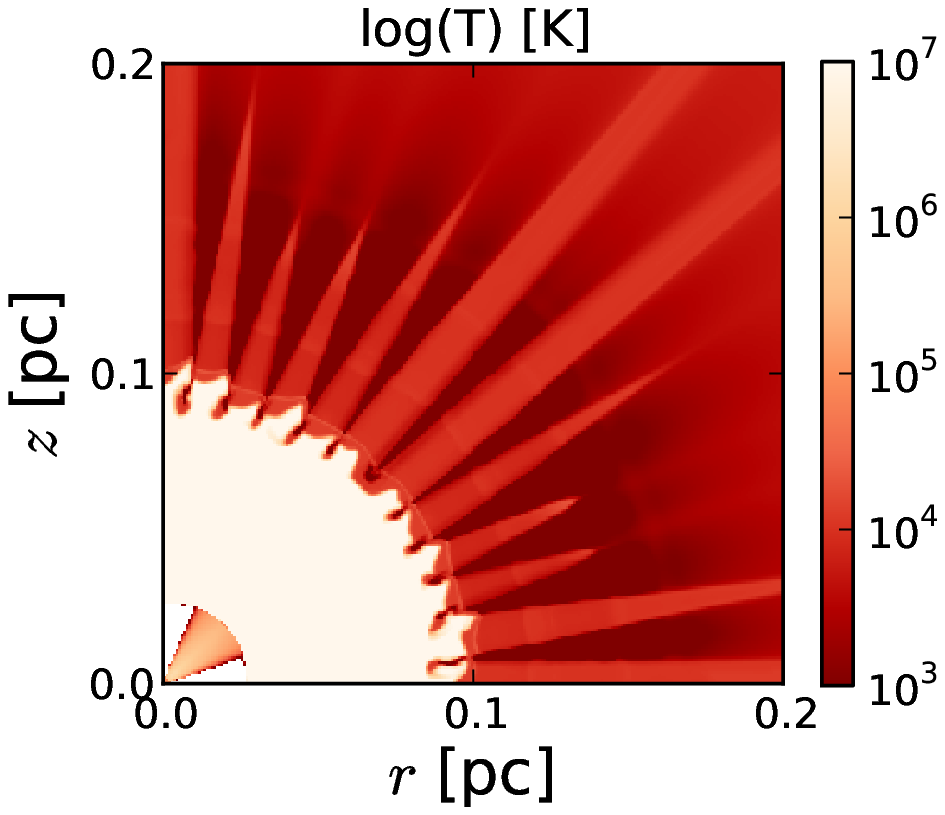}
\caption{Total ion number density (top panels) and temperature (bottom
  panels) in the $r$-$z$ plane as presented in Paper~I. Columns (from
  left to right) correspond to the 1.0-0.569, 1.5-0.597, 2.0-0.633,
  and 2.5-0.677 models without thermal conduction at 6500, 7000, 4500,
  and 1500~yr after the onset of the post-AGB phase, respectively.
}
\label{fig:final_2D}
\end{figure*} 

\begin{figure*}
\includegraphics[width=0.25\linewidth]{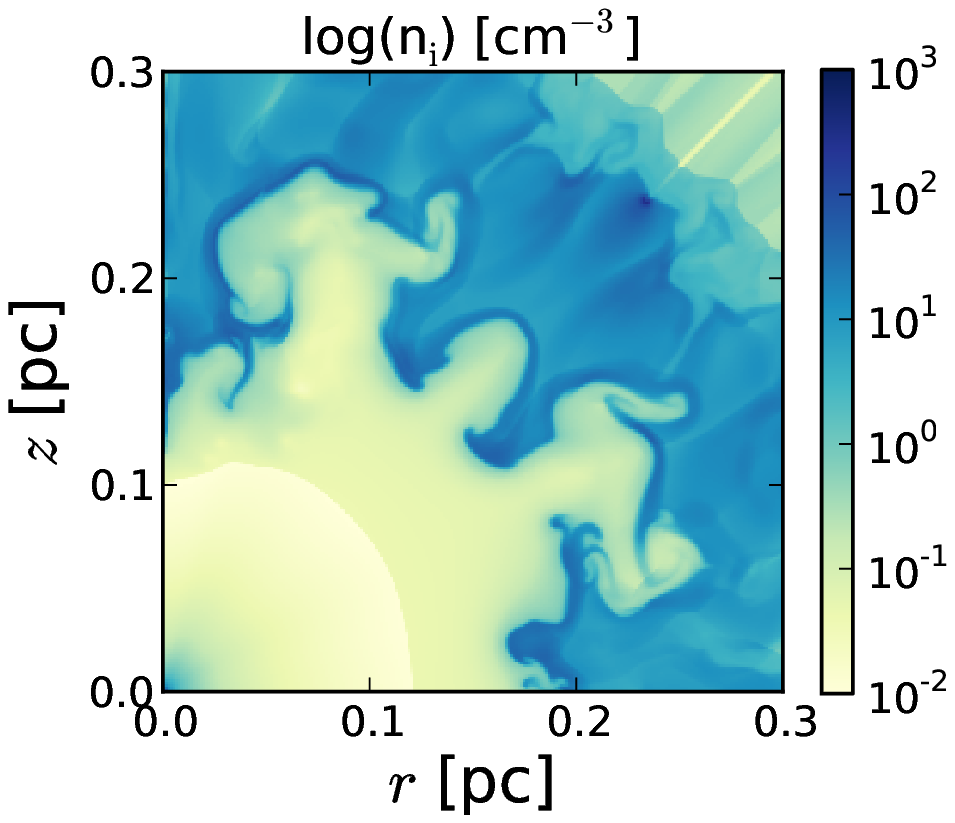}~
\includegraphics[width=0.25\linewidth]{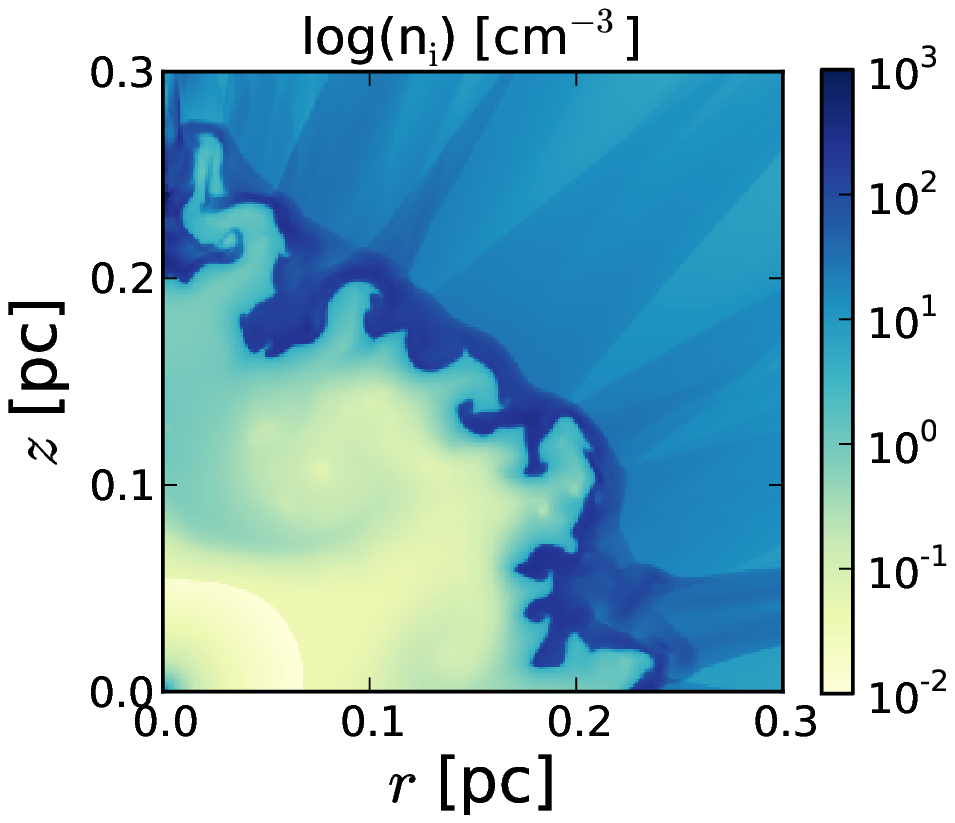}~
\includegraphics[width=0.25\linewidth]{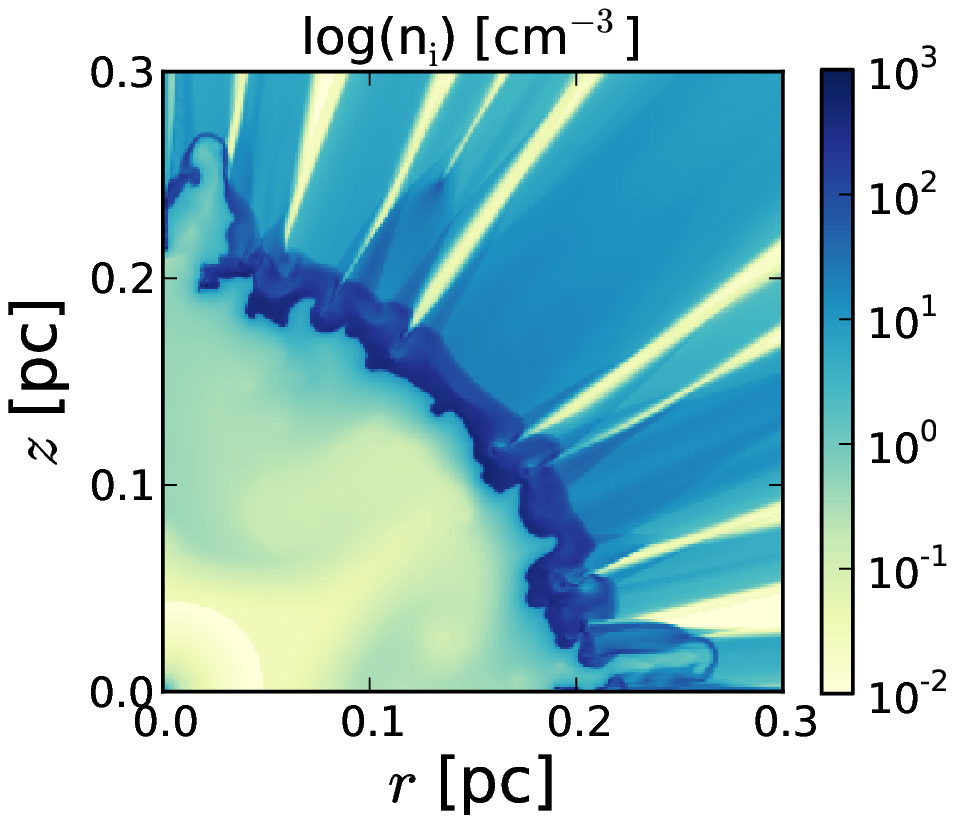}~
\includegraphics[width=0.25\linewidth]{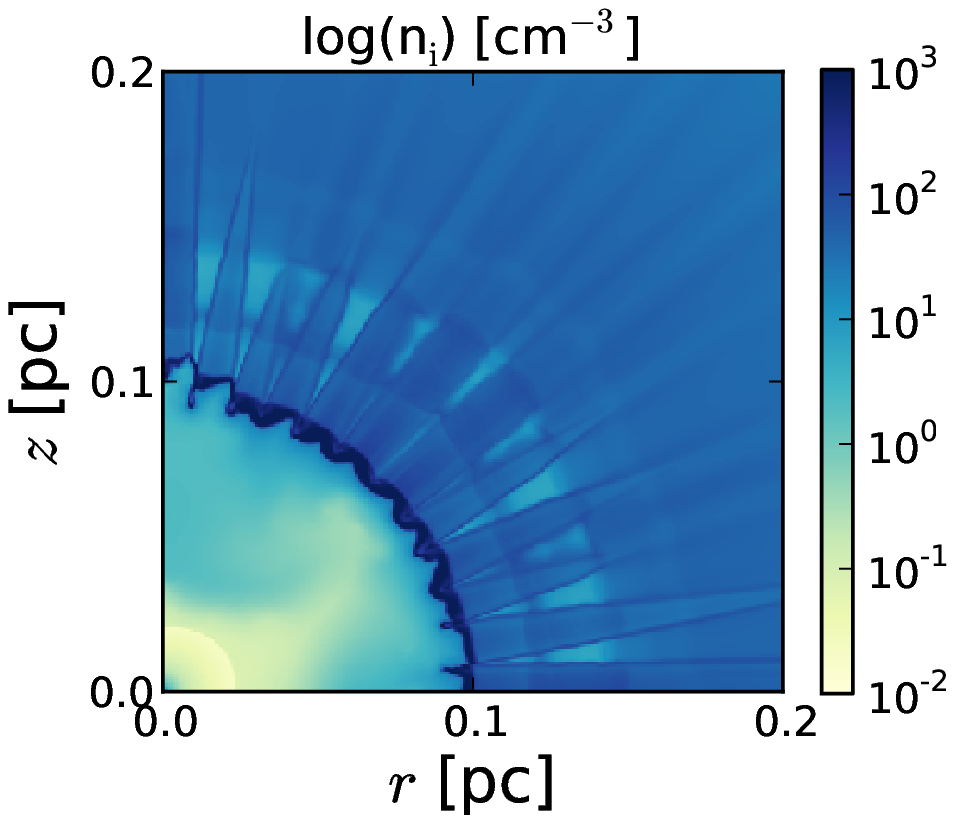}\\
\includegraphics[width=0.25\linewidth]{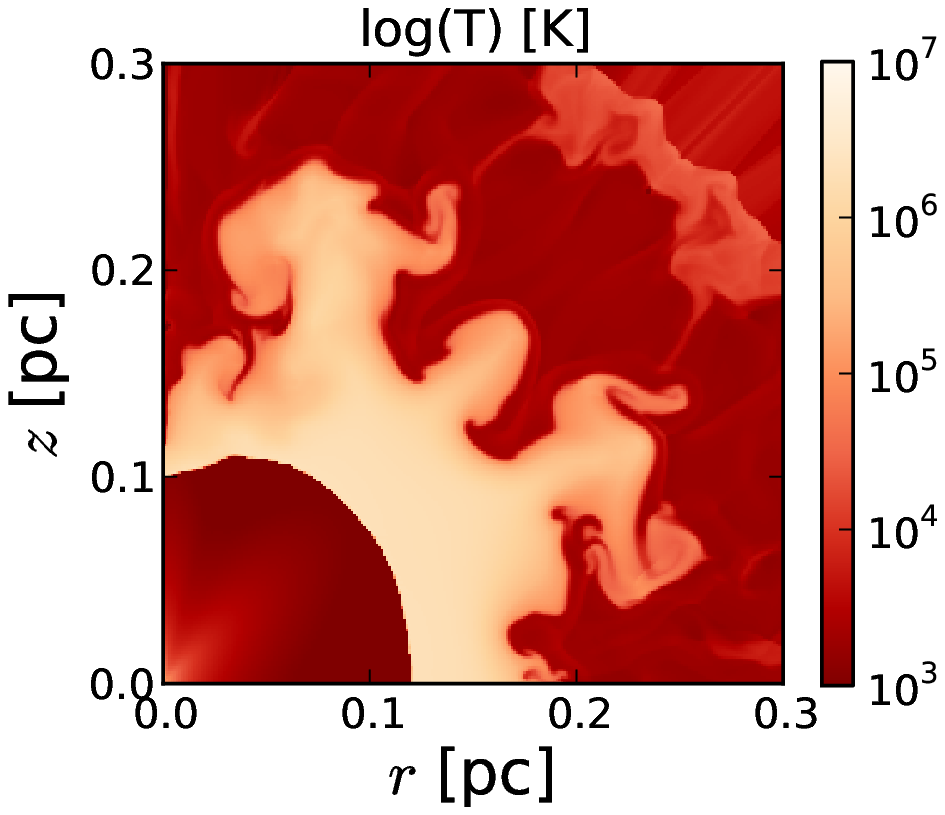}~
\includegraphics[width=0.25\linewidth]{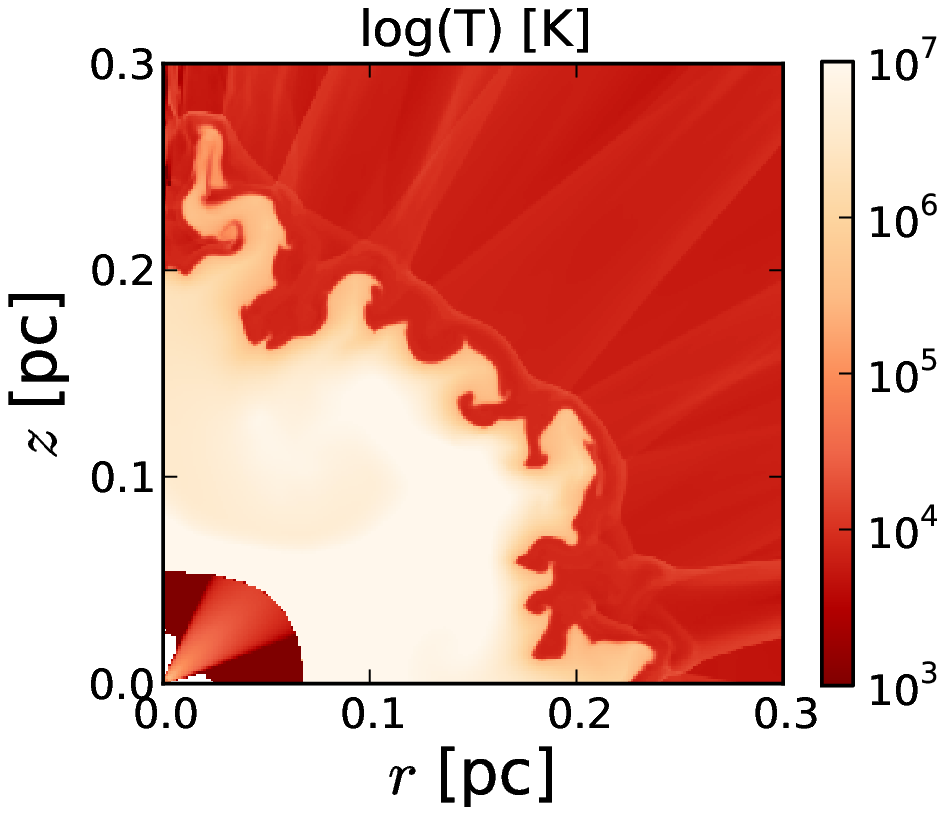}~
\includegraphics[width=0.25\linewidth]{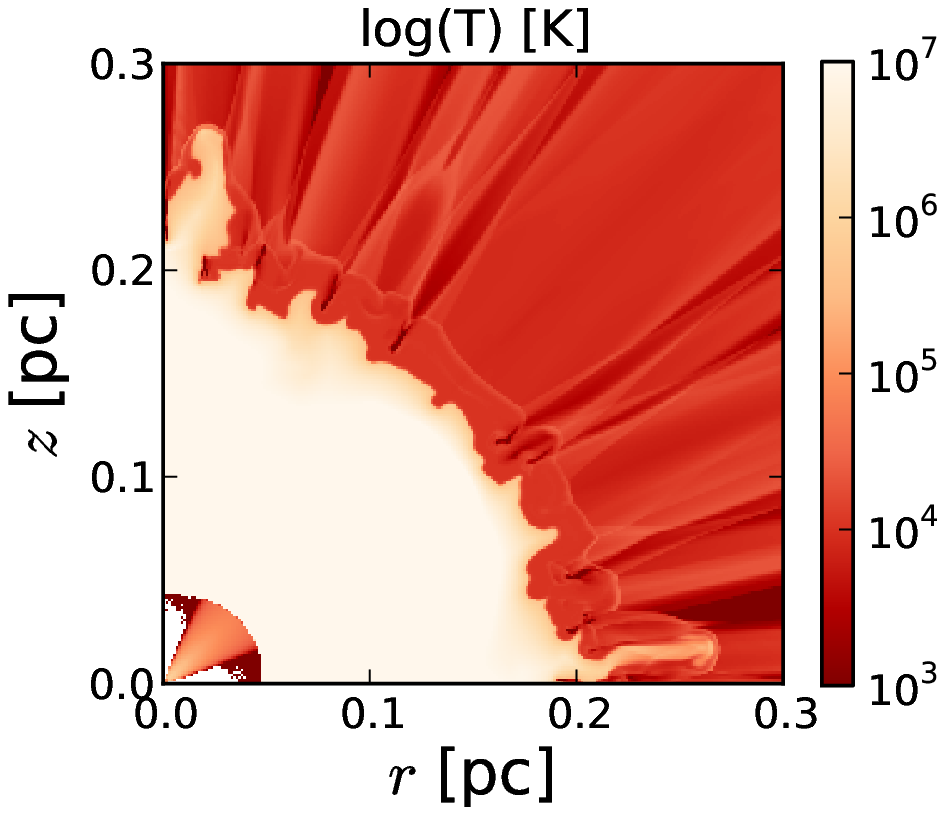}~
\includegraphics[width=0.25\linewidth]{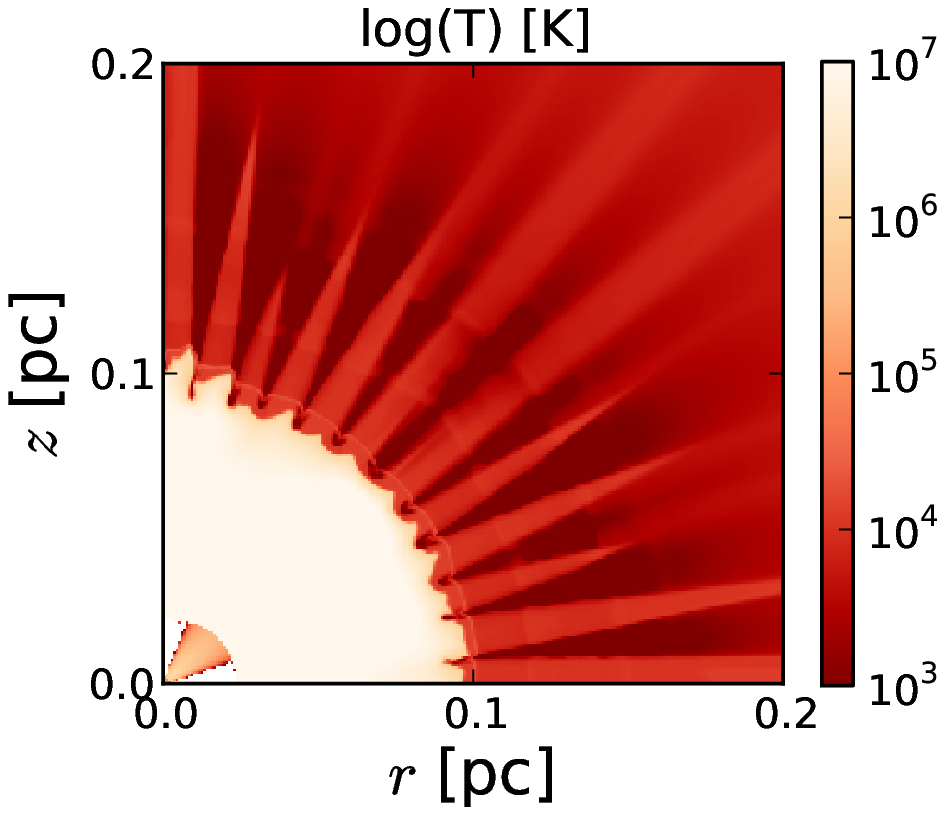}
\caption{Same as Figure~\ref{fig:final_2D} but for cases with thermal
  conduction. Columns (from left to right) correspond to the
  1.0-0.569, 1.5-0.597, 2.0-0.633, and 2.5-0.677 models at 6500, 7000,
  4500, and 1500~yr after the onset of the post-AGB phase,
  respectively.
}
\label{fig:final_2D_2}
\end{figure*}

In \citet[][hereafter Paper I]{Toala2014} we presented high resolution
two-dimensional (2D), radiation-hydrodynamical simulations of the
formation and evolution of hot bubbles inside PNe for different
initial stellar masses. We used the stellar evolution models at solar
metallicities ($Z=0.016$) from \citet{Vassiliadis1993} for the AGB
phase and the corresponding models from \citet{Vassiliadis1994} for
the stellar parameters in the post-AGB phase. We computed the stellar
wind parameters for the post-AGB phase with the WM-basic code
\citep{Pauldrach1986,Pauldrach1987,Pauldrach1994,Pauldrach2001,Pauldrach2012}.
We found that the fast wind/AGB interaction zone is unstable and that
the interface between the hot bubble and the swept-up photoionized
shell corrugates and forms clumps and filaments. Even without thermal
conduction, hydrodynamical ablation and photoevaporation of the dense
clumps leads to turbulent mixing of cooler material into the hot
gas. When thermal conduction is included in the simulations, heat
diffuses from the hot shocked gas in the interior of the bubble into
the surrounding cooler photoionized shell. Dense material from the
shell is heated and expands, and the corrugated nature of the
interface increases the efficiency of this process. In both scenarios,
the net result is a significant amount of material at intermediate
temperatures and densities, {which} originated from the clumps
formed at the unstable interface.

In this paper, we use the results from Paper~I to generate synthetic
X-ray emission, and study the time-variation of the X-ray spectra,
luminosities, and plasma temperatures for models with different
initial stellar masses and models with and without thermal
conduction. The structure of the present paper is as follows: in
Section~\ref{sec:method} we describe the methodology followed, and the
results are presented in Section~\ref{sec:results}. In
Section~\ref{sec:comparobs} we compare our results with
observations. We present the discussion of our results in
Section~\ref{sec:discussion}, and summarize our findings in
Section~\ref{sec:summary}.

\section{Methodology}
\label{sec:method}
\begin{figure*}
\includegraphics[width=0.25\linewidth]{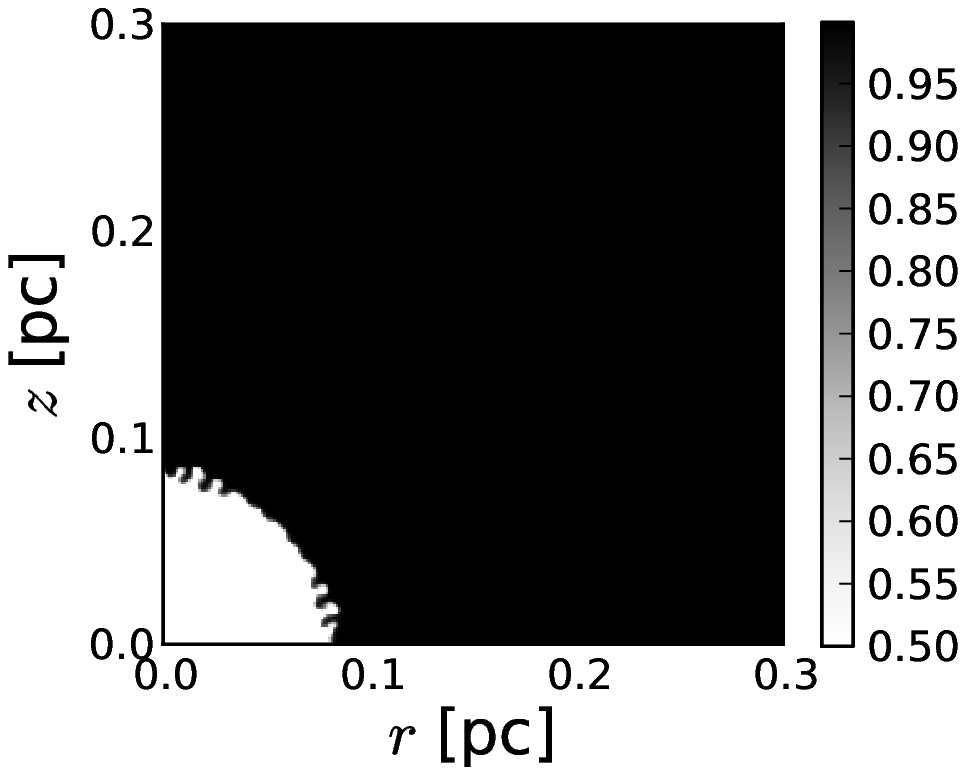}~
\includegraphics[width=0.25\linewidth]{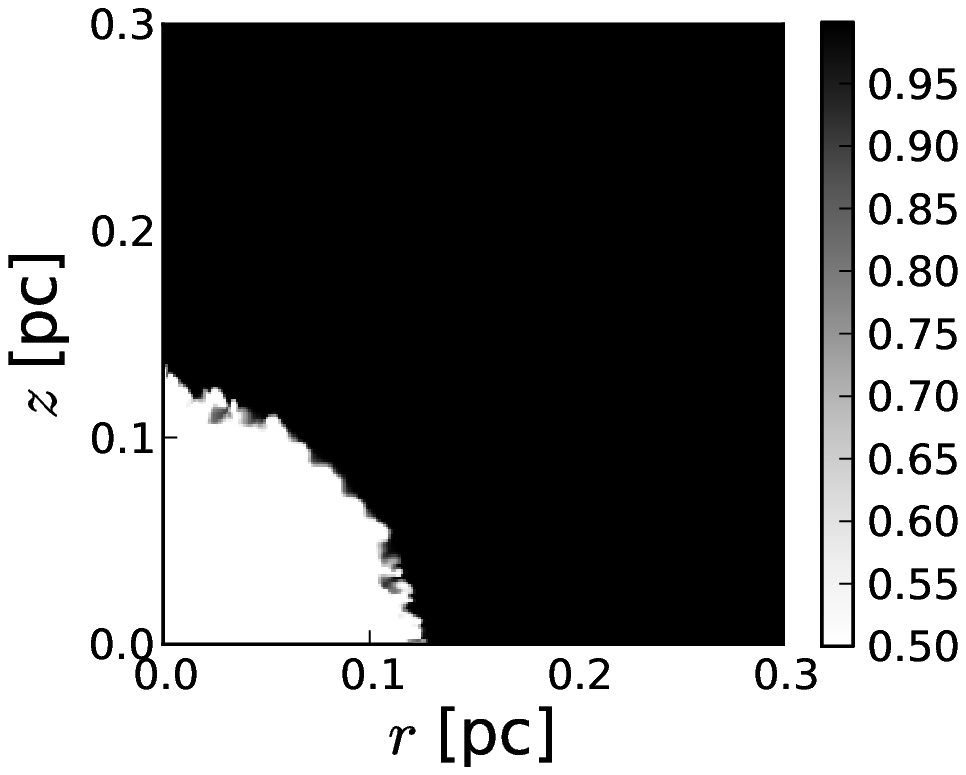}~
\includegraphics[width=0.25\linewidth]{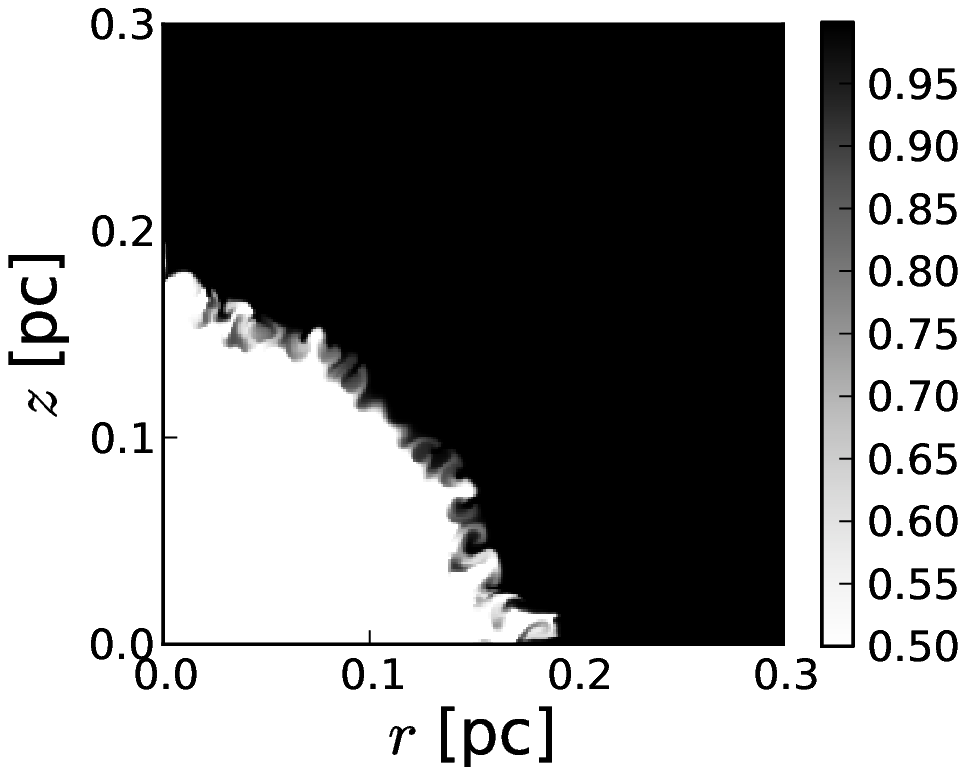}~
\includegraphics[width=0.25\linewidth]{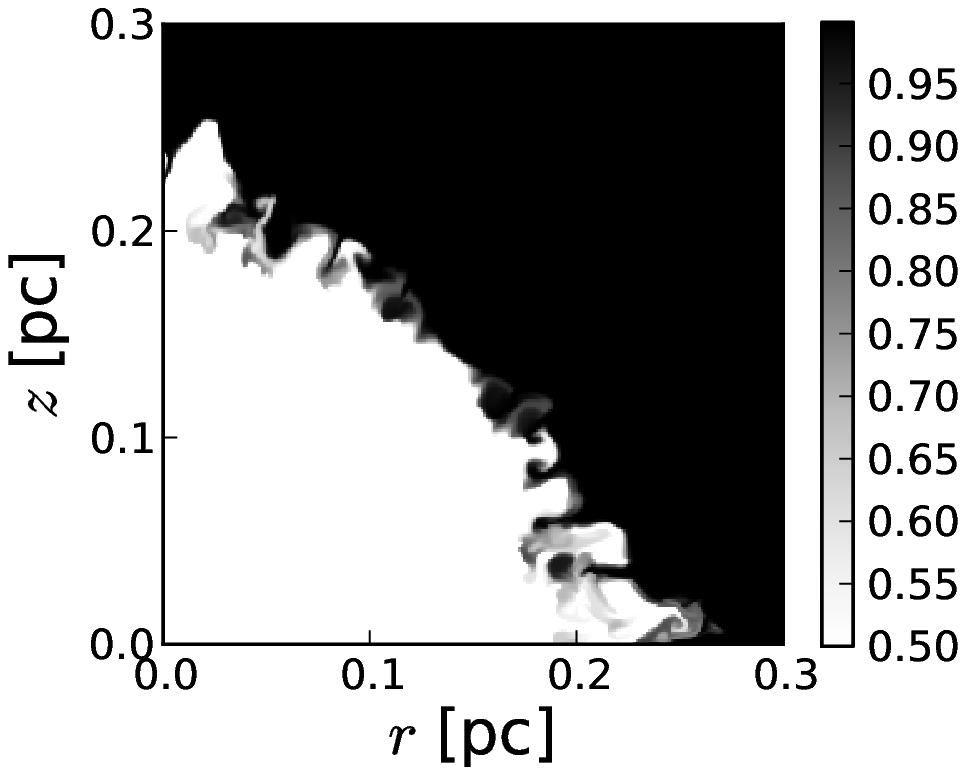}\\
\includegraphics[width=0.25\linewidth]{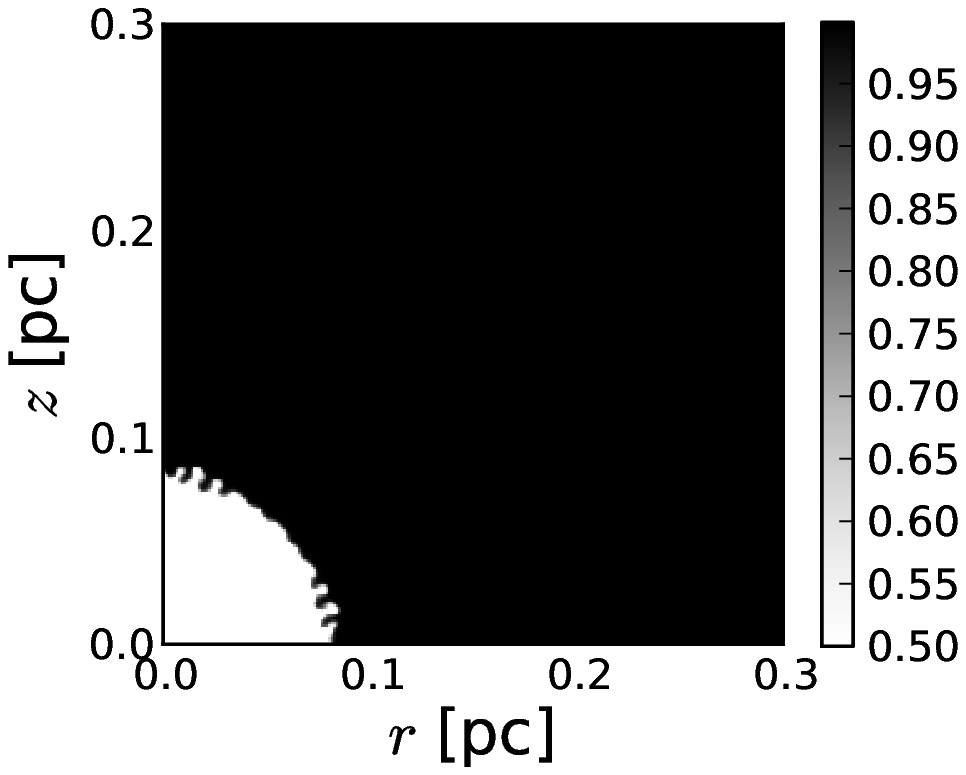}~
\includegraphics[width=0.25\linewidth]{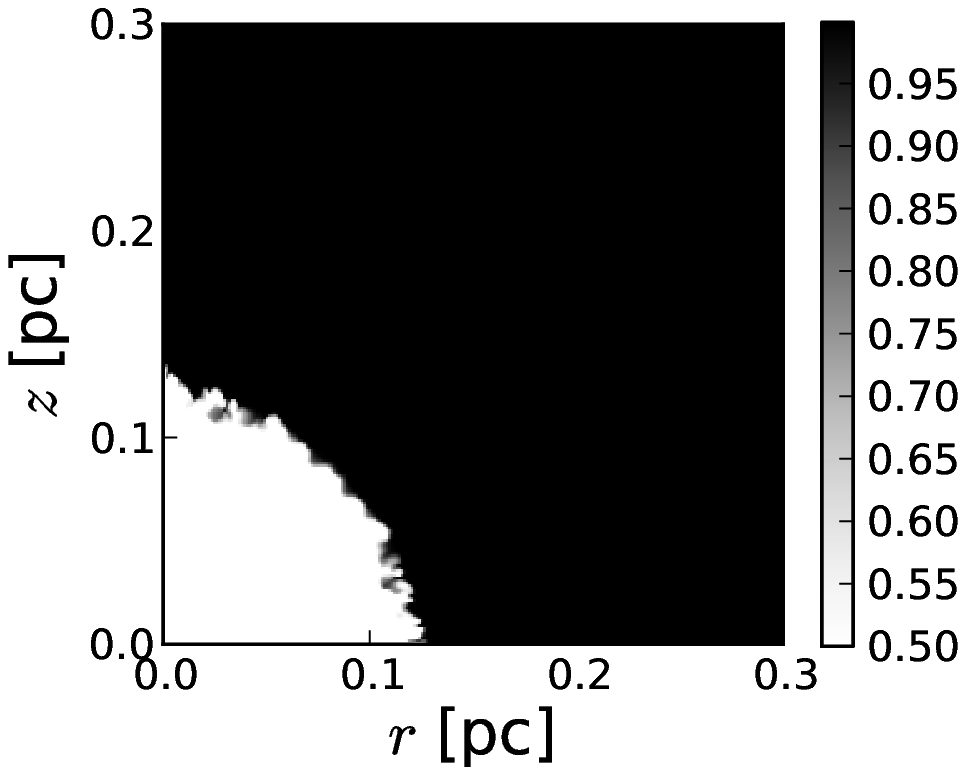}~
\includegraphics[width=0.25\linewidth]{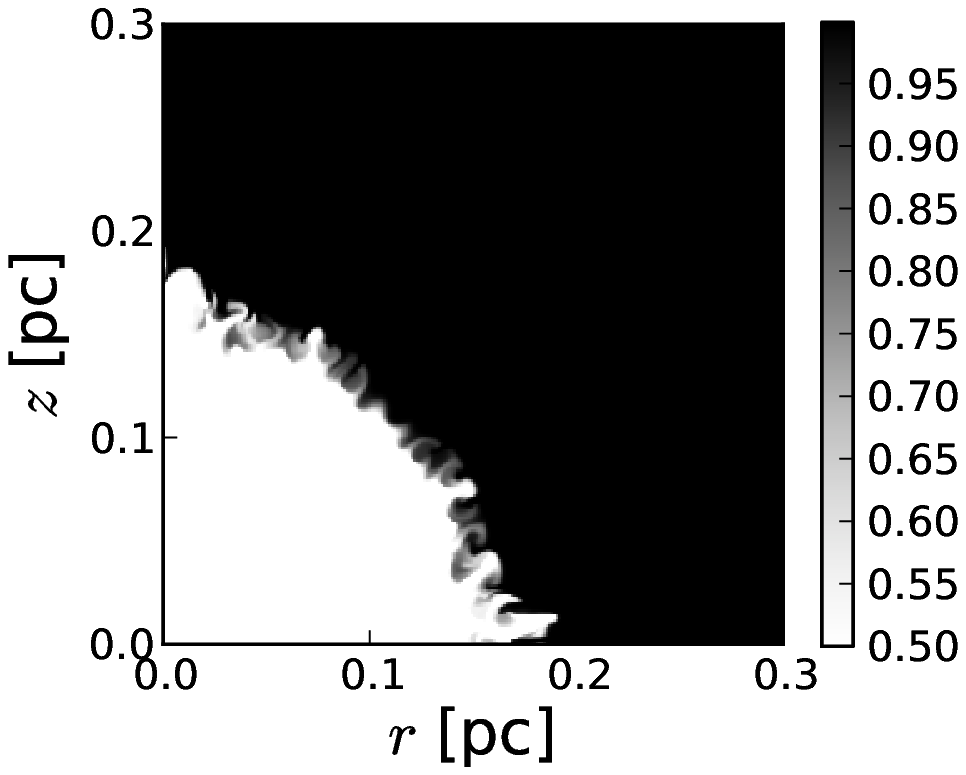}~
\includegraphics[width=0.25\linewidth]{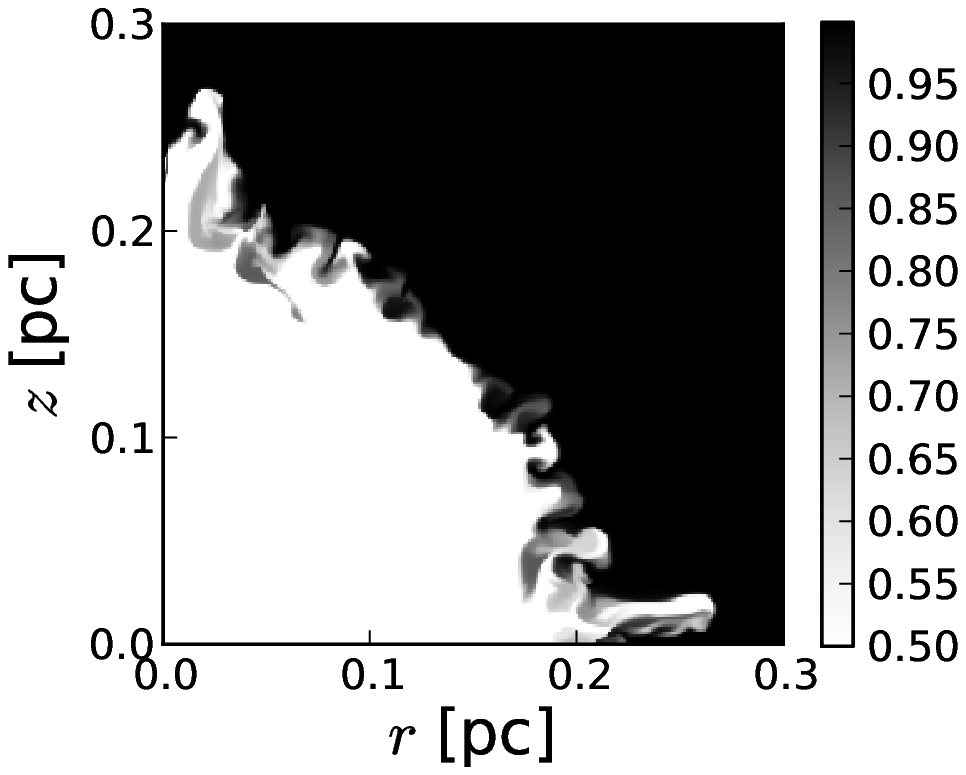}
\caption{Advected scalar $s$ value in the $r$-$z$ plane for the
  2.0-0.633 models. Top row: models without conduction; bottom row:
  models with thermal conduction. From left to right, the panels
  correspond to mean hot bubble radii 0.05, 0.10, 0.15, and 0.20~pc at
  1100, 2100, 3100, and 4100~yr after the onset of the post-AGB phase,
  respectively. Scalar values of $0.5<s<=1$ (shown in greyscale)
  correspond to predominantly nebular (i.e., expelled AGB envelope)
  material, while values of $0<=s<0.5$ (white) correspond to
  predominantly fast stellar wind material.}
\label{fig:scalar}
\end{figure*} 

We start from the 2D, axisymmetric, radiation-hydrodynamic numerical
simulation results of the formation and evolution of hot bubbles in
PNe from Paper I and calculate their corresponding synthetic X-ray
emission. We use the models corresponding to stellar initial masses
(M$_{\mathrm{ZAMS}}$) of 1, 1.5, 2, and 2.5~M$_{\odot}$ and consider
the cases with and without thermal conduction. These models correspond
to final white dwarf (WD) masses of 0.569, 0.597, 0.633, and
0.677~M$_{\odot}$. Models with initial masses of 3.5 and 5~M$_{\odot}$
are not used as they create only ionized-bounded PNe, which never
become optically thin \citep[see][Paper
  I]{Villaver2002b,Perinotto2004}. For simplicity, and consistency
with Paper I, we will refer to the models used here by a combination
of their initial mass and their final WD masses. For example, the
stellar model with initial mass of 1~M$_{\odot}$ and final WD mass of
0.569~M$_{\odot}$ will be labeled as 1.0-0.569. Thus, the labels of
the other models are 1.5-0.597, 2.0-0.633, and 2.5-0.677.

As an illustration of the density and temperature distributions that
resulted from the simulations presented in Paper I\footnote{While
  writing the present paper, we detected an error with our conduction
  subroutine and have recalculated all the models from Paper~I. The
  results are qualitatively similar and we present the new
  calculations here in Figures~\ref{fig:final_2D} and
  \ref{fig:final_2D_2}.}, we show in Figures~\ref{fig:final_2D} and
\ref{fig:final_2D_2} the total ion number density and gas temperature
in the $r$-$z$ plane for all the models, at the point where the hot
bubble in each model has an average radius of $\sim 0.2$~pc, except
for the 2.5-0.677 model which corresponds to the time when the hot
bubble has an averaged radius of 0.1~pc. As described in detail in
Paper~I (see Sections 4.1 and 5 of that paper), instabilities, which
develop early on in the interaction of the fast wind with the nebular
gas, play an important r\^{o}le in the subsequent evolution of the hot
bubble. Dense clumps and filaments resulting from the instabilities in
the swept-up shell present a large surface area. Heat diffusion from
the hot, shocked gas into the cooler clumps and filaments lowers the
temperature throughout the hot bubble and also results in hot
($T_\mathrm{e} \sim 10^6$~K), dense ($n_\mathrm{e} \sim
1.0$~cm$^{-3}$), evaporated nebular material expanding inwards into
the hot bubble (see Fig.~\ref{fig:final_2D_2}). Even in the absence of
conduction, photoevaporation and hydrodynamic ablation of the clumps
and filaments form turbulent mixing layers of shocked, heated nebular
gas ($T_\mathrm{e} \sim \mbox{a few} \times 10^6$~K) at the outer,
corrugated boundary of the hot bubble (see Fig.~\ref{fig:final_2D}).

To understand the evolution of the mixing layers, we use an advected
scalar, $s$, which is assigned the value of $s=1$ in the nebular
(i.e., expelled AGB envelope) material prior to the onset of the fast
wind. The fast wind is injected into the grid with a scalar value of
$s=0$. The scalar enables us to see how far the evaporated nebular
material penetrates into the hot bubble, in the models with
conduction, and the extent of the turbulent mixing layers, in the
models without conduction. Intermediate values of the scalar develop
as a result of mixing within the Eulerian cells of the 2D simulation,
particularly in regions with strong pressure variations, such as
cooling zones, and also in the region of hydrodynamic instabilities
(see Paper~I). Figure~\ref{fig:scalar} shows the evolution of the
scalar for the 2.0-0.633 models with and without thermal
conduction. The final panels in each row of Figure~\ref{fig:scalar}
should be compared to the corresponding panels in
Figures~\ref{fig:final_2D} and \ref{fig:final_2D_2}, when the mean hot
bubble radius is 0.2~pc. Other {initial-final} mass models are
similar.

Figure~\ref{fig:scalar} shows that at early times the distribution of
the scalar in the models with and without conduction is the same, and
shows how hydrodynamic instabilities break up the swept-up dense
shell. As the evolution proceeds, thermal conduction evaporates
material from the corrugated inner wall of the dense shell, which
begins to expand inwards into the hot, shocked wind bubble
(Fig.~\ref{fig:scalar}, lower right-hand panel). On the other hand,
the hydrodynamic mixing layers around the clumps and filaments in the
models without conduction are closely coupled to these structures and
do not expand greatly with time. In the panels of
Figure~\ref{fig:scalar}, computational cells containing mainly fast
stellar wind material ($s\le 0.5$) are coloured white, while those
containing mainly mixed nebular material ($s >0.5$) are shown with a
greyscale. Figure~\ref{fig:scalar} shows that without conduction, the
hot, shocked wind penetrates further into the dense, swept-up nebular
shell.

\begin{figure*}
\includegraphics[width=0.5\linewidth]{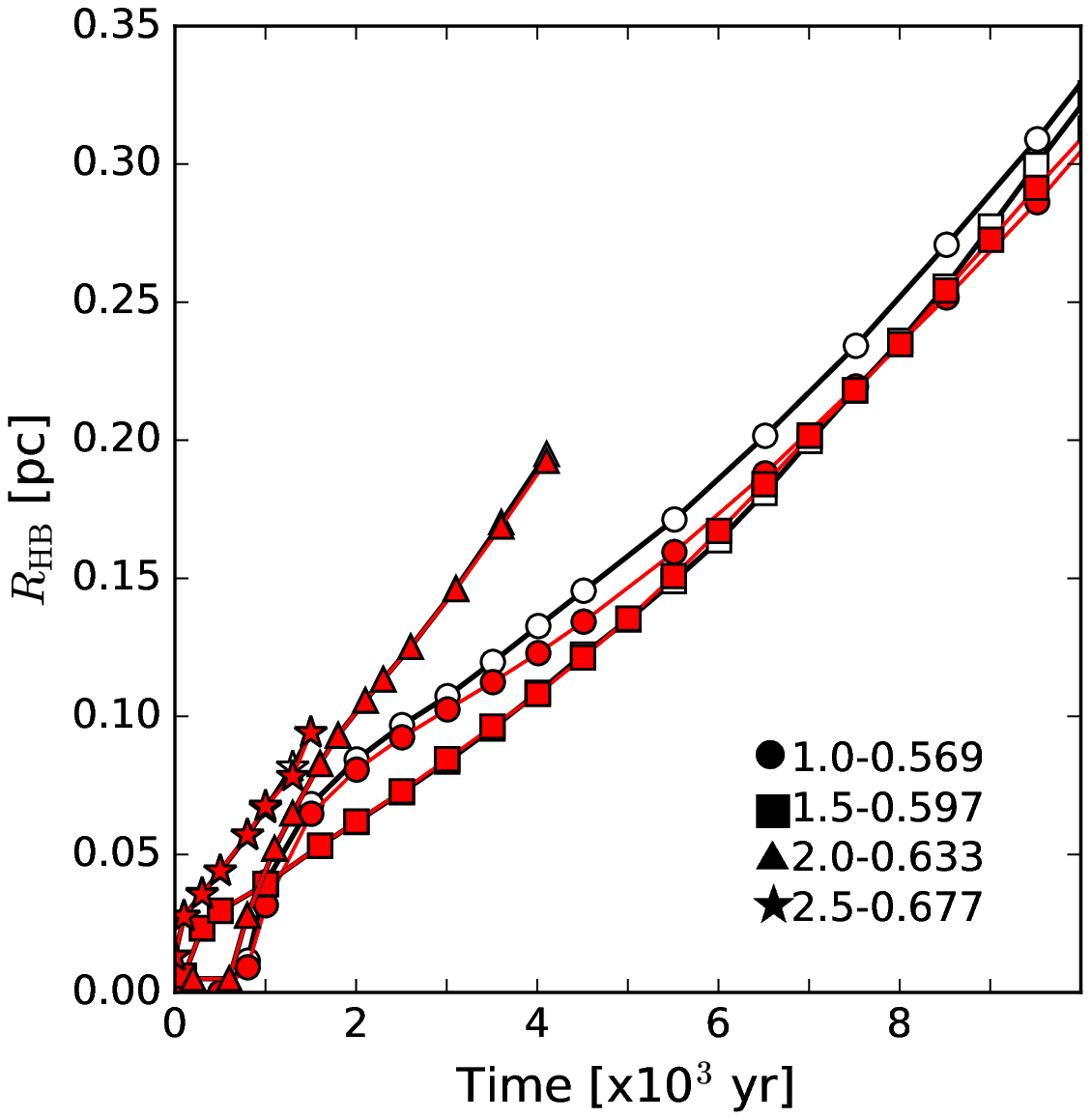}~
\includegraphics[width=0.5\linewidth]{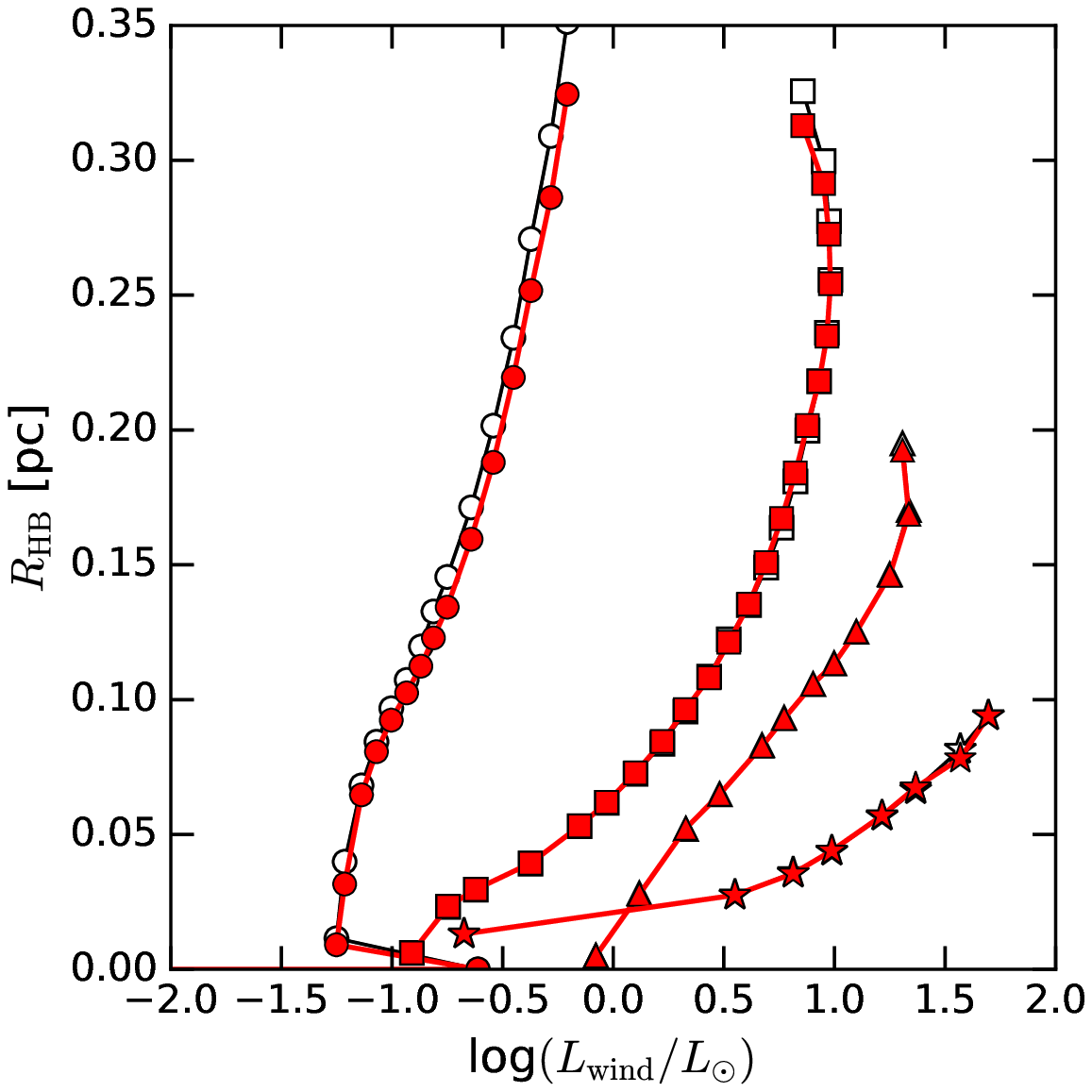}
\caption{Mean radius of the hot bubble ($R_\mathrm{HB}$) as a function
  of time (left panel) and as a function of stellar wind mechanical
  luminosity (right panel) for each stellar model, with and without
  thermal conduction. Filled symbols are for the models that include
  thermal conduction, while open symbols are for the models that do
  not.}
\label{fig:radhottime}
\end{figure*}

The expansion of the hot bubble can be characterized by its mean
radius as a function of time. This is shown in
Figure~\ref{fig:radhottime} for each of the models.  For a given mass
model, the hot bubble radii for the simulations with and without
conduction follow the same expansion except for the lowest mass models
(1.0-0.569). In these particular models, there is more radiative
cooling in the models with conduction and consequently less thermal
energy available to drive the expansion.

The hot bubbles continue to expand even after the stellar wind
mechanical luminosity begins to decrease. Unfortunately, we are not
able to follow the expansion of the 2.0-0.633 and 2.5-0.677 models
much beyond the turnover point in the wind luminosity. This is because
the ram pressure of the fast wind becomes smaller than the thermal
pressure of the hot bubble and the inner shock collapses back down
towards the star. This causes the computational timestep to become
extremely small and the calculation effectively comes to a halt.

A comparison with the expansion of equivalent 1D models, depicted in
Figure~\ref{fig:1D_hotbubble}, shows very similar behaviour. The
lowest mass models (1.0-0.569) expand slightly faster in the 1D case,
while the higher mass models expand slightly slower but there are no
great differences.

\subsection{Differential emission measure}
\label{sec:DEM}
\begin{figure*}
\includegraphics[height=8.3cm]{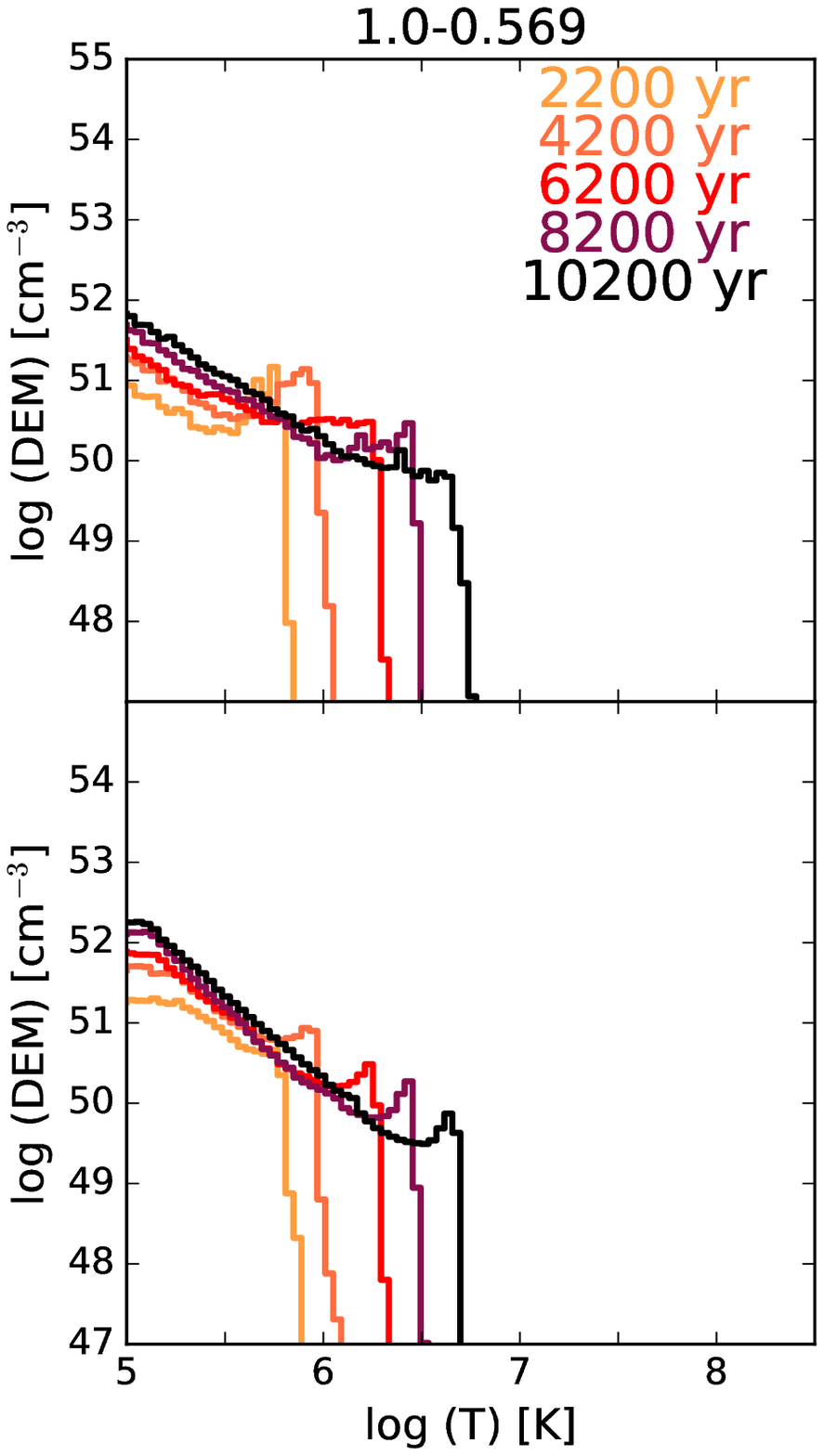}~
\includegraphics[height=8.3cm]{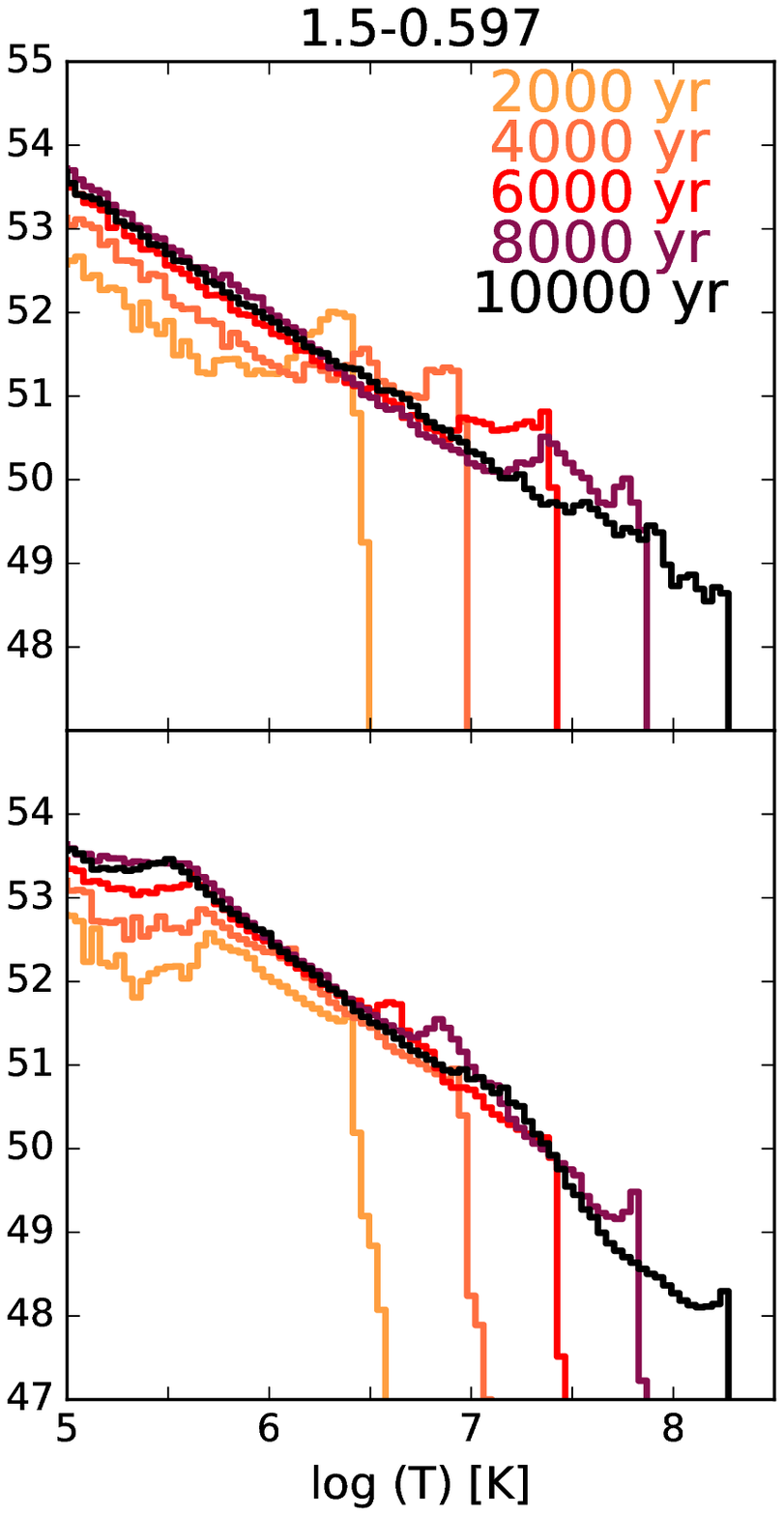}~
\includegraphics[height=8.3cm]{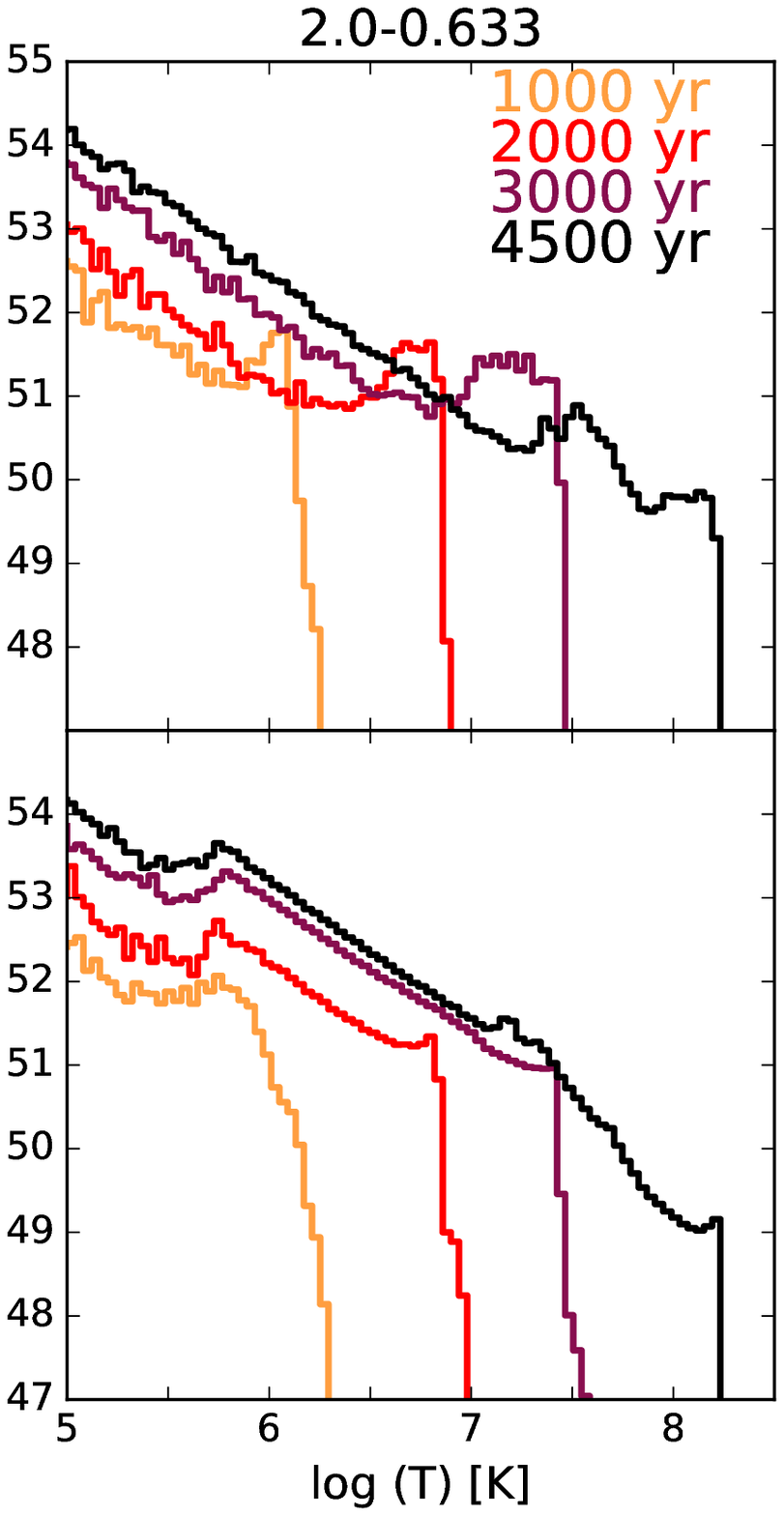}~
\includegraphics[height=8.3cm]{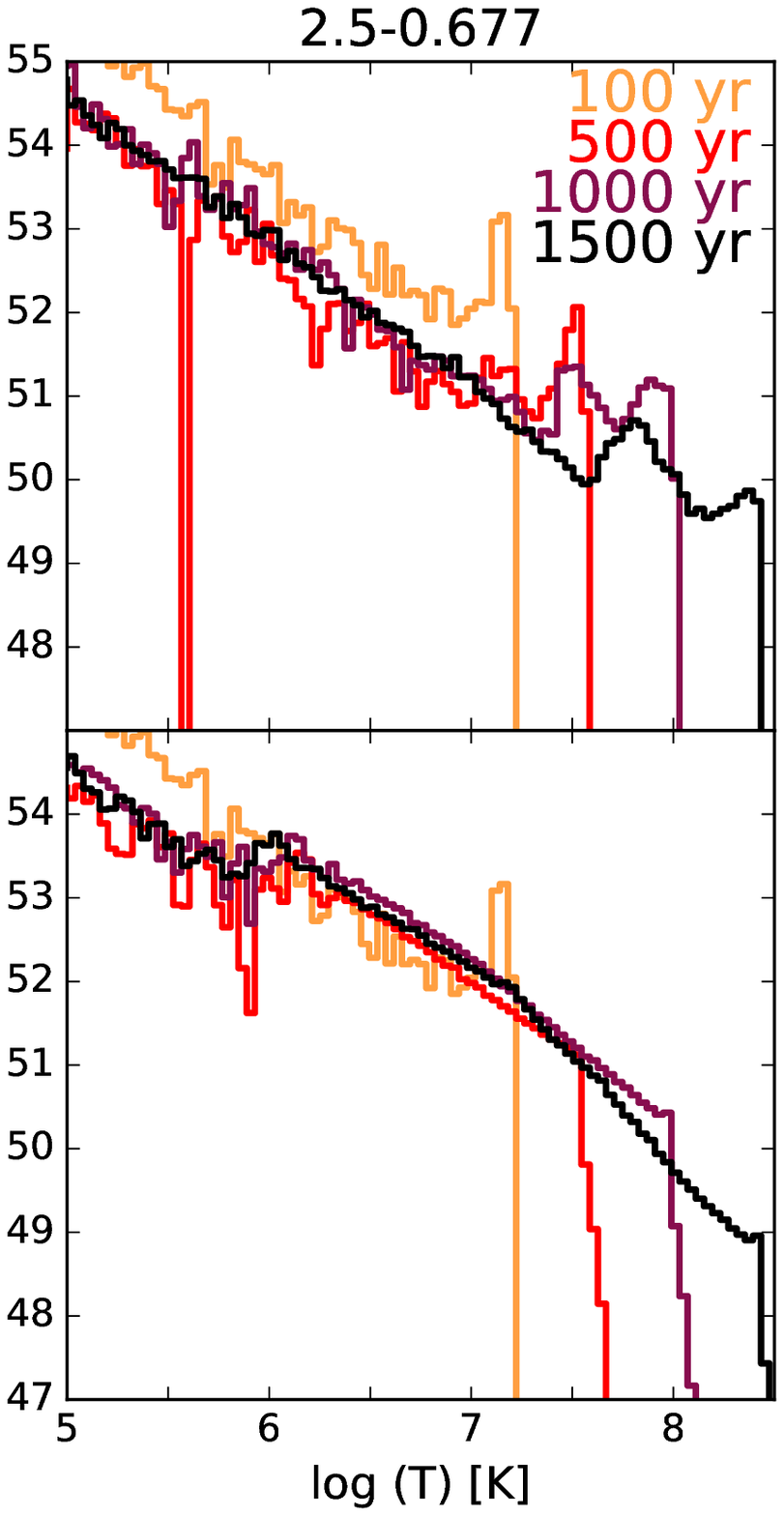}
\caption{Evolution with time of the differential emission measure
  (DEM) of hot bubbles for the four stellar models considered in this
  work. Upper and lower panels correspond to cases without and with
  thermal conduction, respectively. Times are marked with different
  colours.}
\label{fig:DEM_time}
\end{figure*}

The first step towards calculating the X-ray emission from our
simulations is to compute the differential emission measure (DEM),
defined in this work for a given temperature bin as
\begin{equation}
\mathrm{DEM}(T_\mathrm{b}) = \sum_{k, T_k \in T_\mathrm{b}} n_\mathrm{H} n_\mathrm{e} \Delta V_k,
\label{eq:DEM}
\end{equation}
\noindent where $n_{\mathrm{H}}$ and $n_\mathrm{e}$ are the hydrogen
vand electron number density in cell $k$, respectively, $\Delta V_k$ is
the volume of cell $k$ and the sum is performed over cells with gas
temperature falling in the bin whose central temperature is
$T_\mathrm{b}$.  In practice, we use logarithmic binning in the
temperature range $\log(T) = 5$ to $\log(T) = 9$ in intervals of
0.04~dex (i.e., 100 bins). Lower temperatures are not taken into
account in calculating the DEM, since their contribution to the X-ray
emission will be negligible.

As a result of the evolution hot gas inside the PN (i.e., the
temperatures, densities and volume), the DEM histogram of a given
model changes with time. We show in Figure~\ref{fig:DEM_time} the time
evolution of the DEM for all the models used in this work. The maximum
temperature reached by each DEM curve corresponds to the immediate
post-shock temperature at the given time. The increase in the maximum
temperature with time is a consequence of the increasing fast-wind
velocity. Note that all the inner stellar wind shocks in our models
are adiabatic and hence the post-shock temperature dependes only on
the square of the wind velocity and are independent of the shock
position. A different treatment of the conduction could lead to a
radiative inner-shock whose post-shock temperature is position
dependent \citep[e.g.,][]{Steffen2008}.

In particular, the case of the 1.0-0.569 model does not show a
contribution at temperatures above $10^{7}$~K because even after
$\sim10,000$~yr the post-AGB wind velocity is still $<800$~km~s$^{-1}$
and does not produce hotter gas (see Fig.~\ref{fig:wind_parameters} in
Appendix~A). For more massive models, higher stellar wind velocities
are reached on shorter time scales and so their DEM histograms have
significant contributions at temperatures $T\sim10^{8}$~K. It can be
also seen in Figure~\ref{fig:DEM_time} that as the stellar wind
parameters evolve, the relative contribution from the higher
temperatures diminishes because the mass-loss rate, and hence the
post-shock density are falling.

On the other hand, the DEM distributions of the models with thermal
conduction (Fig.~\ref{fig:DEM_time} bottom panels) show a plateau at
lower temperatures and the contribution at higher temperatures for the
late time evolution is lower than in their counterparts without
conduction. This is due to the diffusion of heat from the hot gas into
the surrounding cooler dense shell. In general, the models with
thermal conduction have slightly higher DEM values than those without
thermal conduction, particularly for the $10^6$~K gas. As a result,
the fluxes and X-ray luminosities of the models with thermal
conduction will be higher than in cases without conduction.

For comparison, we show in Appendix~\ref{sec:appb} the DEM
distributions obtained from 1D simulations for the 1.5--0.597 models
with and without conduction. The 1D case without conduction shows a
sparse, discontinuous DEM distribution, quite different to the 2D
case. The 1D case with conduction does not have such a large component
at temperatures $\log T < 6.0$ as its 2D counterpart because the
amount of gas in the conduction layer is smaller (in 2D the surface
area of the conduction layer is much larger due to the clumps and
filaments formed in these simulations).

\subsection{Plasma Temperatures}
\label{sec:plasmatemp}

An important quantity that can be readily obtained from the DEM
distributions is the average temperature of the hot gas
($T_\mathrm{A}$). Observations of diffuse X-ray emission generally
report a characteristic temperature ($T_\mathrm{X}$) for the emitting
gas in PNe, which is obtained by fitting a single temperature model to
the observed spectrum. The average temperature we calculate from our
models should be a reasonable estimate to the observationally derived
characteristic temperature if a single-temperature model is indeed
appropriate (e.g., $T_\mathrm{A} \approx T_\mathrm{X}$).

We define an emissivity weighted averaged temperature,
  $T_\mathrm{A} \equiv \langle T \rangle$,  as:
\begin{equation}
T_\mathrm{A}  = \frac{\int \epsilon(T) \mathrm{DEM}(T) T dT}{\int
  \epsilon(T) \mathrm{DEM}(T) dT} ,
\label{eq:avtemp}
\end{equation}
where $T$ is taken to be the electron temperature and
\begin{equation}
\epsilon(T) = \int_{E_1}^{E_2} \epsilon(T,E) dE 
\label{eq:emiscoef}
\end{equation}
is the emission coefficient integrated over the energy band $E_1 =
0.3$~keV to $E_2 = 2.0$~keV for the standard planetary nebula
abundances used in this work (see Table~1 in
Paper~I). Figure~\ref{fig:emiscoef} shows that, for standard planetary
nebula abundances, the emission coefficient is sharply peaked around
$\log(T) = 6.3$ and drops off very steeply towards lower
temperatures. Since, in general, the DEM(T) rise towards lower T, the
emissivity weighted average temperature will always be close to the
peak temperature of the emission coefficient.

\begin{figure}
\includegraphics[width=1.0\linewidth]{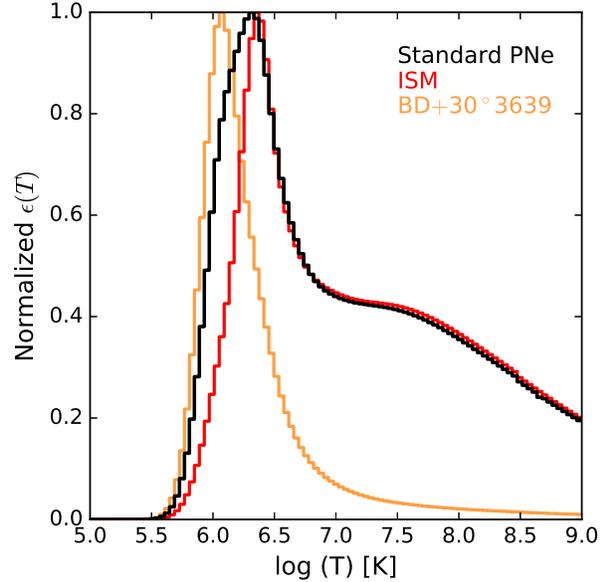}
\caption{Normalized emission coefficient $\epsilon(T)$ (see
  Eq.~\protect\ref{eq:emiscoef}) for the 0.3--2.0~keV energy range
  computed for the standard PN abundance set used in this work (black
  solid line). Other emission coefficients computed for different
  abundance sets are also shown.}
\label{fig:emiscoef}
\end{figure}

\begin{figure}
\includegraphics[width=1.0\linewidth]{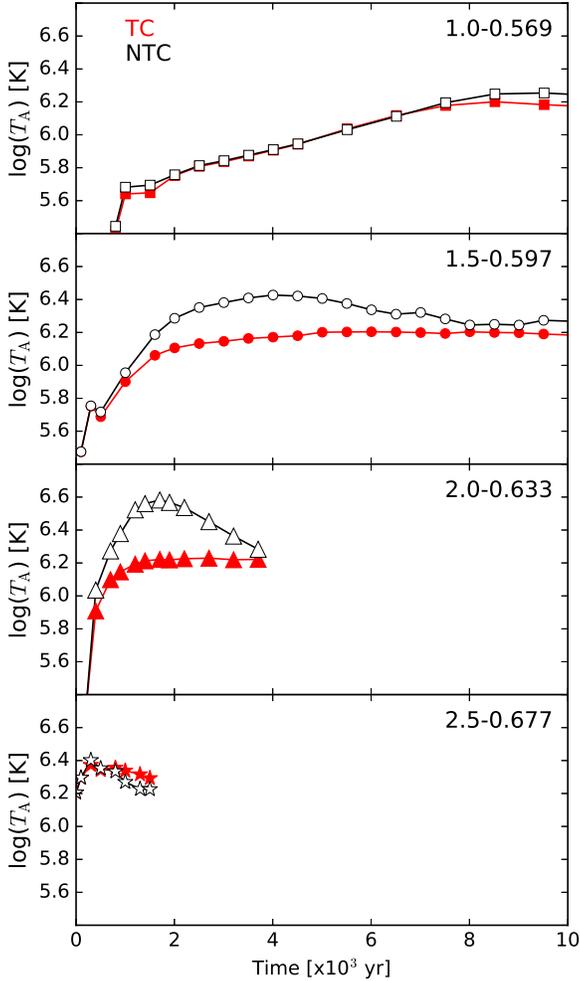}
\caption{Average plasma temperature $T_\mathrm{A}$ in the hot bubbles
  for different models. Filled and open symbols represent models with
  (TC) and without thermal conduction (NTC), respectively. The final
  plotted times correspond to log($T_\mathrm{eff}$)=4.49, 5.12, 5.18,
  and 5.02 for the 1.0-0.569, 1.5-0.597, 2.0-0.633, and 2.5-0.677
  models, respectively.}
\label{fig:aver_temp}
\end{figure}

Figure~\ref{fig:emiscoef} shows why it is not surprising that most
detections of diffuse X-ray emission from extended sources (PNe,
Wolf-Rayet bubbles, and stellar wind bubbles) report soft X-ray
temperatures of at most a few million degrees. For different abundance
sets, the behaviour of the emission coefficient is similar but with
different intensities: a narrow peak between 1--$3\times 10^6$~K that
falls off sharply towards both lower and higher temperatures. Lower
metallicity, such as for the ISM abundance set, moves the emission
coefficient peak towards higher temperatures ($\log T = 6.4$). The
peak is narrower and corresponds to 94\% of that the PN abundance set,
due to the fact that metals such as carbon and oxygen are less
important. Higher metallicity, like that of the [WC]-type central star
of BD$+$30$^{\circ}$3639 \citep[see][]{Marcolino2007}, leads to a
narrow peak at lower temperature ($\log T = 6.1$) and a steep fall off
towards higher temperatures since there is no hydrogen to produce
free-free emission above $\log T = 6.5$. This last case, has
coefficient intensities almost an order of magnitude higher (a factor
of 9.5) than the standard PN abundance set.

The average temperatures, $T_\mathrm{A}$, as a function of time for
all the models are plotted in Figure~\ref{fig:aver_temp}. As expected,
the average temperature is always close to the peak temperature of the
emission coefficient. Moreover, the average temperature values are in
agreement with those derived from observations of hot bubbles in PNe,
which generally fall in the range (1--3)$ \times 10^6$~K
\citep[see][]{Ruiz2013}. In general, we find that without conduction
(Fig.~\ref{fig:aver_temp} - open symbols), the average temperature
rises and then slowly decreases for all stellar models. The 2.0-0.633
model without conduction reaches the highest average plasma
temperature $T_\mathrm{A}\sim 4\times 10^6$~K of all the models
studied.
 
When thermal conduction is included, the average plasma temperature
rises steadily throughout the evolution, until a constant value of
$\log(T_\mathrm{A}) \sim 6.2$ (i.e., $T_\mathrm{A} \sim 1.6\times
10^6$~K) is reached, except for the highest mass model
(2.5-0677). This high-mass model evolves so quickly that the average
plasma temperature behaves in a similar way to its counterpart without
conduction.

The lowest mass model, 1.0-0.569, has an average temperature that
increases very slowly with time. This is because the CSPN fast wind
velocity in this model increases slowly as a function of time and even
after 10,000~yrs has only reached $\sim 600$~km~s$^{-1}$ (see
Fig.~\ref{fig:wind_parameters}). Consequently, even at the latest
times in the simulation there is little gas at or above the peak
temperature of the emission coefficent (see Fig.~\ref{fig:DEM_time})
and the average temperature derived from this model climbs slowly.

\begin{figure*}
\includegraphics[width=1.0\linewidth]{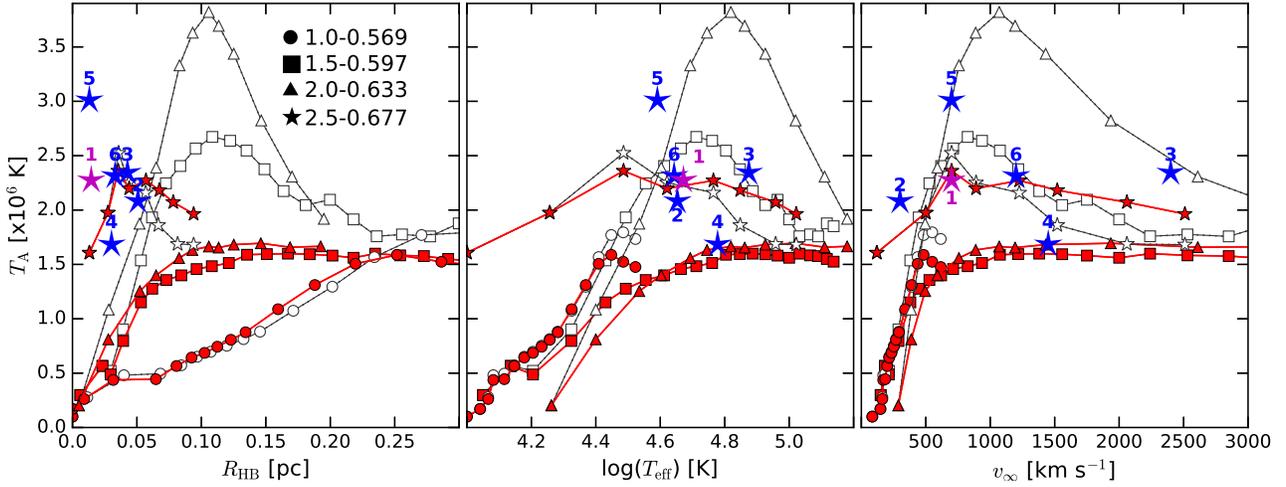}
\caption{Average plasma temperature as a function of hot bubble radius
  (left panel), stellar effective temperature (middle panel) and
  stellar wind velocity (right panel) for different models. Filled and
  open symbols are for the models with and without thermal conduction,
  respectively. The stars, labeled from 1 to 6, correspond to
  observations of BD$+$30$^{\circ}$3639, NGC\,2392, NGC\,3242,
  NGC\,6543, IC\,418, and NGC\,6826 respectively (see Table~1).}
\label{fig:avt-rad}
\end{figure*}

\begin{table*}
\caption{X-ray properties of selected PNe}
\centering
\begin{tabular}{clccccccccl}
\hline\hline\noalign{\smallskip}
\multicolumn{1}{c}{Label}&
\multicolumn{1}{l}{Object}&
\multicolumn{1}{c}{Hot Bubble}&
\multicolumn{1}{c}{distance}&
\multicolumn{1}{c}{$L_\mathrm{X}^\mathrm{a}$}&
\multicolumn{1}{c}{$F_\mathrm{X}^\mathrm{a}$}&
\multicolumn{1}{c}{$S_\mathrm{X}^\mathrm{a}$}&
\multicolumn{1}{c}{$kT_\mathrm{X}^\mathrm{a}$}&
\multicolumn{1}{c}{$T_\mathrm{eff}^\mathrm{b}$}&
\multicolumn{1}{c}{$v_{\infty}^\mathrm{b}$}&
\multicolumn{1}{l}{References$^\mathrm{e}$}\\
\multicolumn{1}{c}{}&
\multicolumn{1}{l}{}&
\multicolumn{1}{c}{Radius}&
\multicolumn{1}{c}{}&
\multicolumn{1}{c}{}&
\multicolumn{1}{c}{}&
\multicolumn{1}{c}{}&
\multicolumn{1}{c}{}&
\multicolumn{1}{c}{}&
\multicolumn{1}{c}{}&
\multicolumn{1}{l}{}\\
\multicolumn{1}{c}{}&
\multicolumn{1}{l}{}&
\multicolumn{1}{c}{(arcsec)}&
\multicolumn{1}{c}{(kpc)}&
\multicolumn{1}{c}{(cgs)}&
\multicolumn{1}{c}{(cgs)}&
\multicolumn{1}{c}{(cgs)}&
\multicolumn{1}{c}{(keV)}&
\multicolumn{1}{c}{(kK)}&
\multicolumn{1}{c}{(km~s$^{-1}$)}&
\multicolumn{1}{l}{}\\
\hline
\noalign{\smallskip}
1 & BD$+$30$^{\circ}$3639$^\mathrm{c}$ & 2.5 & 1.20 & 7.4$\times$10$^{32}$ & 4.3$\times$10$^{-12}$ & 2.2$\times$10$^{-13}$ & 0.20 & 47 & 700  & (1,2,3) \\
2 & NGC\,2392            & 8.2 & 1.28 & 1.8$\times$10$^{31}$ & 9.2$\times$10$^{-14}$ & 4.3$\times$10$^{-16}$ & 0.18 & 45 & 300  & (4,5) \\
3 & NGC\,3242            & 8.8 & 1.00 & 7.3$\times$10$^{30}$ & 6.1$\times$10$^{-14}$ & 2.5$\times$10$^{-16}$ & 0.20 & 75 & 2400 & (6,7,8) \\
4 & NGC\,6543            & 6.2 & 1.00 & 1.0$\times$10$^{32}$ & 8.0$\times$10$^{-13}$ & 6.5$\times$10$^{-15}$ & 0.15 & 50 & 1450 & (5,7,9) \\
\hline
5 & IC\,418$^\mathrm{d}$  & 2.2 & 1.20 & 8.4$\times$10$^{29}$ & 4.9$\times$10$^{-15}$ & 3.1$\times$10$^{-16}$ & 0.26 & 39 & 700  & (4,6,7) \\
6 & NGC\,6826$^\mathrm{d}$& 5.3 & 1.30 & 2.0$\times$10$^{30}$ & 9.9$\times$10$^{-15}$ & 1.1$\times$10$^{-16}$ & 0.20 & 44 & 1200 & (4,6,7) \\
\hline
\hline
\end{tabular}
\begin{list}{}{}
\item{$^{\mathrm{a}}L_\mathrm{X}$, $F_\mathrm{X}$, $S_\mathrm{X}$, and
  $kT_\mathrm{X}$ are the X-ray luminosity, unabsorbed flux, surface
  brightness, and plasma temperature as obtained from
  observations. $L_\mathrm{X}$, $F_\mathrm{X}$, $S_\mathrm{X}$ are
  presented in cgs units, that is, (erg~s$^{-1}$),
  (erg~s$^{-1}$~cm$^{-2}$), and (erg~s$^{-1}$~cm$^{-2}$~arcsec$^{-2}$),
  respectively.}
\item{$^{\mathrm{b}}T_\mathrm{eff}$ and $v_{\infty}$ are the stellar
    effective temperature and terminal wind velocity of the CSPN.}
\item{$^\mathrm{c}$PNe with [WC]-type CSPN.}
\item{$^\mathrm{d}$PNe with no strict fit to the data \citep[see][]{Ruiz2013}}
\item{$^\mathrm{e}$References: (1) \citet{Yu2009}, (2) \citet{Li2002},
  (3) \citet{Marcolino2007}, (4) \citet{Ruiz2013}, (5)
  \citet{Herald2011}, (6) \citet{Ruiz2011}, (7) \citet{Pauldrach2004},
  (8) \citet{Frew2008} and (9) \citet{Chu2001}.}
\end{list}
\label{tab:observations}
\end{table*}

The result that models with thermal conduction show very uniform
values of the mean temperature, both as a function of time and between
models for different initial stellar masses suggests that the
properties of the conduction layer are essentially independent of the
stellar parameters in our models. The variation and higher mean
temperatures of the models without conduction can be explained by
looking at the DEM plots (Fig.~\ref{fig:DEM_time}). The important
factor is the relative value of the DEM at higher temperatures to the
DEM around $\log(T) = 6$, which indicates the importance of the hotter
gas to the mean temperature. At early times, the increase in stellar
wind velocity, relatively high mass-loss rates and small volume of the
hot bubble mean that high temperature gas is a major contributor to
the DEM in the models without conduction, while at later times,
although the maximum temperature is higher, the contribution to the
DEM in these models is relatively smaller due to lower densities and
geometrical dilution and because the turbulent mixing layer around the
filaments consists mainly of lower temperature gas.

It is also interesting to compare the average plasma temperature
($T_\mathrm{A}$) as a function of other parameters, e.g., hot bubble
radius, stellar effective temperature, and terminal wind velocity.
Figure~\ref{fig:avt-rad} (left-hand panel) shows that small bubbles
($R_\mathrm{HB}<0.05$~pc) can have a variety of mean plasma
temperatures, $0.4\times 10^6 < T_\mathrm{A} < 2.6\times 10^6$~K,
depending on the stellar mass model. On the other hand, large bubbles
($R_\mathrm{HB} > 0.2$~pc) all have temperatures in the narrow range
$1.5\times 10^6 < T_\mathrm{A} < 1.8\times 10^6$~K \citep[see figure~6
  in][]{Ruiz2013}. In our simulations, average temperatures above
$T_\mathrm{A} > 2.5\times10^6$~K are only found for models without
conduction and intermediate hot bubble radii. Furthermore, the
comparisons with effective temperature and terminal wind velocity
(Fig.~\ref{fig:avt-rad}, central and right-hand panels) show that the
largest values of the average temperature in the models without
conduction are not associated with either the maximum stellar
effective temperature or the peak wind power (see also
Fig.~\ref{fig:wind_parameters}).

We include in Fig.~\ref{fig:avt-rad} the corresponding derived
characteristic temperatures for the X-ray observations of objects
listed in \citet{Ruiz2013} (numbered star symbols; see Table~1 and
references therein). One of these points (numbered 1) corresponds to
the PN BD$+$30$^{\circ}$3639, which is included as it correponds to
the best-quality observation among all reported PNe to date even
though it has a [WC]-type CSPN \citep[e.g.,][]{Yu2009}. Other PNe with
[WR]-type CSPN have been left outside\footnote{Note that we do not
  include in our table NGC\,7009 and NGC\,7027. The analysis of the
  X-ray emission reported so far for NGC\,7009 was performed with {\it
    XMM-Newton} observations \citep[]{Guerrero2002} and did not have
  the spatial resolution to separate the CSPN emission from the
  diffuse X-ray emission \citep[see figures 3 and 4
    in][]{Kastner2012}. On the other hand, we do not use the derived
  X-ray properties of NGC\,7027 since extinction within the nebula
  precludes a clear analysis of the hot bubble.}. We remark that our
estimated surface brightness for IC\,418, NGC\,2392, and NGC\,6826
differ from those listed by \citet{Ruiz2013} for a factor of
$\sim$4. The authors used erroneously the diameters instead of the
radii of the PNe to calculate the surface brightness (M.A.\,Guerrero
private communication). It is also important to bear in mind that the
X-ray properties of IC\,418 and NGC\,6826 were obtained without
performing a rigorous fit to the data. Other X-ray properties of the
PNe listed in our table have been updated.

Figure~\ref{fig:avt-rad} (centre panel) shows that our models with and
without conduction span the range of mean plasma temperatures derived
from observations, when plotted against the stellar effective
temperature or stellar wind velocity. On the other hand, the
observational points appear to favour the highest mass models when the
size of the hot bubble is taken into consideration, since these
correspond to the smallest bubbles. We note that all of the brightest
observed PNe showing diffuse X-rays have bubbles with radii
$R_\mathrm{HB} < 0.05$~pc.

In Appendix~\ref{sec:appb} we show the average temperatures of the 1D
simulations of the 1.5--0.597 models with and without conduction both
as a function of time and as a function of stellar effective
temperature (see Fig.~\ref{fig:1D_avtemp}). The average temperatures
of the 1D simulations without conduction climb to $T_\mathrm{A} \sim
10^7$~K, very different to the 2D case, because there is very little
gas around the temperature corresponding to the peak in the emission
coefficient (see Fig.~\ref{fig:1D_DEM}). The 1D simulations with
thermal conduction reach slightly higher average temperatures than
their 2D counterparts, because the thermal conduction layer is not so
extensive in the 1D models due to the absence of corrugations.

\subsection{Spectra}
\label{sec:spec1}

The synthetic X-ray emission for each model at each time is computed
using the extensively tested {\sc chianti} database and software
package \citep[Version 7.1.3;
][]{OriginalChianti,Dere2009,Landi2013}. The calculation performed by
{\sc chianti} includes the contributions due to lines and to
free-free, free-bound, and two-photon continua. We used the updated
default ionization equilibrium ion fractions included in the latest
version of {\sc chianti}\footnote{see:
  http://www.chiantidatabase.org/cug.pdf} \citep[see][and references
therein]{Landi2013}. The abundances used to compute the spectra are
those defined for a standard PN composition in Cloudy version 13.0
\citep{Ferland2013}, for consistency with Paper~I. A spectrum is
generated for each of the temperature bins and weighted by the
appropriate DEM value. The complete spectrum is simply the sum of all
these individual contributions.

All synthetic spectra are computed in the 0.3--2.0~keV energy range,
corresponding to 6--40~\AA\, in the wavelength range, with a spectral
bin of 0.01~\AA\, and the spectral line FWHM is assumed to be 0.1~\AA.

\subsection{Absorption}
\label{sec:absorp}
To facilitate direct comparison with observations, the intrinsic
spectra produced with the {\sc chianti} software should be corrected
for absorption by neutral material along the line of sight. Extinction
is particularly important for energies $< 0.5$~keV and can reduce the
spectral intensity by an order of magnitude or more.

We use the the photoelectric absorption cross sections from
\citet{Balucinska1992} and standard ISM abundances. We consider two
values for the atomic hydrogen column density: a moderate value
$N_\mathrm{H} = 8\times 10^{20}$~cm$^{-2}$, and a high value
$N_\mathrm{H} = 5\times 10^{21}$~cm$^{-2}$. These values are within
the range derived from observations \citep[e.g.,][]{Ruiz2013}. Since
the spectra of PNe are generally quite soft, extinction plays an
important r\^{o}le in the shape of the spectrum at lower
energies. Even so, there are some PNe for which an important count
rate has been measured at low energies \citep[e.g., NGC\,6543 and
BD\,$+$30$^{\circ}$3639;][]{Kastner2000,Chu2001,Yu2009}.

Finally, the absorbed spectrum should be convolved with the
instrumental response matrices. Convolution with the
\textit{XMM-Newton} EPIC-pn and \textit{Chandra} ACIS instrumental
responses and effective areas can be done within the {\sc chianti}
software after the absorption has been applied.

\section{Results}
\label{sec:results}

\subsection{Synthetic X-ray spectra}
\label{sec:synthetic}

\begin{figure}
\includegraphics[width=1.0\linewidth]{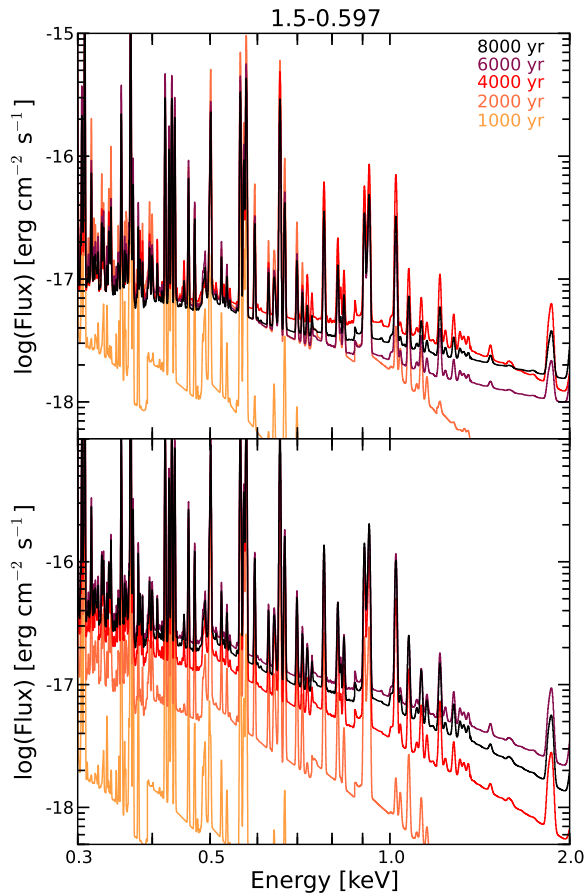}
\caption{Time evolution of the spectra in the 0.3-2.0~keV energy range
  for the 1.5-0.597 model at different times after the onset of the
  post-AGB phase. Top: Without thermal conduction. Bottom: With
  thermal conduction. Colours represent different times.}
\label{fig:spectra1}
\end{figure}

As an example of the resultant synthetic X-ray spectra obtained with
{\sc chianti} using the procedure described in \ref{sec:spec1},
Figure~\ref{fig:spectra1} shows total intrinsic (i.e., non-absorbed)
spectra for the 1.5-0.597 model with and without thermal conduction at
different times after the onset of the post-AGB phase. The spectra are
plotted in the energy range 0.3--2.0~keV as this corresponds to the
range generally presented in observational studies of the diffuse
X-rays from PNe \citep[e.g.,][and references therein]{Ruiz2013}. All
spectra show a great number of spectral lines but there are noticeable
differences between models with and without thermal conduction.

The case without thermal conduction (Figure~\ref{fig:spectra1}, top
panel) shows an evolution in the slope of the continuum with
time. This follows from the time evolution of the DEM distributions
seen in Fig.~\ref{fig:DEM_time} and the average temperature seen in
Fig.~\ref{fig:aver_temp}. We have seen how, for the 1.5-0.597 model,
the average temperature increases gradually to reach a maximum around
4000~yrs after the end of the AGB stage and then decreases. The
changing average temperature is responsible for the changing slope of
the continuum. Furthermore, from Fig.~\ref{fig:DEM_time} we see that
the general trend is for the values of the DEM distribution to
increase with time. This is responsible for the general increase in
the fluxes as time progresses.

\begin{figure}
\includegraphics[width=1.0\linewidth]{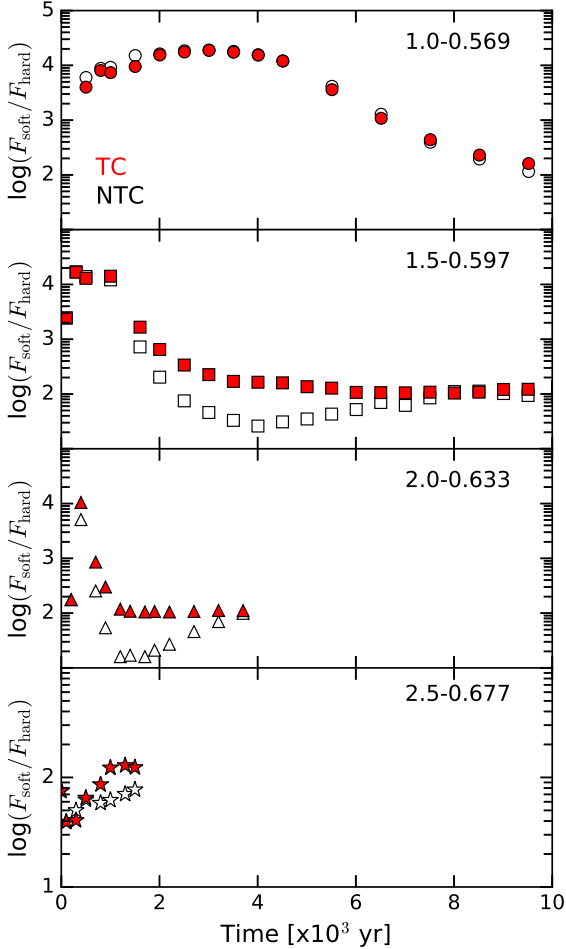}
\caption{Variation of the ratio of soft (0.3-1.0~keV) to hard
  (1.0--2.0~keV) X-ray fluxes with time for all models. Filled and
  open symbols represent results with (TC) and without thermal
  conduction (NTC), respectively.}
\label{fig:lum_ratio}
\end{figure} 

On the other hand, the average temperature for the 1.5-0.597 model
with conduction remains constant after 3000~yrs (see
Fig.~\ref{fig:aver_temp}). For this reason, the slope of the continuum
does not change in Fig.~\ref{fig:spectra1}. The fluxes increase
because the DEM values increase as time advances.

The results for the 1D simulations of the 1.5--0.597 model are given
in Appendix~\ref{sec:appb}. Although the simulated spectra for the 1D
cases with thermal conduction show similar behaviour to their 2D
counterparts, the results for the cases without conduction are very
different. Due to the very high average temperature ($T_\mathrm{A}
\sim 10^{7}$~K), the spectral shape of the 1D results without
conduction is essentially flat in the 0.3--2.0~keV energy range.

Another way of looking at the evolution of the spectral shape is to
consider the hardness ratio. This is the ratio of the soft-band
(0.3--1.0~keV) luminosity to the hard-band (1.0--2.0~keV) luminosity.
In Figure~\ref{fig:lum_ratio}, we show the variation of the hardness
ratio as a function of time for all models. The hardness ratio
reflects the behaviour of the average temperature, $T_\mathrm{A}$ (see
Fig.~\ref{fig:aver_temp}): for lower average temperatures the spectrum
will be softer. The models with conduction generally have softer
spectra than those without (except for the highest mass model
2.5-0.677), which is consistent with their average temperatures being
lower. The hardness ratios of the models with conduction show less
variation than those without conduction because the conduction layer
dominates the X-ray emissivity and is weighted toward lower
temperatures (see Fig.~\ref{fig:DEM_time}). For the models without
conduction, the spectra are hardest when the average temperature is
highest, which occurs while the hot bubble is still reasonably compact
and the DEM distribution has an important contribution from high
temperature gas.

At early times, the soft band dominates because the plasma average
temperatures are low as a consequence of the low stellar wind
velocity. As time goes on, the wind velocity, and hence the plasma
average temperature, increases and the contribution to the hard X-ray
band becomes more important. The models without conduction reach
higher average temperatures than the models with conduction, and this
is evident by the low values of the hardness ratio
$F_\mathrm{soft}/F_\mathrm{hard}$, which correspond exactly to the
peak in $T_\mathrm{A}$ seen in Figure~\ref{fig:aver_temp}. On the
other hand, for the most massive model 2.5-0.677, the stellar wind
evolves so quickly that the properties of the mixing layer dominate
the spectra from early times.

\subsection{Luminosities}
\label{sec:lum}
\begin{figure*}
\includegraphics[height=9.1cm]{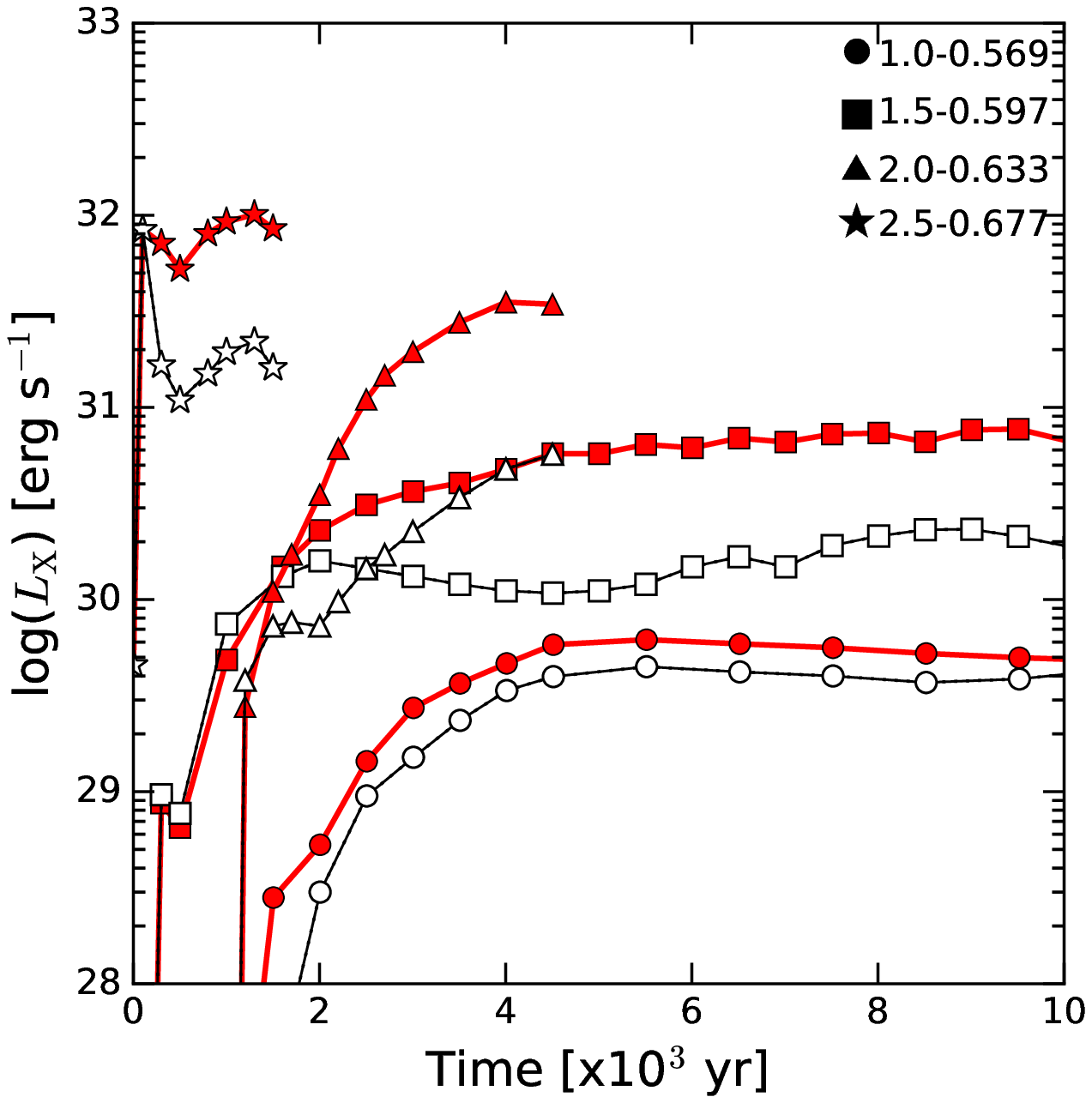}~
\includegraphics[height=9.1cm]{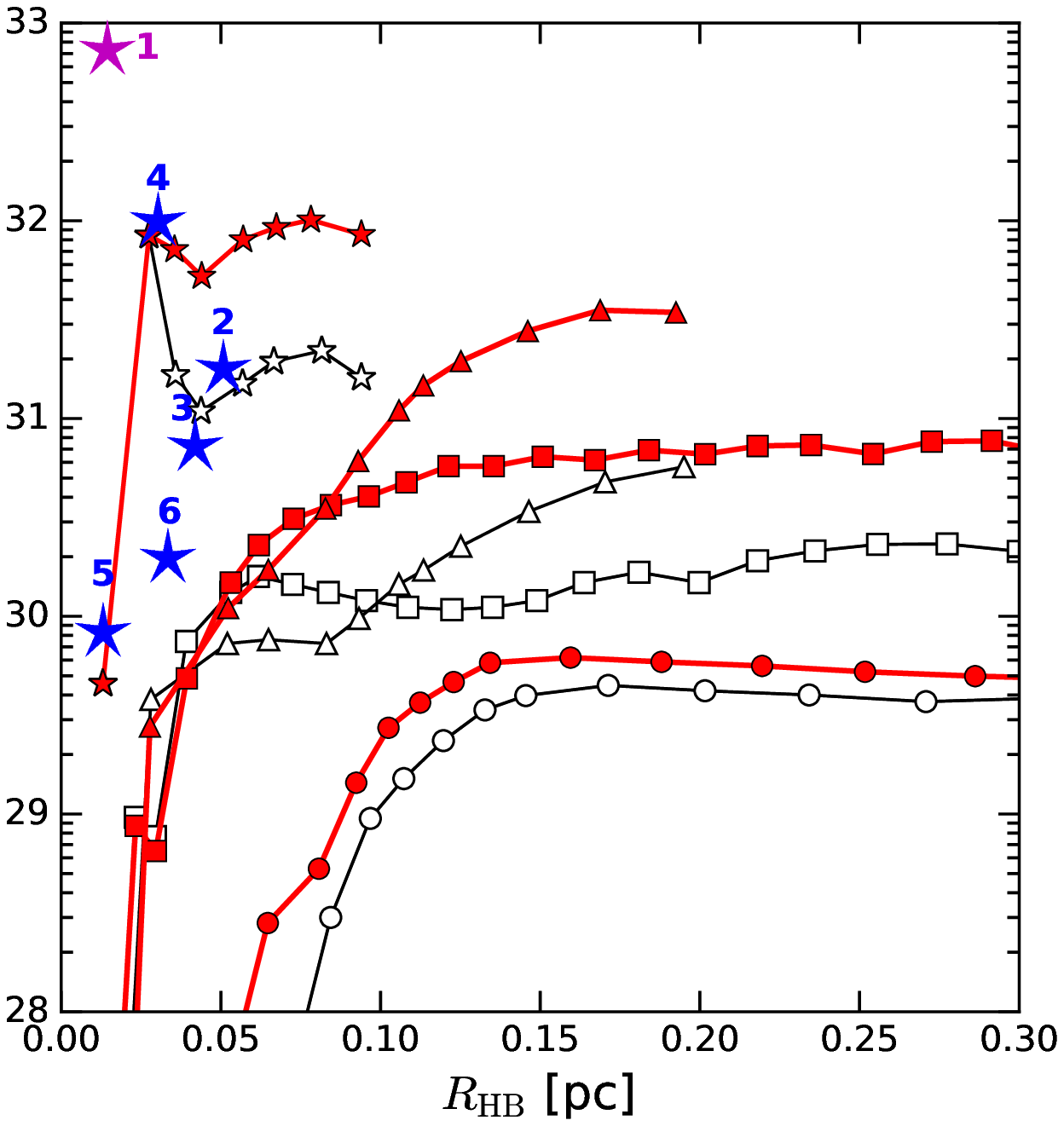}\\
\includegraphics[height=9.1cm]{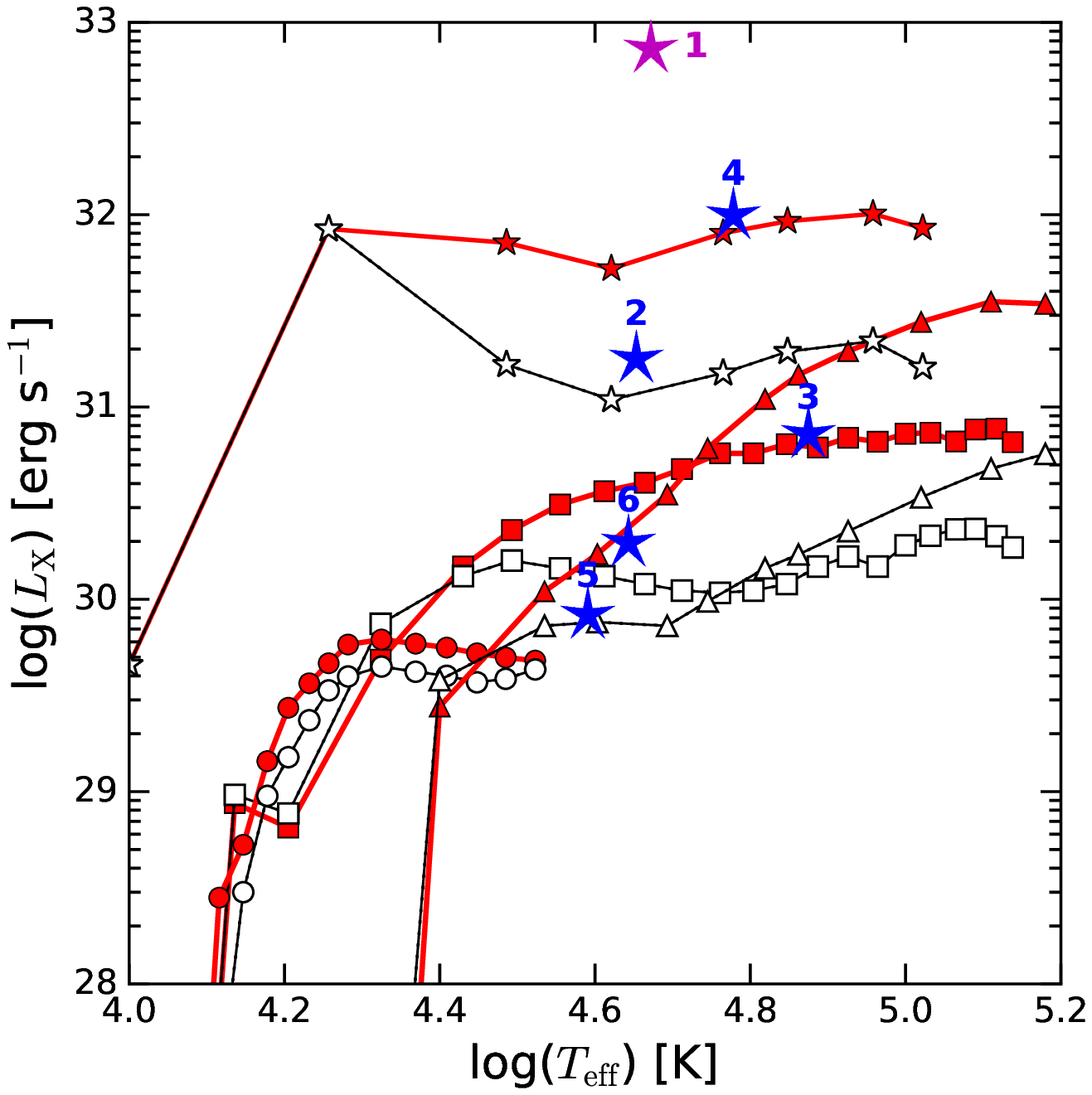}~
\includegraphics[height=9.1cm]{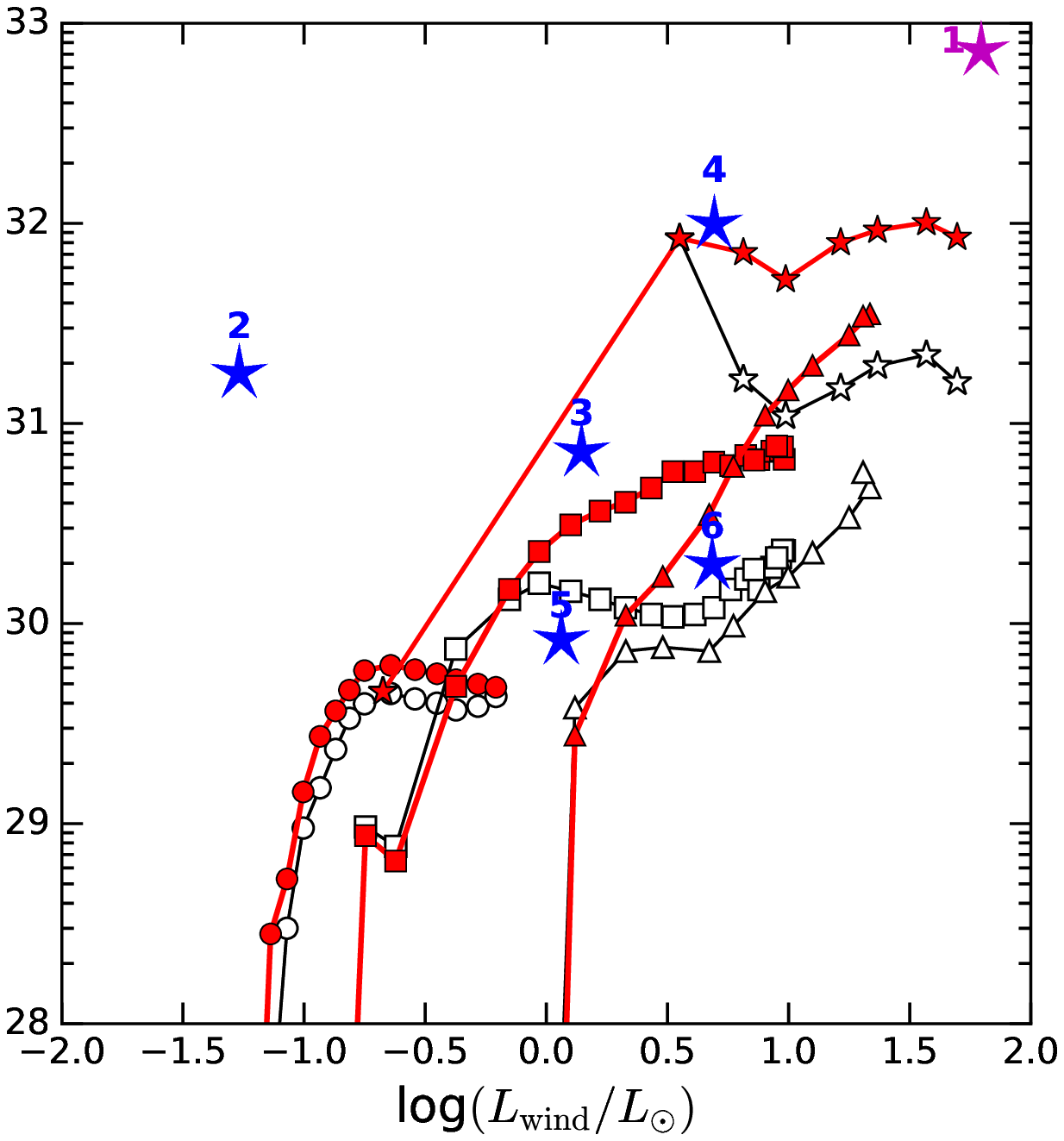}
\caption{Evolution of the X-ray luminosity in the 0.3--2.0~keV energy
  range as a function of time (upper left panel), hot bubble radius
  (upper right panel), stellar effective temperature (lower left
  panel), and stellar wind mechanical luminosity (lower right
  panel). Filled and open symbols represent models with and without
  thermal conduction, respectively. Stars, labeled from 1 to 6,
  correspond to observations of BD$+$30$^{\circ}$3639, NGC\,2392,
  NGC\,3242, NGC\,6543, IC\,418, and NGC\,6826 respectively (see
  Table~1).}
\label{fig:lum_time}
\end{figure*}

The X-ray luminosities, $L_{\mathrm{X}}$, in the 0.3--2.0~keV energy
band are found by integrating the synthetic spectra over this energy
range. In Figure~\ref{fig:lum_time} top left panel we show the time
evolution of the X-ray luminosity for all models used in the present
paper. The first thing to notice is that all models result in a wide
range of X-ray luminosities within the X-ray luminosity range reported
by observations (see Table~\ref{tab:observations}). As anticipated,
models with thermal conduction have higher X-ray luminosities (by up
to an order of magnitude in the case of the two most massive models)
compared to the corresponding models without thermal conduction.

In all our models, the luminosity behaviour seen in
Figure~\ref{fig:lum_time} can be understood with reference to the DEM
histograms shown in Figure~\ref{fig:DEM_time}. In particular, the DEM
value around $T \sim T_\mathrm{A} \sim 1.5 \times 10^6$~K shows that
for the lower mass models 1.0-0.569 and 1.5-0.597, the luminosity
rises at first then stays more-or-less constant as the hot bubble
continues to expand for the cases both with and without
conduction. For the 2.0-0.633 models DEM value at the average plasma
temperature rises continually for the duration of the simulation. The
highest mass models 2.5-0.677 have high luminosity even from early
times.

The luminosity as a function of hot bubble radius
(Fig.~\ref{fig:lum_time}, upper right panel) shows that the less
massive models 1.0-0.560 and 1.5-0.597 show little variation in their
luminosities over the radius range $0.05 < R_\mathrm{HB} <
0.2$~pc. Only the most massive model (2.5-0.677) is capable of
producing very small ($R_\mathrm{HB} < 0.05$~pc), highly X-ray
luminous bubbles. Unfortunately, we cannot follow the evolution of
these models at larger sizes. This figure also plots the
observationally derived luminosities and radii for the six objects
listed in Table~\ref{tab:observations}. Interestingly, all these
objects have very small diffuse X-ray bubbles.

\begin{figure*}
\includegraphics[height=9.1cm]{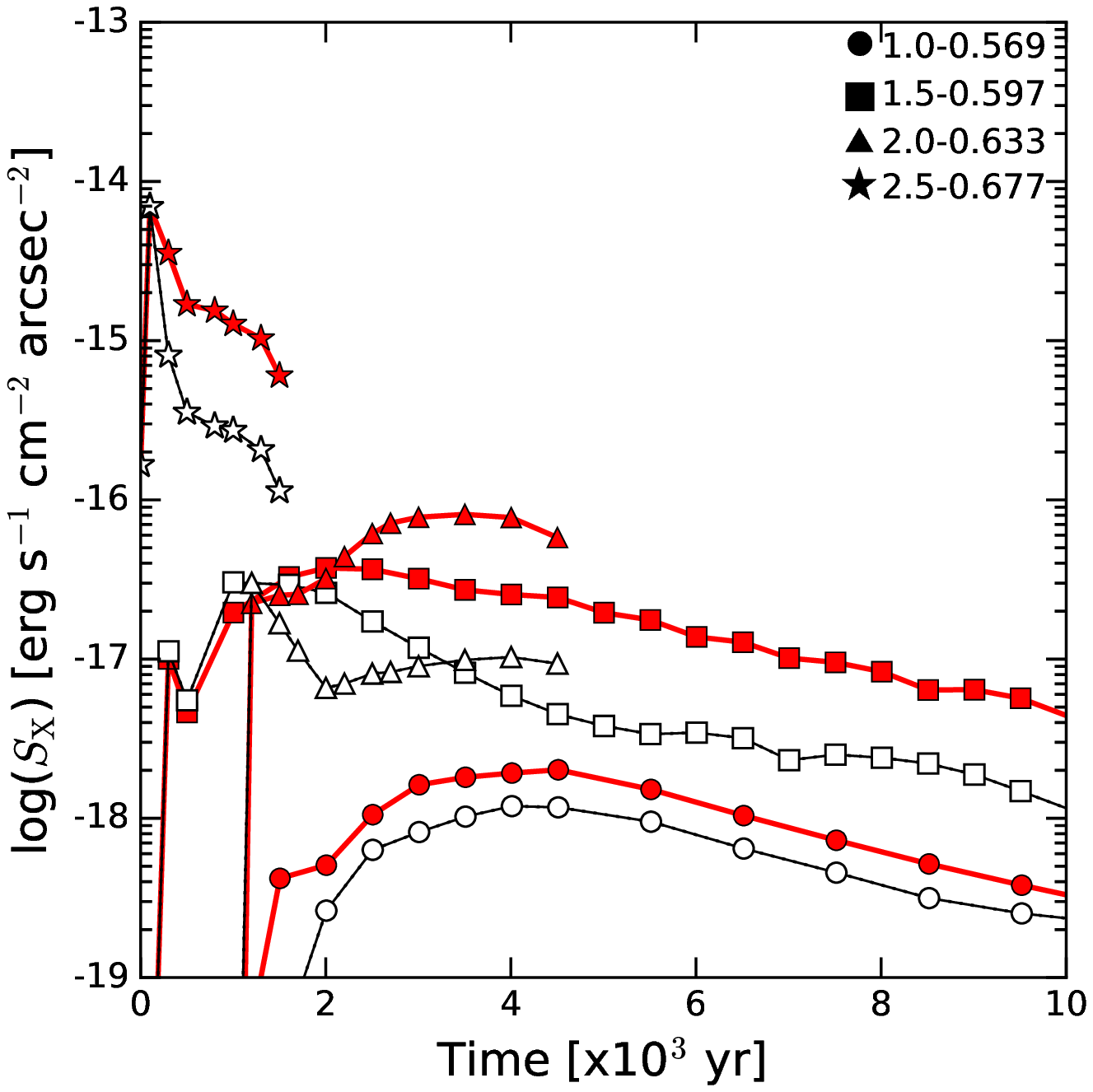}~
\includegraphics[height=9.1cm]{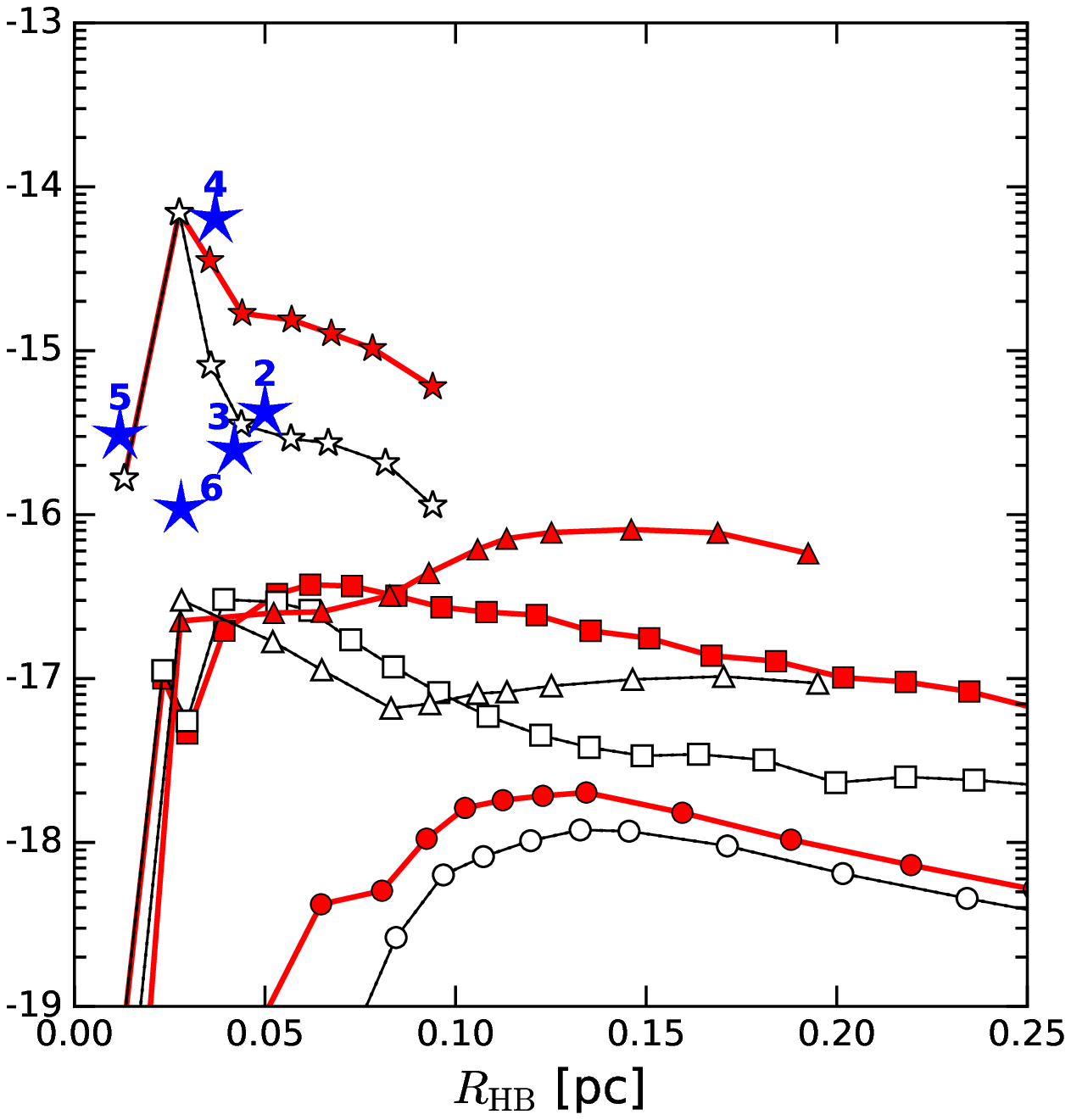}\\
\includegraphics[height=9.1cm]{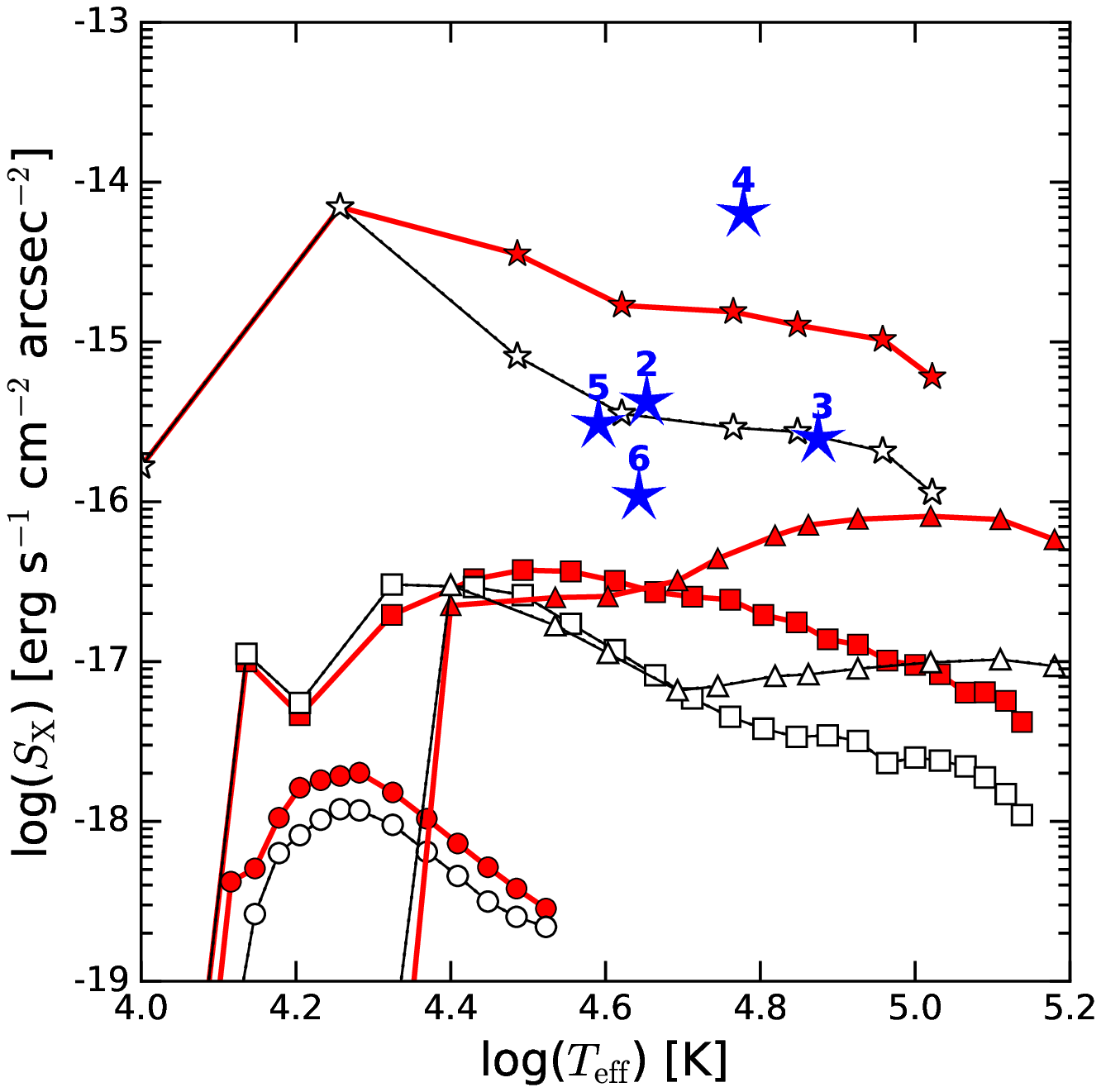}~
\includegraphics[height=9.1cm]{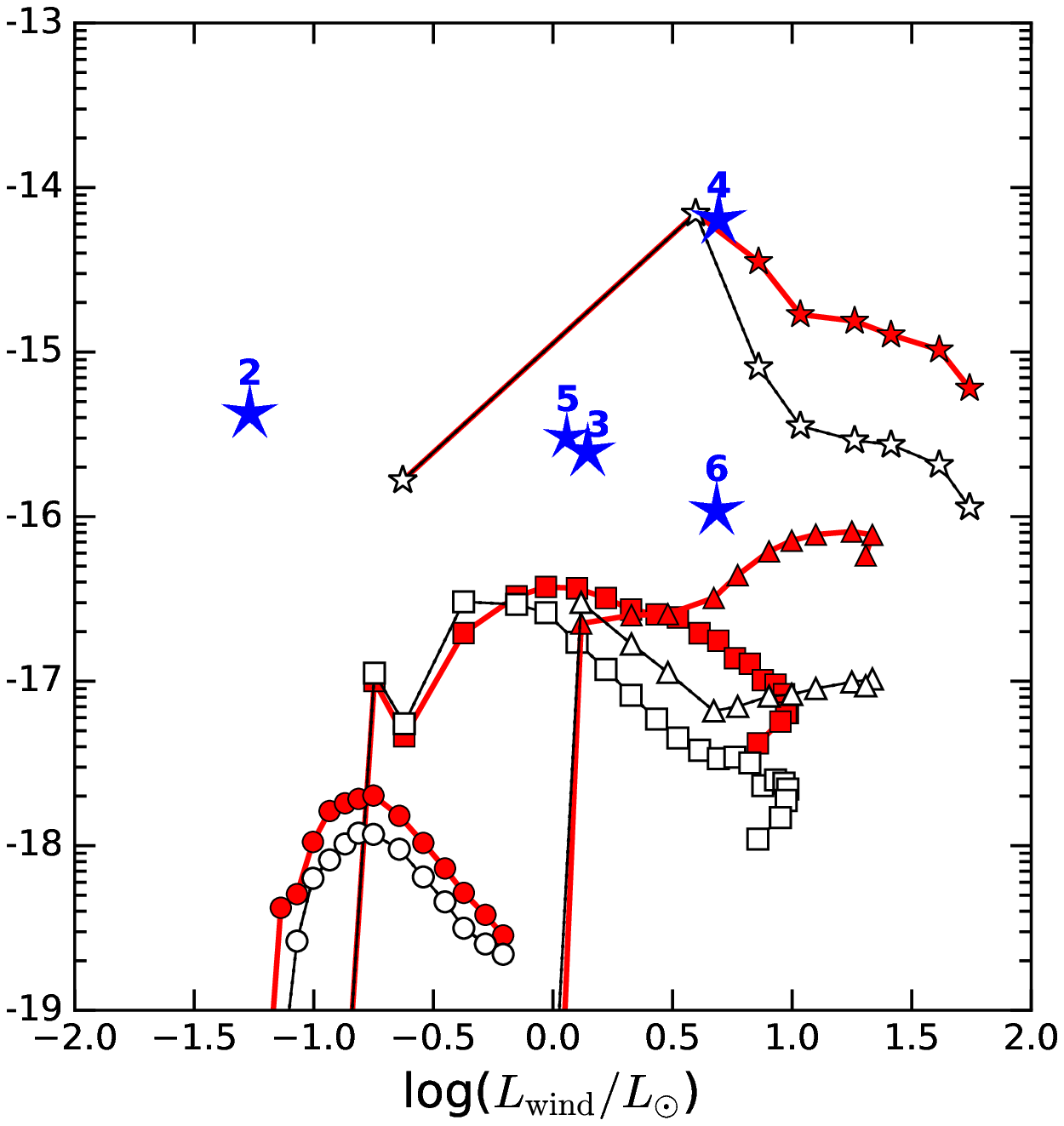}
\caption{Evolution of the X-ray averaged surface brightness in the
  0.3--2.0~keV energy range for all models as a function of time
  (upper left panel), hot bubble radius (upper right panel), stellar
  effective temperature (lower left panel), and stellar wind
  mechanical luminosity (lower right panel). Filled and open symbols
  represent models with and without thermal conduction,
  respectively. Stars, labeled from 2 to 6, correspond to observations
  of NGC\,2392, NGC\,3242, NGC\,6543, IC\,418, and NGC\,6826
  respectively (see Table~1).}
\label{fig:sx_time}
\end{figure*}

As a function of the CSPN effective temperature
(Fig.~\ref{fig:lum_time}, lower left panel), all models show a steady
increase in X-ray luminosity with $T_\mathrm{eff}$. With the exception
of Object~1 (BD$+$30$^{\circ}$3639), our set of models encompasses the
full range of observed luminosities and corresponding effective
temperatures listed in Table~\ref{tab:observations}. Even models
without thermal conduction are capable of producing the luminosities
of several observed objects. However, other derived parameters,
particularly the radius of the X-ray emitting bubble, should be taken
into account when determining the best-fit model.  Our models do not
proceed beyond the maximum $T_\mathrm{eff}$ of each CSPN (see
Fig.~\ref{fig:wind_parameters}) for the technical reasons described
earlier and so none of our models correspond to objects whose CSPN is
descending the white dwarf cooling track.  The 1.0-0.569 model is
included for completeness, even though it is unlikely to correspond to
any object detected with current instruments since the predicted
diffuse X-ray luminosities are low ($L_\mathrm{X} <
10^{30}$~erg~s$^{-1}$).

Finally, Fig.~\ref{fig:lum_time} lower-right panel shows the X-ray
luminosity as a function of stellar wind mechanical luminosity for all
the models. {In this figure it should be borne in mind that the
observational points have poorly determined mass-loss rates and hence
the corresponding wind mechanical luminosities are also uncertain. In
particular, Object~2 (NGC~2392) has a very weak stellar wind whose
velocity is difficult to determine but which has been estimated as
$v_{\infty}\sim$300~km~s$^{-1}$ \citep{Herald2011}.

Since we know the radius of the hot bubble as a function of time for
each model (Fig.~\ref{fig:radhottime}), we can also calculate the
expected averaged surface brightness. The surface brightness as a
function of time, hot bubble radius, CSPN effective temperature, and
stellar wind mechanical luminosity are plotted in
Figure~\ref{fig:sx_time}. In comparison to the equivalent luminosity
curves (see Fig.~\ref{fig:lum_time}), the surface brightness shows
more constant values in time periods where the luminosity is
increasing. This is because the increase in the bubble size offsets
the increase in luminosity. The values we obtain for the (unabsorbed)
surface brightness are within the range reported by the objects in
Table~\ref{tab:observations} but only for the most massive model
2.5-0.677. Only this model can produce small, bright X-ray
bubbles. All our other models produce low surface-brightness objects.
More extensive observational campaigns, such as {\sc ChanPlaNS} and
follow-up observations, should improve the detections of low
surface-brightness objects.

\begin{figure}
\includegraphics[width=\linewidth]{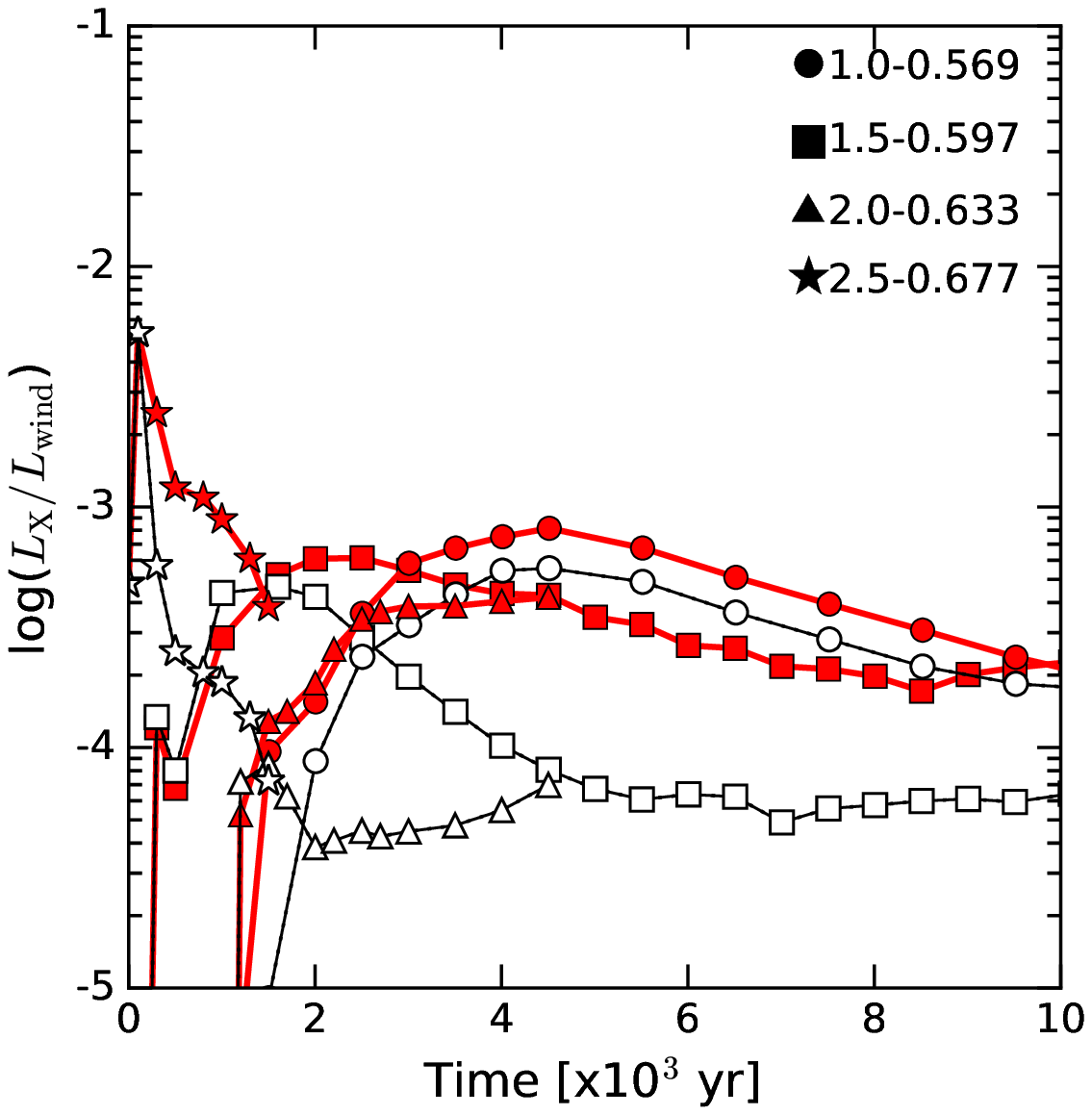} 
\includegraphics[width=\linewidth]{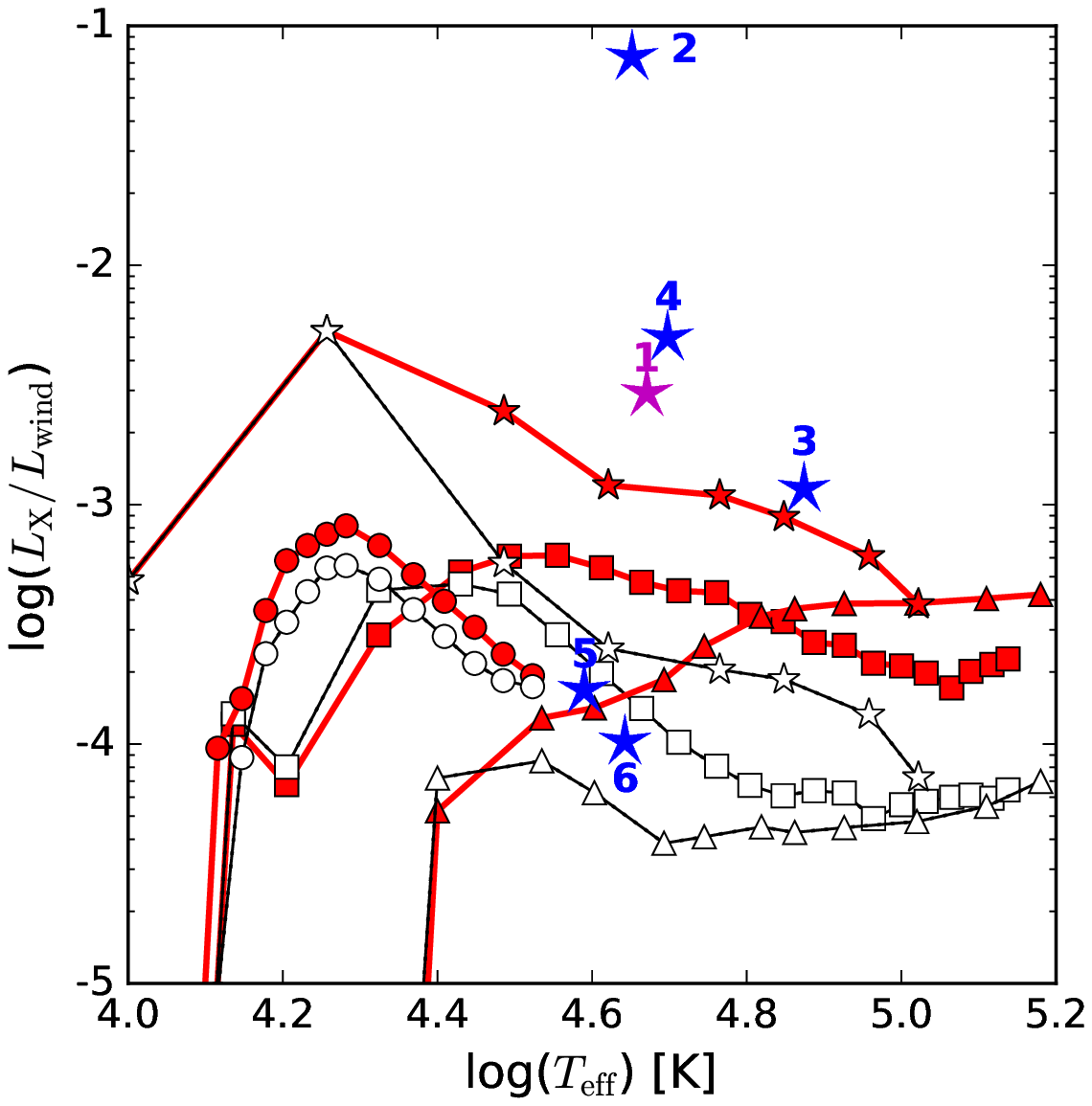} 
\caption{Evolution of the X-ray luminosity as a fraction of the CSPN
  wind mechanical luminosity for all models. Top panel:
  $L_\mathrm{X}/L_\mathrm{wind}$ as a function of time. Bottom panel:
  $L_\mathrm{X}/L_\mathrm{wind}$ as a function of central star
  effective temperature. Filled and open symbols represent models with
  and without thermal conduction, respectively.}
\label{fig:lx_lw_time}
\end{figure} 

Finally, we plot the X-ray luminosity in the energy range 0.3--2.0~keV
as a fraction of the stellar wind mechanical luminosity (see
Fig.~\ref{fig:lx_lw_time}) against time and also as a function of CSPN
effective temperature. Generally, less than 1~\% of the stellar wind
mechanical luminosity is converted into X-ray luminosity This is
consistent with the results presented by \citet{Steffen2008} for their
efficient conduction models, and with the handful of observations that
support their models. Initially the models with and without conduction
convert the similar fractions of stellar wind mechanical luminosity to
soft X-ray luminosity. However, as time proceeds, the models with
conduction convert a factor of up to an order of magnitude more
stellar wind luminosity to X-rays. This is because the X-rays
generally come from the vicinity of the mixing layer, where conditions
are more uniform and to some extent immune from rapid changes in the
stellar wind parameters.

\subsection{Contribution of the shocked fast wind}
\label{sec:shocked}

In Section~\ref{sec:method} we described the use of an advected scalar
$s$ in our numerical simulations (see also Paper~I). As shown before,
this scalar is useful for delimiting the different fluid contributions
to the bubble structure (see Fig.~\ref{fig:scalar}). With this
parameter, we can separate the contributions of the shocked fast
stellar wind and the evaporated/mixed nebular material\footnote{The
  nebular material surrounded the star at the onset of the fast wind
  stage and was expelled from the stellar envelope during the previous
  AGB stage in a dense, slow wind.} to the total X-ray luminosity. In
the mixing region, the scalar takes intermediate values between $s =
0$ (fast wind) and $s = 1$ (nebular material). For simplicity, we use
$s = 0.5$ as the value that distinguishes the two fluids, i.e. $0 < s
< 0.5$ is dominated by fast wind material, while $0.5 < s < 1$ is
dominated by nebular material.

As described in Section~\ref{sec:DEM}, the first step in calculating
the contribution of the shocked fast stellar wind to the total X-ray
luminosity is to compute the DEM for material corresponding to $0 < s
< 0.5$. The resultant DEMs, together with the {\sc chianti} software,
were used to compute spectra, and hence the X-ray luminosities in the
0.3--2.0~keV energy range, $L_\mathrm{X,0-0.5}$, corresponding to the
shocked fast stellar wind gas (see Sections~\ref{sec:spec1} and
\ref{sec:lum}).

\begin{figure*}
\includegraphics[width=1.\linewidth]{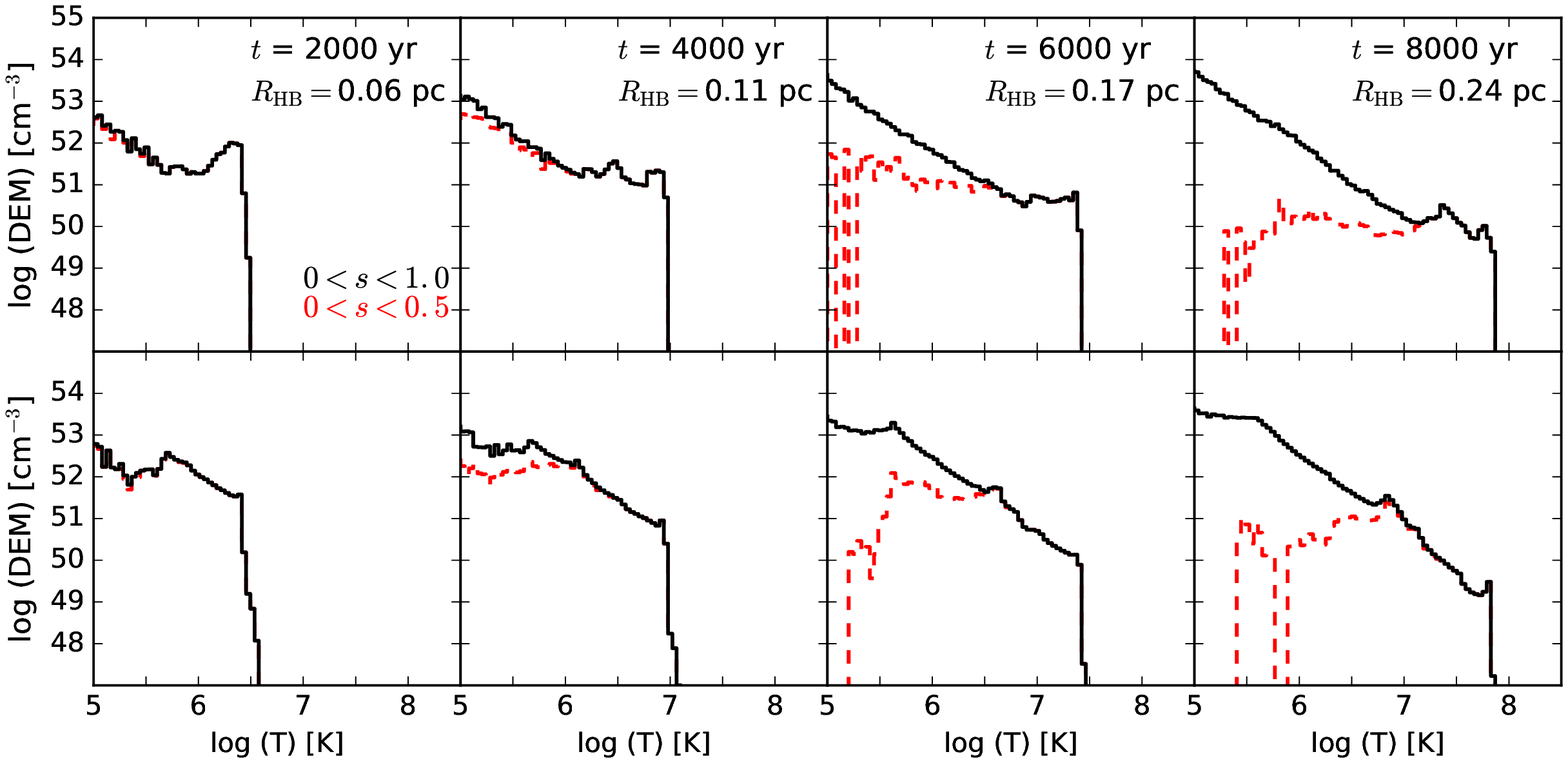}
\caption{Evolution of DEM for two different ranges of the scalar
  values for the 1.5-0.597 model. Solid (black) lines correspond to
  DEM computed over the complete scalar range ($0<s<1$) while the
  dashed (red) lines are the results of taking only the shocked fast
  stellar wind contribution ($0<s<0.5$). Upper and lower panels show
  the results without and with thermal conduction, respectively. Times
  and the corresponding radius of the hot bubble are marked on each
  column.}
\label{fig:DEM_time_0_5} 
\end{figure*}

Figure~\ref{fig:DEM_time_0_5} shows the time evolution of the DEMs
computed for the scalar range ($0<s<0.5$) together with those
calculated for the total scalar range ($0<s<1$) for the 1.5-0.597
model with and without thermal conduction. At the earliest times ($t <
2000$~yrs), the shocked fast wind is the dominant contributor to the
DEM at all temperatures $T>10^5$~K. The photoionized nebular gas has
temperatures $T \sim 10^4$~K.

As time progresses, the contribution of the nebular gas to the X-ray
luminosity increases due to the expansion of the shell and the
incorporation of new material into the mixing and conduction
layers. In contrast, the contribution of the hot shocked wind remains
fairly constant, since the wind velocity increases with time and so
shocks to higher and higher temperatures, which do not contribute to
the diffuse X-ray luminosity. At late times ($t > 6000$~yrs) the
nebular material dominates the DEM distributions in both cases, except
at the very highest temperatures ($\log T > 7$); it is the main
contributor in the turbulent mixing layer for the models without
conduction and it is the main contributor to the conduction layer in
the other set of models.

Figure~\ref{fig:lum_time_0_5} shows the ratio of the X-ray luminosity
of the shocked fast stellar wind material to the total X-ray
luminosity (in the 0.3--2.0~keV energy range:
$L_\mathrm{X,0-0.5}/L_\mathrm{X}$) as a function of time for all
stellar models. This figure demonstrates that at early times the
shocked stellar wind gas dominates the total X-ray luminosity for all
cases. This is because neither the turbulent mixing regions nor the
conduction layers nor any other possible mechanism for heating the
nebular material have had enough time to develop. As time progresses,
the $L_\mathrm{X,0-0.5}/L_\mathrm{X}$ ratio decreases because while
the fast stellar wind contribution to the diffuse X-ray luminosity in
the 0.3--2.0~keV energy band remains roughly constant, the total
diffuse X-ray luminosity increases due to the incorporation of new
material into the mixing layers that accompanies the expansion of the
hot bubble. The small differences in the
$L_\mathrm{X,0-0.5}/L_\mathrm{X}$ ratio seen between models with and
without conduction for a given stellar mass are due to the X-ray
luminosity difference (typically up to a factor of 4) between both
sequences, as seen in Figure~\ref{fig:lum_time}. These differences
tend to be larger with increasing stellar mass.

\begin{figure}
\includegraphics[width=1.\linewidth]{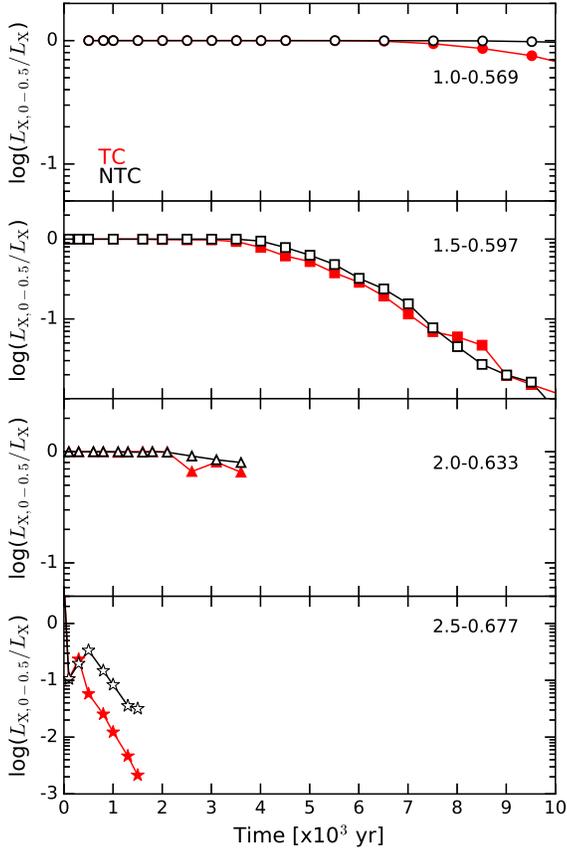}
\caption{Evolution of the X-ray luminosity with time for gas with
  scalar values $0 < s < 0.5$ (shocked fast stellar wind) compared to
  the total X-ray luminosities {in the 0.3-2.0~keV energy
    range}. Filled and open symbols represent models with (TC) and
  without thermal conduction (NTC), respectively.}
\label{fig:lum_time_0_5}
\end{figure} 

Finally, we remark that the luminosity ratios for the 1.0-0.569 models
remain almost constant for the cases with and without conduction, thus
demonstrating that all X-ray properties dervied from these models are
due almost entirely to the shocked fast stellar wind material.

\section{Comparison with observations}
\label{sec:comparobs}

\subsection{Correction for absorption and convolution with
  instrumental response matrices}
\label{sec:abconvol}

The intrinsic spectra shown in Figure~\ref{fig:spectra1} need to be
corrected for interstellar absorption and convolved with the
instrumental response matrices before they can be compared with
observations. As an illustration, Figure~\ref{fig:spectra2} shows the
effects of absorption on the intrinsic spectrum of the 1.5-0.597 model
without thermal conduction at 8000~yr. Results are shown for both
moderate ($N_{\mathrm{H}}=8\times10^{20}$~cm$^{-2}$) and high
($N_{\mathrm{H}}=5\times10^{21}$~cm$^{-2}$) neutral hydrogen column
densities. The principal emission lines are identified in this figure,
showing clearly how carbon and nitrogen dominate the low-energy
spectrum for the abundances used in this work, while oxygen and neon
emission lines are important at high
energies. Figure~\ref{fig:spectra2} shows how the most intense
X-ray-emitting part of the spectrum is severely affected by
extinction. For example, the emission below 0.4~keV is diminished by
more than two orders of magnitude.

\begin{figure}
\includegraphics[width=1.0\linewidth]{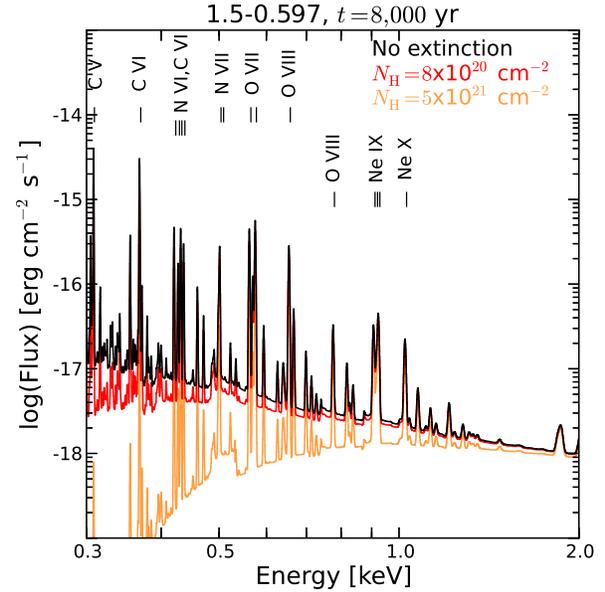}
\caption{Synthetic spectra generated using the {\sc chianti} software
  for the 1.5-0.597 model without thermal conduction at 8000~yr of
  post-AGB evolution with different interstellar absorbing column
  densities. {The principal emission lines are identified.}}
\label{fig:spectra2}
\end{figure}

\begin{figure}
\includegraphics[width=1.0\linewidth]{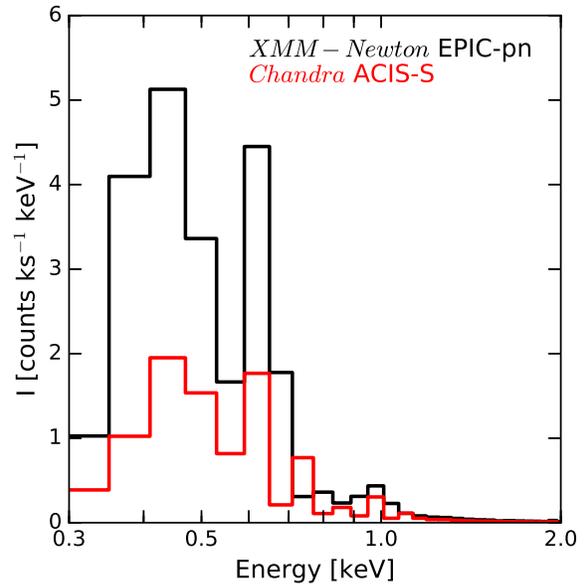}
\caption{Synthetic spectra for the 1.5-0.597 model at 8000~yr after
  the onset of the post-AGB wind without thermal conduction (see
  Fig.~\ref{fig:spectra2}). The spectra have been convolved with the
  \textit{Chandra} (red) and \textit{XMM-Newton} EPIC-pn camera with
  the medium optical filter (black) response matrices using spectral
  bins of width 60~eV and assuming an absorption column density of
  $N_{\mathrm{H}}=8\times10^{20}$~cm$^{-2}$.}
\label{fig:spectra3}
\end{figure}

Figure~\ref{fig:spectra3} shows the result of convolving the absorbed
spectrum shown in Figure~\ref{fig:spectra2} (moderate absorption) with
the response matrices from the \textit{Chandra} ACIS-S and also
\textit{XMM-Newton} EPIC-pn X-ray cameras\footnote{In the case of the
  EPIC-pn we assumed the medium optical filter.}, binned with a
resolution of 60~eV. Note that most of the rich detail of the spectral
lines is lost and only a few features can be identified in the
spectra. For example, there are indications of the C\,{\sc vi} line at
0.37~keV, C\,{\sc vi} and the N\,{\sc vi} He-like triplet at
$\sim0.43$~keV and the O\,{\sc vii} triplet at $\sim0.58$~keV together
with a possible contribution from O\,{\sc viii} at $\sim 0.65$~keV.
Both the binning and the convolution with the instrumental response
matrices combine contributions from various lines in a single
bin. Recall that these spectra were generated using the {\sc Cloudy}
standard PN abundance set. If other abundances are used the relative
line strengths will be different.

Even though \textit{XMM-Newton} possesses a higher sensitivity towards
lower energies, the following analysis will focus on spectra convolved
with the \textit{Chandra} ACIS-S response matrices, in order to
facilitate direct comparison with current and future results of the
\textit{Chandra} Planetary Nebula Survey ({\sc ChanPlaNS};
\citealp{Kastner2012}, \citealp{Freeman2014}, \citealp{Montez2015}).

Figure~\ref{fig:specsample} shows examples of the final absorbed and
convolved spectra for cases with and without thermal conduction for
the 1.5-0.597 models at a selection of different times in the post-AGB
evolution. For the earliest time (left-most panels), corresponding to
2000~yr after the onset of the fast wind, results are presented for
both moderate ($N_\mathrm{H}=8\times10^{20}$~cm$^{-2}$) and high
($N_\mathrm{H}=5\times10^{21}$~cm$^{-2}$) neutral hydrogen absorbing
column densities. These figures illustrate very clearly the quenching
effect that a high interstellar neutral absorbing column density has
on these very soft spectra at energies below 0.6~keV: all evidence of
the N\,{\sc vi} line has disappeared and it is difficult to
distinguish between different models.

It can also be seen in Figure~\ref{fig:specsample} that the count rate
generally increases with time, simply reflecting the increase in the
DEM with time of the 1.5--0.597 models (see Fig.~\ref{fig:DEM_time}),
which also results in the time increase in the total luminosity (see
Fig.~\ref{fig:lum_time}). Moreover, the spectra for the cases with
conduction have higher count rates, by up to a factor of 4, than their
counterparts without conduction. This is entirely consistent with the
relative DEM values shown in Fig.~\ref{fig:DEM_time}.

For the models without conduction (Fig.~\ref{fig:specsample}, upper
panels), the change in the spectral shape from 2000 to 4000 to
8000~yrs reflects the changing DEM of these models (see
Fig.~\ref{fig:DEM_time}). At 2000~yrs, the hottest gas has a
temperature $\log T \sim 6.3$ and is shocked fast stellar wind
material. This gas is the main contributor to the diffuse X-ray
emission since it coincides with the peak of the emission coefficient
(see Fig.~\ref{fig:emiscoef}).  At 4000~yrs the average temperature of
the gas reaches its peak (see Fig.~\ref{fig:aver_temp}), although the
main contribution to the diffuse X-ray emission still comes from gas
with temperature around $\log T \sim 6.3$, at which temperature the
turbulent mixing layer gas is the main component
(cf.\ Fig.~\ref{fig:DEM_time_0_5}). Higher temperature gas now also
contributes to the X-ray emission, and this is reflected in the shape
of the spectrum.  At the latest time shown (8000~yrs), although the
average temperature is very similar to that at 2000~yrs (see
Fig.~\ref{fig:aver_temp}), the main contributor to the X-ray emission
is the lower temperature gas in the turbulent mixing layer. The main
peaks in the absorbed, convolved spectra correspond to emission
lines. Higher temperature gas has an increased O\,{\sc vii} ($\sim
0.58$~keV) to N\,{\sc vi} ($\sim 0.43$~keV) line ratio.

For the 1.5--0.597 models with thermal conduction
(Fig.~\ref{fig:specsample}, lower panels), the shape of the DEM does
not change much from 2000 to 4000 to 8000~yrs, but it does increase in
magnitude (see Fig.~\ref{fig:DEM_time} and \ref{fig:spectra1}). This
is simply a consequence of the expansion of the hot bubble and the
resulting larger emission measure of gas at soft X-ray emitting
temperatures.


It is worth mentioning here that, even though we are not tailoring the
abundances or morphology to any particular PN,
Fig.~\ref{fig:specsample} shows similar spectral features to {\it
  Chandra} observations reported by \citet{Chu2001},
\citet{Gruendl2006}, and \citet{Ruiz2013} for the PNe NGC\,6543,
NGC\,7026 and NGC\,2392, although differences in the binning make a
direct comparison difficult.  In this paper, a single set of
abundances has been used for both fast stellar wind and nebular
components but it should be appreciated that if different abundance
sets are used for the two components, then this would also have an
effect on the evolution of the spectral shape.

\begin{figure*}
\includegraphics[width=0.32\linewidth]{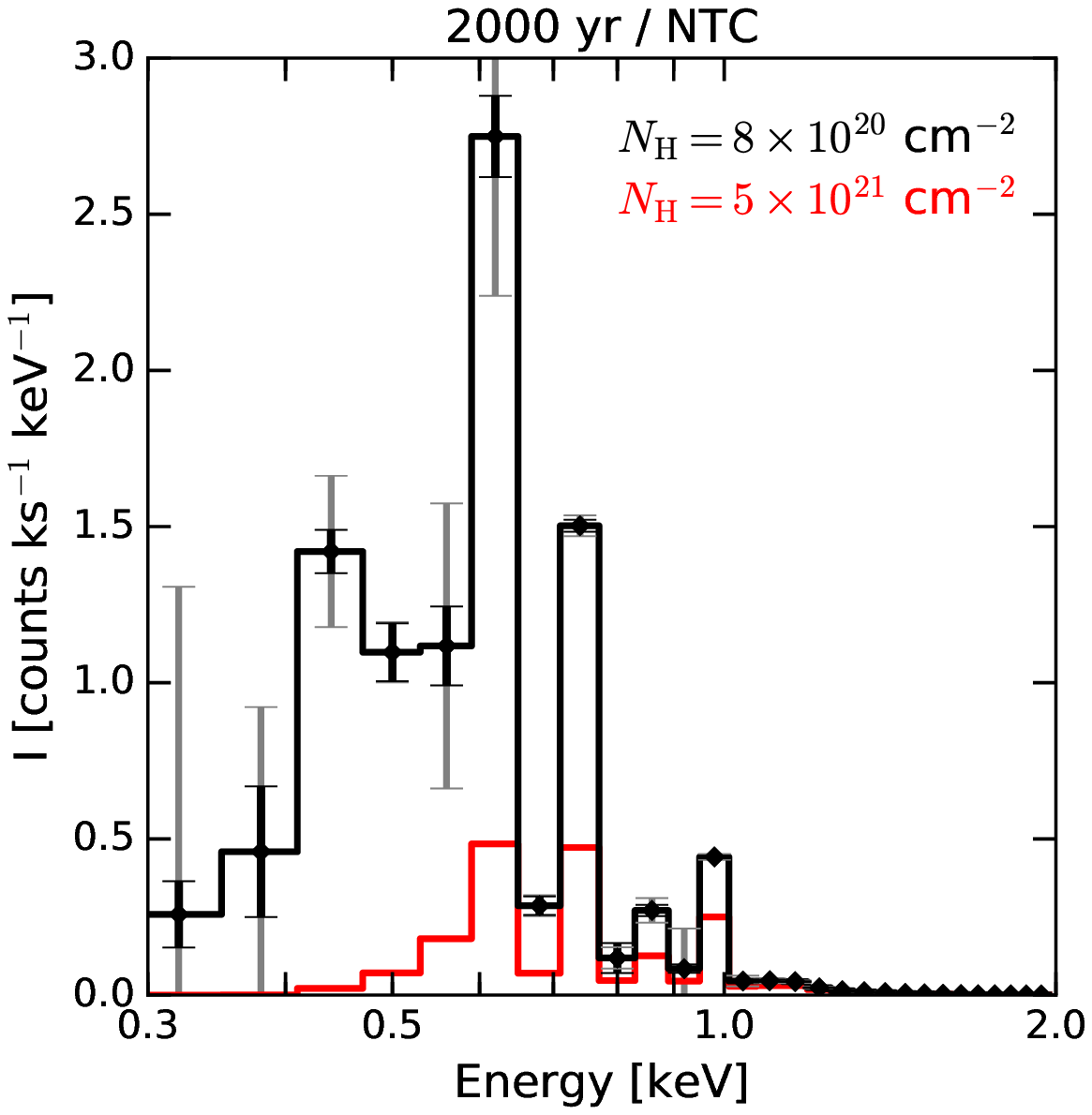}~
\includegraphics[width=0.32\linewidth]{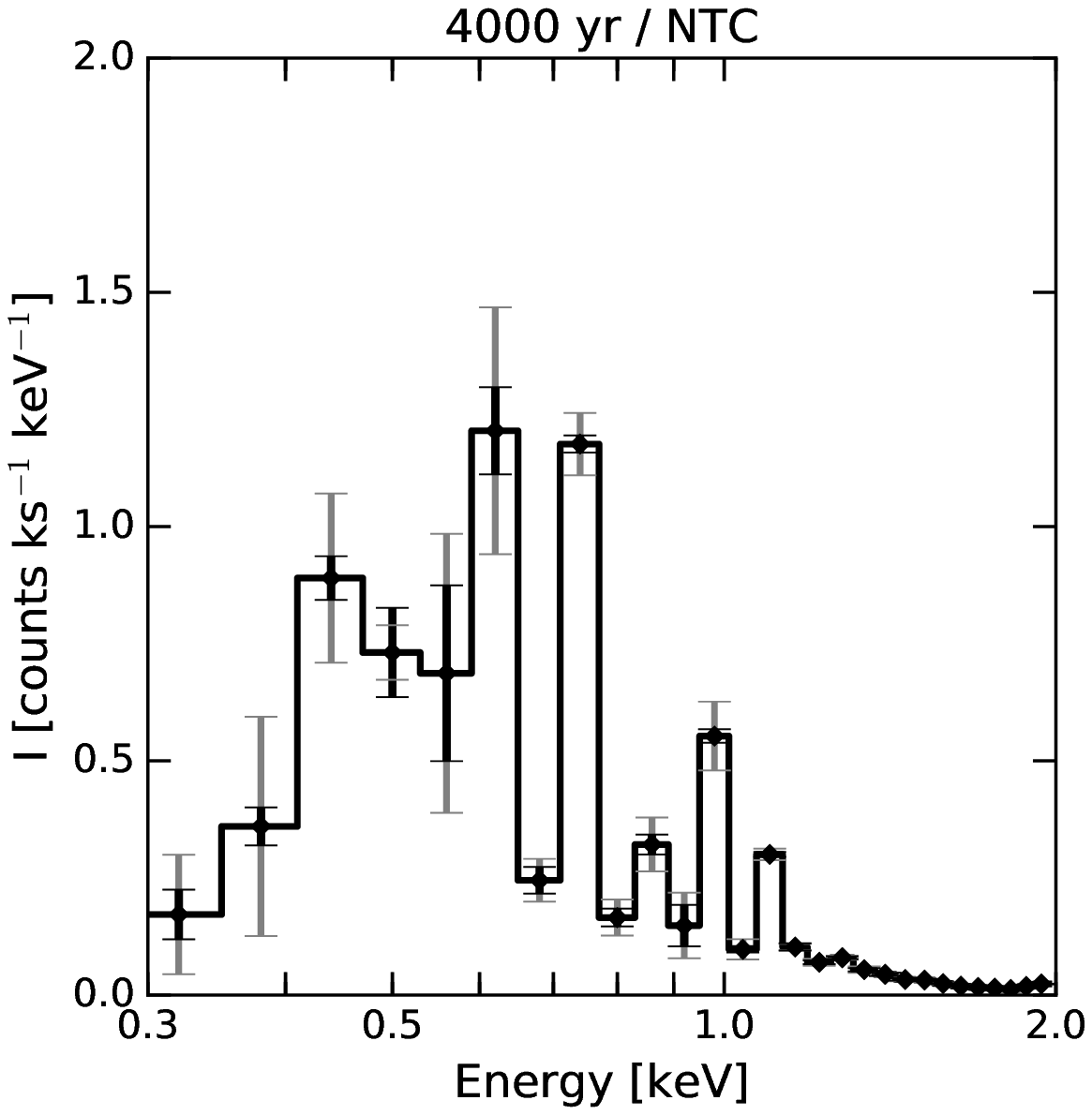}~
\includegraphics[width=0.32\linewidth]{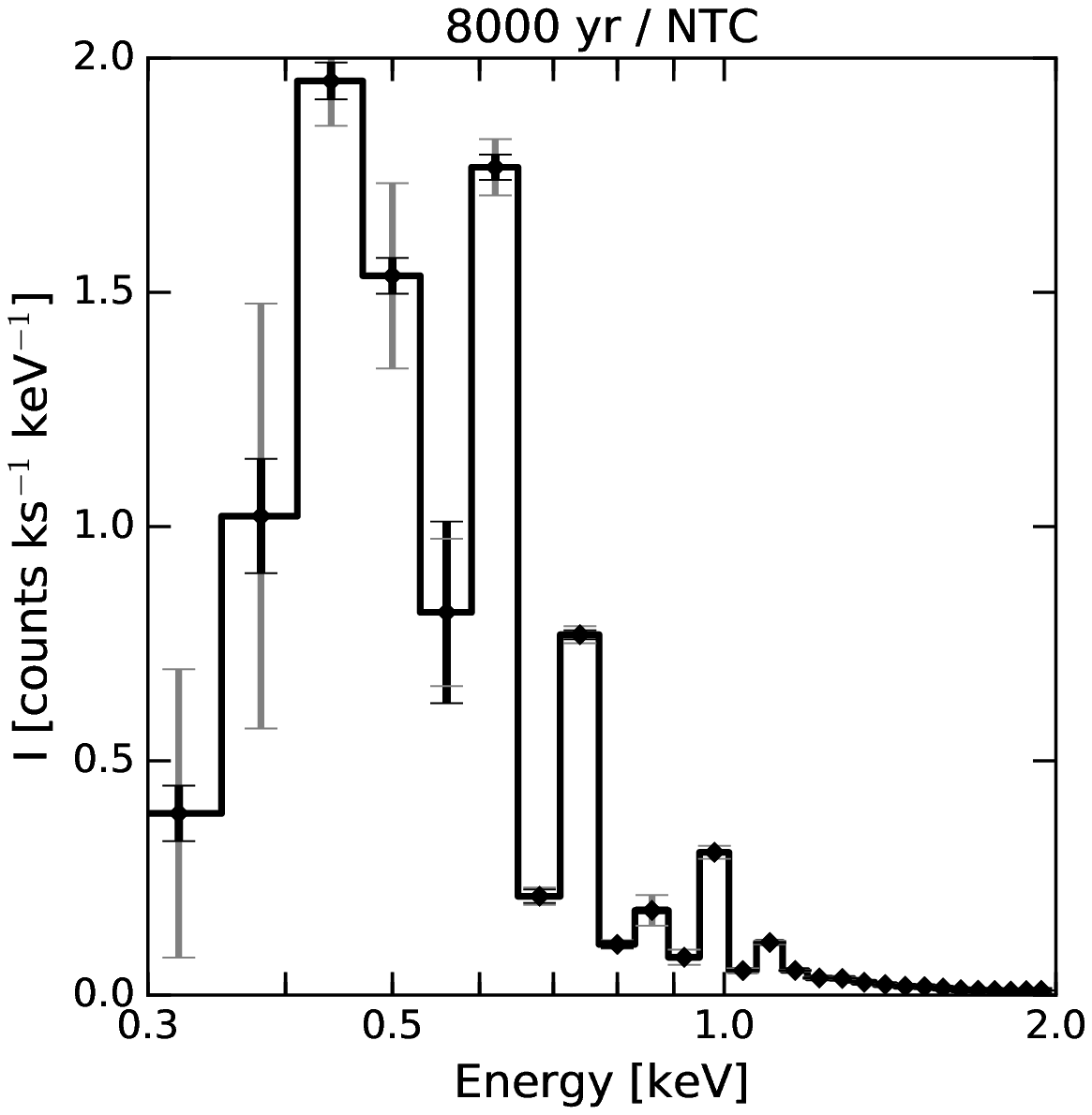}\\
\includegraphics[width=0.32\linewidth]{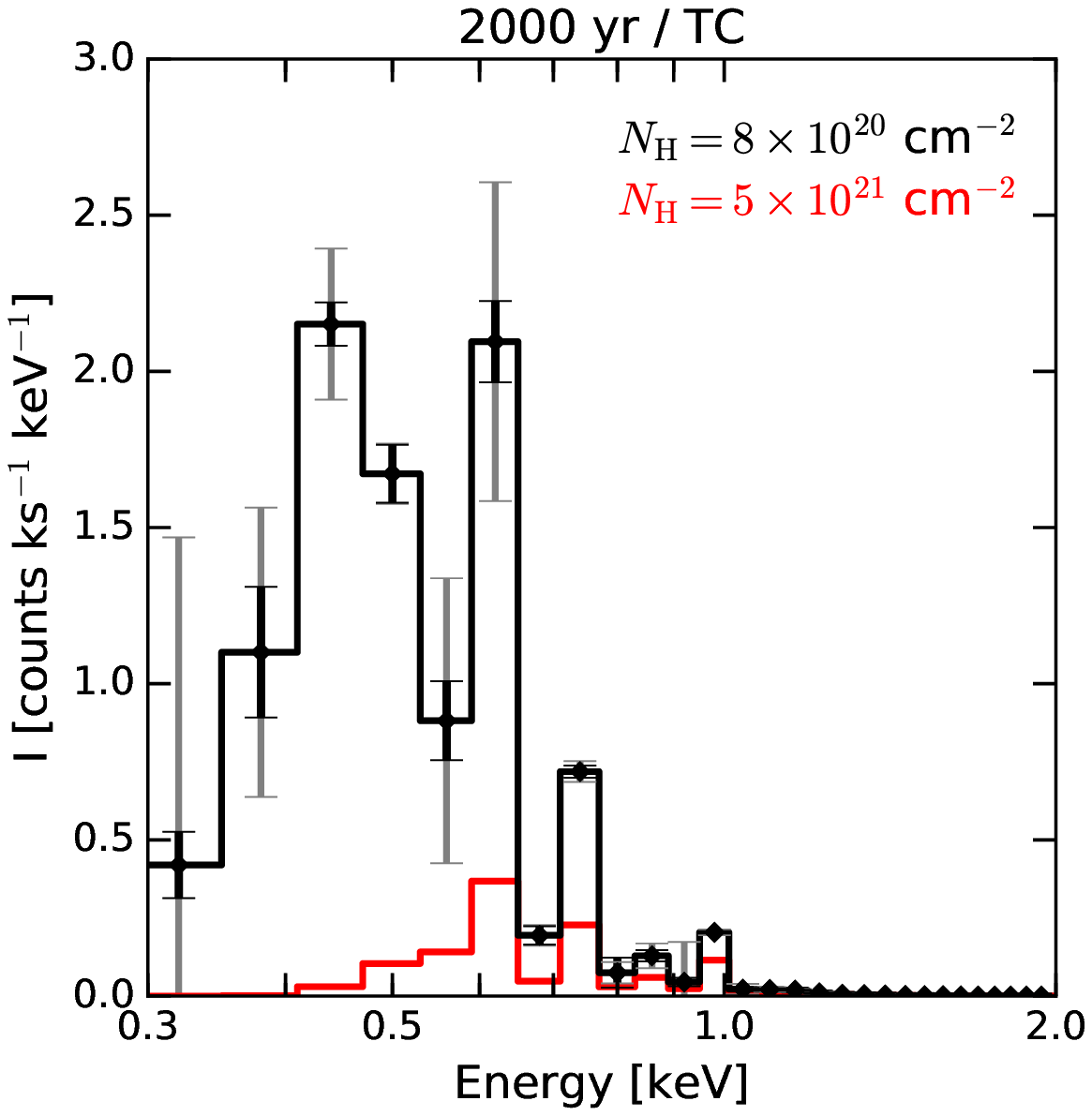}~
\includegraphics[width=0.32\linewidth]{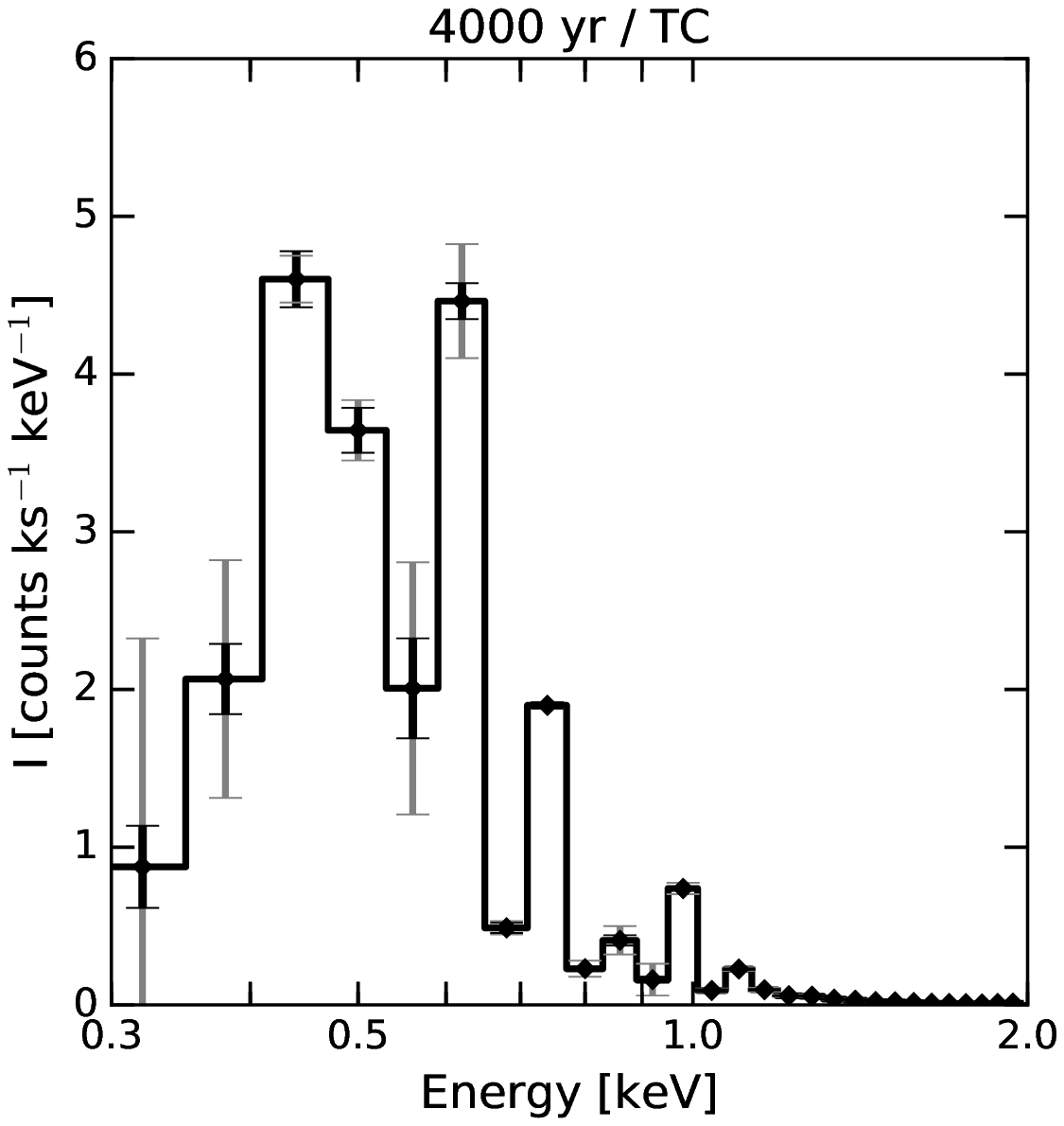}~
\includegraphics[width=0.32\linewidth]{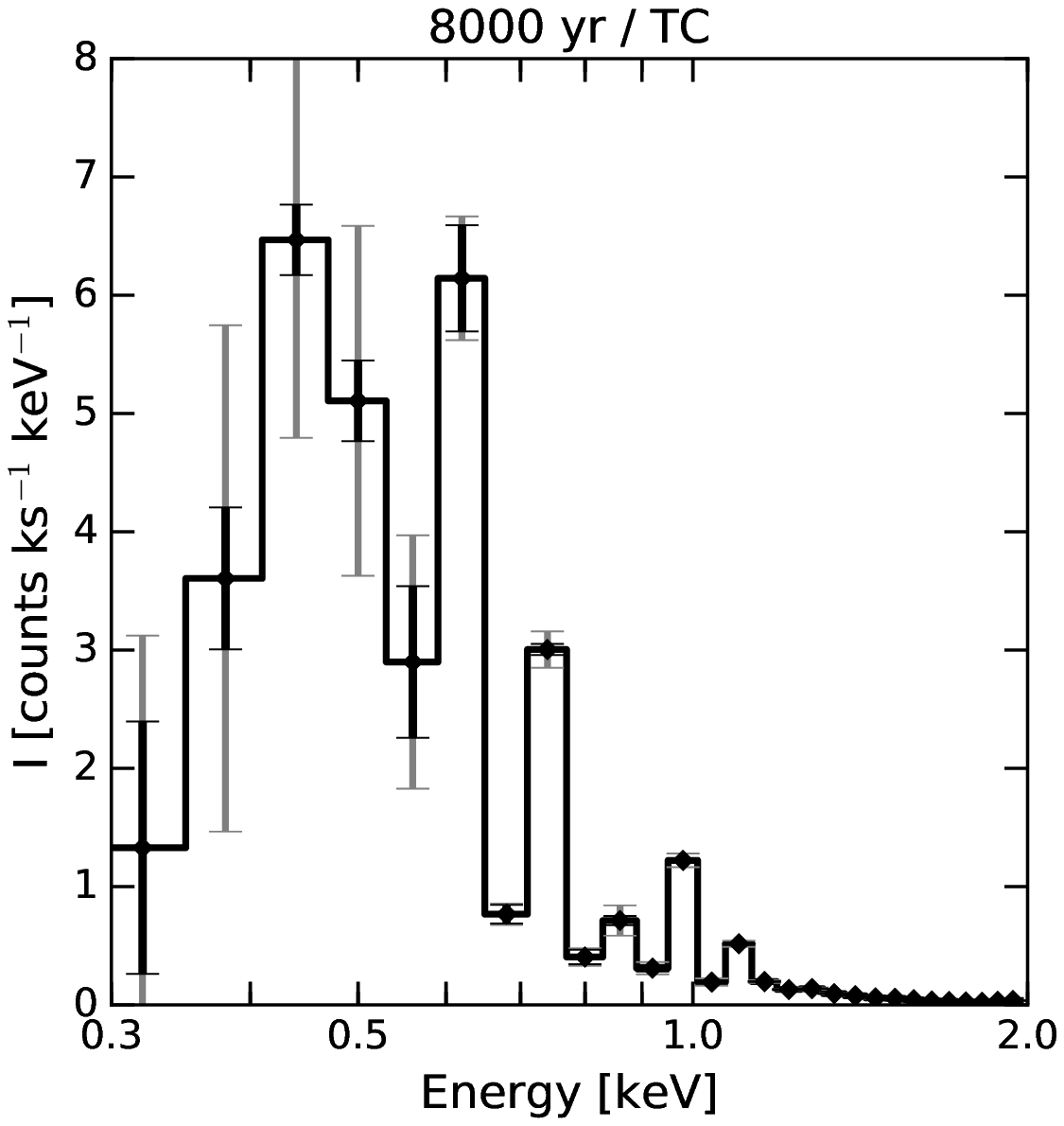}
\caption{Synthetic X-ray spectra for the 1.5-0.597 models with (TC)
  and without thermal conduction (NTC) at 2000, 4000 and 8000~yr after
  the onset of the post-AGB phase. Black and red lines represent two
  values of the absorbing neutral hydrogen column densities: moderate
  $8\times10^{20}$~cm$^{-2}$, and high $5\times10^{21}$~cm$^{-2}$,
  respectively. Grey and black error bars correspond to spectral
  sampling of 50 and 200 counts, respectively (see text for
  details). All spectra were convolved with the \textit{Chandra}
  ACIS-S response matrices using a 60~eV bin width.}
\label{fig:specsample}
\end{figure*}

\subsection{Sampled spectra}
\label{sec:sample} 

It should be borne in mind that the absorbed, convolved, synthetic
spectra, such as those presented in Figure~\ref{fig:specsample} (solid
lines), correspond to ideal spectra, i.e., as if they were obtained by
\textit{infinite}-exposure time observations. A real observation lasts
a finite amount of time. It is possible to mimic finite-time
observations by sampling the synthetic spectrum to imitate photons
(counts) in a real observation.

In order to simulate an observation, the cumulative absorbed spectrum
is randomly sampled in the 0.3--2.0~keV energy range, according to the
desired photon count number.  The process was repeated $10^{4}$ times
in order to generate independent selections of a given photon count
number. The standard deviations for 50 and 200 photon count samples
are shown in Fig.~\ref{fig:specsample} by grey and black error bars,
respectively, where we consider only the case of a moderate absorbing
column density ($N_\mathrm{H}=8\times10^{20}$~cm$^{-2}$).

Figure~\ref{fig:specsample} shows that observations with 50 counts are
not enough to give a fair description of the true spectral shape when
the binning is 60~eV. A coarse binning would be recomended for such a
low number of total counts. Hence the physical properties of hot
bubbles in PNe derived from such low count-rate spectra will not be
very reliable. Unfortunately, there are several PNe that have not been
observed with more than 50 counts due to their reduced exposure times
\citep[see][]{Kastner2012,Ruiz2013,Freeman2014}. On the other hand,
200-count spectra give a reasonable description of the spectral
shape. Hence the physical properties derived from these spectra will
be more representative of the real conditions in the hot bubble gas.

This is a very interesting subject. In addition, different abundances
in the nebular and fast stellar wind gas could be taken into account
in the spectral calculation and the absorbing column density,
$N_\mathrm{H}$, could be varied. Such a detailed analysis will be
pursued in a subsequent paper.

\subsection{X-ray surface brightness}
\label{sec:surfbright}
As an additional tool in the comparison with observations, we can
create surface brightness profiles of the soft X-ray emission from the
hot bubbles. For this, we have integrated the X-ray intensity in the
0.3--2.0~keV energy band, taking into account the absorption by the
neutral interstellar medium. Furthermore, all profiles must be
convolved with the spatial response of the telescope, which is a
combination of the instrumental PSF and smoothing kernel. This can be
simulated by using a Gaussian profile.

Figure~\ref{fig:surface_b} (top panel) presents the resulting
normalized surface brightness profiles from the 1.5-0.597 models with
and without thermal conduction at the time when the hot bubble has a
mean radius of 0.2~pc (see Figs.~\ref{fig:final_2D} and
\ref{fig:final_2D_2}). An absorbing neutral hydrogen column density of
$N_{\mathrm{H}}=8\times10^{20}$~cm$^{-2}$ was assumed. The normalized
surface brightness profiles would be described as limb-brightened for
the graphs shown in the top panel. However, the radial positions of
the peaks of the profiles are not at the edge of the bubbles, as might
be expected, but at 65--75\% of the radius. This is because the soft
X-ray emission produced by cooler gas, which is found around the edge
of the hot bubble, is more affected by absorption than the X-ray
emission from hotter gas (see, e.g., Figure~\ref{fig:spectra2}) as
described by \citet{Steffen2008}.  We note that the unsmoothed surface
brightness profiles from both models present structure in their inner
regions. This is due to the large-scale non-radial motions in these
simulations, which drag material from the shell into the inner part of
the hot bubble, as seen in middle panels of Figures~\ref{fig:final_2D}
and \ref{fig:final_2D_2}.

The other panels in Figure~\ref{fig:surface_b} show the results
obtained from applying Gaussians with different FWHM to the same
profiles shown in the top panel. Each value of the FWHM represents a
fraction of the effective radius of the hot bubble. The effect of
applying a heavy smoothing process (higher FWHM) can give the
appearance that the X-ray-emitting gas is distributed with a maximum
at the center of the hot bubble, i.e., has a flatter surface
brightness distribution. This smoothing process may be unavoidable in
low-count-number detections of PNe.

\begin{figure}
\includegraphics[width=1\linewidth]{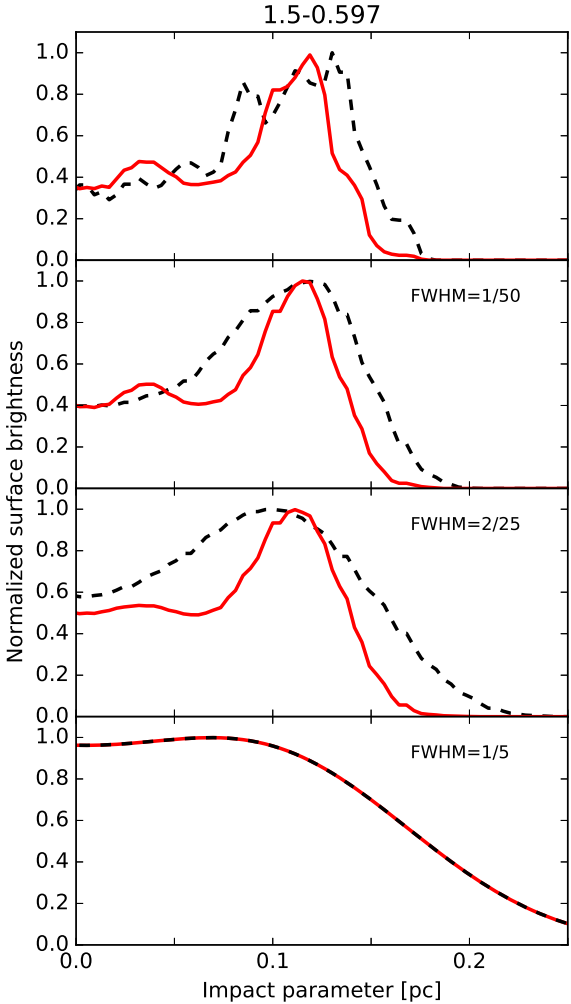}
\caption{Normalized surface brightness profiles in the 0.3--2.0~keV
  energy range for the 1.5-0.597
  model for the time when the rim shell is $\sim0.2$~pc with
  interstellar absorption of $N_{\mathrm{H}}=8\times10^{20}$~cm$^{-2}$
  for different Gaussian convolutions. Dashed (black) lines represent
  the results without thermal conduction and solid (red) lines the
  results with thermal conduction. Each value of the FWHM represents a
  fraction of the effective radius of the hot bubble.}
\label{fig:surface_b}
\end{figure}

\section{Discussion}
\label{sec:discussion}

Our work in this paper confirms that thermal conduction can be an
important physical process for the production of soft, diffuse X-ray
emission in PNe. However, instabilities, which form in the swept-up
shell as the fast stellar wind accelerates into the expelled envelope
(i.e., nebular) material, are also of great importance especially if
conduction is suppressed or absent. The clumps and filaments in the
broken shell can act as sources for photoevaporated flows and
hydrodynamic mixing layers. The gas in the turbulent mixing layers can
be shock-heated to averaged temperatures of $T_\mathrm{A} \sim
10^6$~K. We find that the characteristic temperature of the X-ray
emitting gas reported by observations is well approximated by the mean
emissivity weighted temperature of the hot bubble ($T_\mathrm{A}
\approx T_\mathrm{X}$; see Fig.~\ref{fig:avt-rad}). The combination of
the emissivity coefficient, which for planetary nebula abundances
peaks about $\log T \sim 6.3$ and the differential emission measure
distributions, which are highest for $\log T \leqslant 6$, results in
a mean temperature precisely within the observed range. The
hydrodynamic mixing layer results give higher temperatures than the
thermal conduction results.

For the four initial stellar masses considered in this paper, the
brightest X-ray objects are the shortest lived, for example the X-ray
luminosity of the 2.5-0.677 model fades within 2000~yrs, while the
luminosity of the 1.0-0.569 model, although low, continues at the same
level for more than 10,000~yrs. However, the surface brightness of the
longer-lived models decreases with time as the size of the hot bubble
increases, and this could influence the detection of these
objects. Our numerical results are also consistent with the sizes of
PNe detected in diffuse X-rays: the bright, short-lived X-ray emission
comes from hot bubbles smaller than $R_\mathrm{HB}\sim 0.1$~pc and
corresponds to the most massive stars, while larger ($R_\mathrm{HB}>
0.1$~pc), less bright hot bubbles are produced by the more slowly
evolving lower-mass objects. Our models would seem to predict far more
large, low brightness objects than are actually observed.  The lower
limits for the X-ray luminosity and surface brightness of these models
correspond to the cases without conduction and are an order of
magnitude smaller. It is possible that we are observationally limited
by the current generation of X-ray satellites. Detections of surface
brightness levels below
10$^{-17}$~erg~s$^{-1}$~cm$^{-2}$~arcsec$^{-2}$ will be achieved with
the Athena$+$ telescope \citep[][]{Nandra2013}.

We find that at early times all of our models predict that the X-ray
emission will be dominated by the shocked fast wind material but at
intermediate and later times the nebular contribution dominates. This
is consistent with the previous theoretical work of
\citet{Akashi2007}, who showed that diffuse X-ray emission can come
from a shocked stellar wind whose velocity is rapidly increasing in
the first 1000~yrs of post-AGB evolution.

Although we assume in our models a single set of abundances for both
the nebular and CSPN fast wind material, in reality these {\bf could}
be different, with the CSPN abundances being metal-enriched due to the
expulsion of the hydrogen-rich envelope
\citep{Georgiev2008,Morisset2009}. Thus, the change in the relative
importance of the hot, shocked stellar wind gas and the heated nebular
material seen in Figure~\ref{fig:lum_time_0_5} should also manifest
itself as an evolution of the abundances responsible for the X-ray
spectra of these objects. For instance, at early times we could expect
the abundances to be dominated by hot, shocked stellar wind material,
while at later times nebular abundances will determine the spectral
characteristics. This is more relevant for the case of PNe with
[WR]-type nuclei and has been recently explored by \citet{Sandin2016}.

There is some observational evidence that very young PNe do indeed
show X-ray spectra that can be well fitted as a hot plasma with
stellar abundances, suggesting that the stellar wind dominates the
X-ray emission in these objects. Even though it is a short-lived
phase, and even shorter in the most massive cases, there is one
observed object, BD$+$30$^{\circ}$3639, that requires stellar
abundances, since C and Ne are enhanced. \citet{Yu2009} analyzed {\it
  Chandra} Low Energy Transmission Gratings (LETG) spectra in
conjunction with the ACIS-S data. The high-quality LETG/ACIS-S spectra
enabled the authors to conclude unambiguously that the diffuse X-ray
emission in this PN has abundances close to those of the CSPN. The
second brightest X-ray PN, NGC\,6543, on the other hand, does not show
any evidence for enhanced C and Ne, and so nebular abundances are more
appropriate for this object \citep{Guerrero2015}, although higher
signal-to-noise observations are required to confirm this
statement. Both observations are fully consistent with our numerical
models, particularly since the hot bubble in NGC\,6543 is larger than
that of BD$+$30$^{\circ}$3639 and it is assumed to be an older object.

In the presentation of our results, we have taken care to include
observational effects so that our models can be compared directly
with, for example, the results from the {\sc ChanPlaNS} survey. In
particular, we show that the radial surface brightness profile that is
obtained from a given hot bubble depends very much on the size of the
hot bubble combined with the instrumental PSF with which the models
are convolved. The same model can give a limb-brightened or flat
surface brightness profile depending on the FWHM of the Gaussian
profile of the PSF. Indeed, observations of the same object with
\textit{XMM-Newton} and \textit{Chandra} can give different radial
surface brightness profiles, such as is the case with NGC\,3242 (see
\citealp{Ruiz2011},\citealp{Kastner2012}). This is because the PSF of
the EPIC \textit{XMM-Newton} X-ray detectors is not able to resolve
structures with the same level of detail as the back-illuminated
ACIS-S CCD3 of the \textit{Chandra} telescope. Given that the
\textit{Chandra} observations of diffuse X-rays in PNe are only
marginally resolved in the majority of cases (small angular size) and
in many cases have a low count rate, it is unwise to attribute a
particular morphology to the X-ray emitting gas.

In a similar spirit, we have simulated finite-time observations by
sampling the ideal spectrum a fixed number of times. From these
experiments, we conclude that observed spectra with a low total number
of counts do not reliably describe the spectral shape and hence should
not be used to derive the physical parameters (temperature,
abundances, absorbing column density of neutral gas) of the observed
hot bubble. We suggest that at least 200 total counts are needed per
object.

At the earliest times, the hot bubble in our models does not occupy a
large fraction of the computational grid. For this reason, the results
for the smallest bubbles are not as detailed as at later times, when
the bubbles are larger and the structures are resolved with more
computational cells. As part of our ongoing investigations into the
production of diffuse X-ray emission in planetary nebulae, we will
study the earliest stages of the hot bubble evolution at higher
numerical resolution in future work. This is particularly important
since the brightest and best studied observed X-ray PN have small
radii ($R_\mathrm{HB} < 0.1$~pc). Increasing the numerical and
temporal resolution at the earliest stages of high initial stellar
mass models ($2.0 M_\odot < M < 2.5 M_\odot$) will also permit us to
characterise the fragmentation of the swept-up shell due to
instabilities in the wind-wind interaction region. It would also be
interesting to include a larger number of stellar models in this mass
range.

Finally, although our results with thermal conduction are consistent
with a full range of observed parameters and trends, it is worth
mentioning that even when magnetic fields are present in a PN and
conduction can be assumed to be suppressed, soft diffuse X-ray
emission from the hot bubbles can be expected in the majority of cases
due to the turbulent mixing layers. \citet{Stute2006} have pointed out
that even small magnetic fields of the order of $\mu$G would be able
to suppress conduction by thermal electrons in the direction
perpendicular to the magnetic field lines. Recent, low-resolution,
spectropolarimetric observations of a sample of CSPN in a variety of
different PN morphologies do not find significant evidence for surface
magnetic fields in the central stars \citep{Steffen2014}. Any magnetic
field in the PN itself must therefore be already present in the
expelled envelope. Polarization of thermal molecular lines of CO and
SiO in the envelopes of AGB stars \citep{Vlemmings2012} and also maser
studies of these objects \citep{Vlemmings2007} show that magnetic
fields are generally present at this stage of evolution. Dust
polarization studies of two protoplanetary nebulae show well-organized
magnetic fields aligned with molecular outflows in these objects
\citep{Sabin2014}. Indeed, magnetic fields are one of the possible
mechanisms for shaping bipolar nebulae. Polarized dust also provides
evidence for magnetic fields in the bipolar PNe NGC\,6302, NGC\,6537,
and NGC\,7027 \citep{Sabin2007}, however, this sort of planetary
nebula is least likely to show diffuse X-ray emission
\citep{{Kastner2012},{Freeman2014}}. {Thus, fossil magnetic fields
could be present in the nebular gas surrounding the CSPN at the onset
of the fast wind. The high incidence of diffuse X-ray detections in
PNe ($\sim 30\%$; \citealp{{Kastner2012},{Freeman2014}}) therefore requires
explanations that do not necessarily require thermal conduction.}

\section{Summary and conclusions}
\label{sec:summary}

In this paper, we have presented calculations of the synthetic diffuse
X-ray emission from hot bubbles in PNe from the results of 2D axisymmetric
high-resolution, radiation-hydrodynamic simulations presented in
\citet[][]{Toala2014}. The numerical results correspond to models with
initial masses (M$_{\mathrm{ZAMS}}$) of 1, 1.5, 2, and
2.5~M$_{\odot}$, which have final WD masses of 0.569, 0.597, 0.633,
and 0.677~M$_{\odot}$, respectively. The density and temperature
information from the numerical simulations are used to compute the
differential emission measure (DEM) distribution of the gas and study
its evolution with time. The DEM distributions were used to compute
synthetic spectra in the 0.3--2.0~keV energy range making use of {\sc
  chianti} software. This allows us to study the time evolution of the
spectra of each model. Luminosities were obtained from integrating the
spectra. In addition, the use of an advected scalar in the numerical
simulations enables us to separate out the contribution of the shocked
fast stellar wind material from that of the swept-up nebular gas
to the X-ray emission.

Our main findings can be summarized as follows:
\begin{enumerate}[(i)]

\item Instabilities created in the wind-wind interaction zone {at the
  contact discontinuity between the fast stellar wind and the nebular
  gas} lead to the formation of clumps and filaments whose large
  surface area results in extensive turbulent mixing layers or regions
  of thermally evaporated nebular material having densities and
  temperatures intermediate between the tenuous hot, shocked fast wind
  plasma and the dense photoionized nebular material. This is the gas
  responsible for producing the soft X-ray emission, whose spectra and
  luminosities agree nicely with observations. Cases in which thermal
  conduction is included have higher fluxes and luminosities than
  cases with purely hydrodynamic mixing layers.

\item The differential emission measure distributions (DEMs) from
  models without thermal conduction are different to those from models
  with conduction. When there is no conduction, the DEM distribution
  falls monotonically from high DEM values at low plasma temperatures
  to low DEM values a high plasma temperatures. In models with
  thermal conduction, heat diffusion transfers heat from the hottest
  gas to the gas at the interface between the hot bubble and the
  swept-up nebular shell and, increases the DEM values for all
  intermediate temperatures. A plateau of roughly constant DEM value
  forms at temperatures $T <10^6$~K, which corresponds to nebular gas
  in the mixing layer.

\item The emissivity weighted mean plasma temperatures of all our
  models lie within derived values of the temperature of observed
  X-ray-emitting gas in PNe. This is a consequence of a sharply peaked
  emission coefficient and DEM distributions that increase towards
  lower temperatures. Models with thermal conduction show very uniform
  values of the mean temperature, both as a function of time and
  between models for different stellar masses. This suggests that the
  properties of the conduction layer are nearly independent of the
  stellar parameters. The mean temperatures of the models without
  conduction show more variation and reach higher values {at early
    times. They fall off to temperature levels obtained for models
    with thermal conduction once the mixing layer begins to dominate.}

\item At early times, the diffuse X-ray emission is provided by
  shocked fast stellar wind material. For the models with thermal
  conduction, the contribution of the evaporated nebular material to
  the diffuse X-ray emission becomes dominant ($>50\%$) well before
  peak luminosity is reached. For the models without thermal
  conduction, nebular material heated in the turbulent mixing layers
  can dominate the diffuse X-ray emission around peak luminosity.
  This suggests that stellar abundances should be appropriate for the
  diffuse X-ray emission from very young, compact PNe, but that X-ray
  emission from older PNe would be better fit with nebular abundances.

\item Our results suggest that the higher initial mass stellar models
  $M > 2.0 M_\odot$ produce a better fit to the current detected
  objects since they produce small, X-ray luminous bubbles. It would
  be worthwhile to investigate these models at higher spatial and
  temporal resolution and with a fuller coverage of the initial
  stellar mass parameter space.

\item Using sampled spectra to represent finite time observations, we
  have shown that 200 counts or more are necessary to give a reliable
  description of the spectral shape, from which the physical
  properties of the hot plasma in PNe can be derived. The spectra from
  our models with thermal conduction are all very similar, but spectra
  from models without conduction are much more varied.

\item The radial surface brightness profile of an observed planetary
  nebula has as much to do with the instrumental PSF and whether the
  angular size of the object is resolved, as with the actual
  distribution of the X-ray emitting gas. For this reason, it is
  unreasonable to discard models that do not produce flat X-ray
  surface brightness profiles. Once a limb-brightened theoretical
  radial surface brightness profile is smoothed to the same resolution
  as the observations, it too can appear to be flat.

\end{enumerate}

\section*{Acknowledgments}
We are thankful to our referee, Dr.\,Detlef Sch\"{o}nberner, who
performed an outstanding job that helped improve our results and the
clarity of this paper. We would like to thank M.\,A.\,Guerrero and
W.\,J.\,Henney for fruitful discussions during the realization of this
work. SJA and JAT acknowledge financial support through PAPIIT project
IN101713 from DGAPA-UNAM (Mexico). JAT thanks support by CSIC JAE-Pre
student grant 2011-00189 (Spain). JAT was also supported by AYA
2011-29754-C03-02 of the Spanish Ministerio de Econom\'{i}a y
Competitividad (MINECO) co-funded by FEDER funds.

\appendix
\section{Stellar parameters}
\label{sec:appa} 
In this appendix we present the stellar wind parameters, that is,
terminal wind velocity ($v_{\infty}$) and mass-loss rate ($\dot{M}$),
and also the ionizing photon flux ($Q_\mathrm{H}$) for the different
models used in the present paper. Note that very similar figures have
been presented in Paper~I. As mentioned in Paper~I, all of our stellar
wind parameters and ionization photon fluxes were computed using the
stellar atmosphere code {\sc WM-Basic} \citep[see][and references
  therein]{Pauldrach2012}. Figure~\ref{fig:wind_parameters} also
presents the evolution of the mechanical luminosity as a function of
the effective temperature of each model.

In all our simulations the zero point of evolution $t$=0 corresponds
to log($T_\mathrm{eff}$)=4 at the onset of the post-AGB phase, as
defined by the stellar evolutionary models of
\citet[][]{Vassiliadis1994} (see Paper~I).

\begin{figure*}
\includegraphics[width=.45\linewidth]{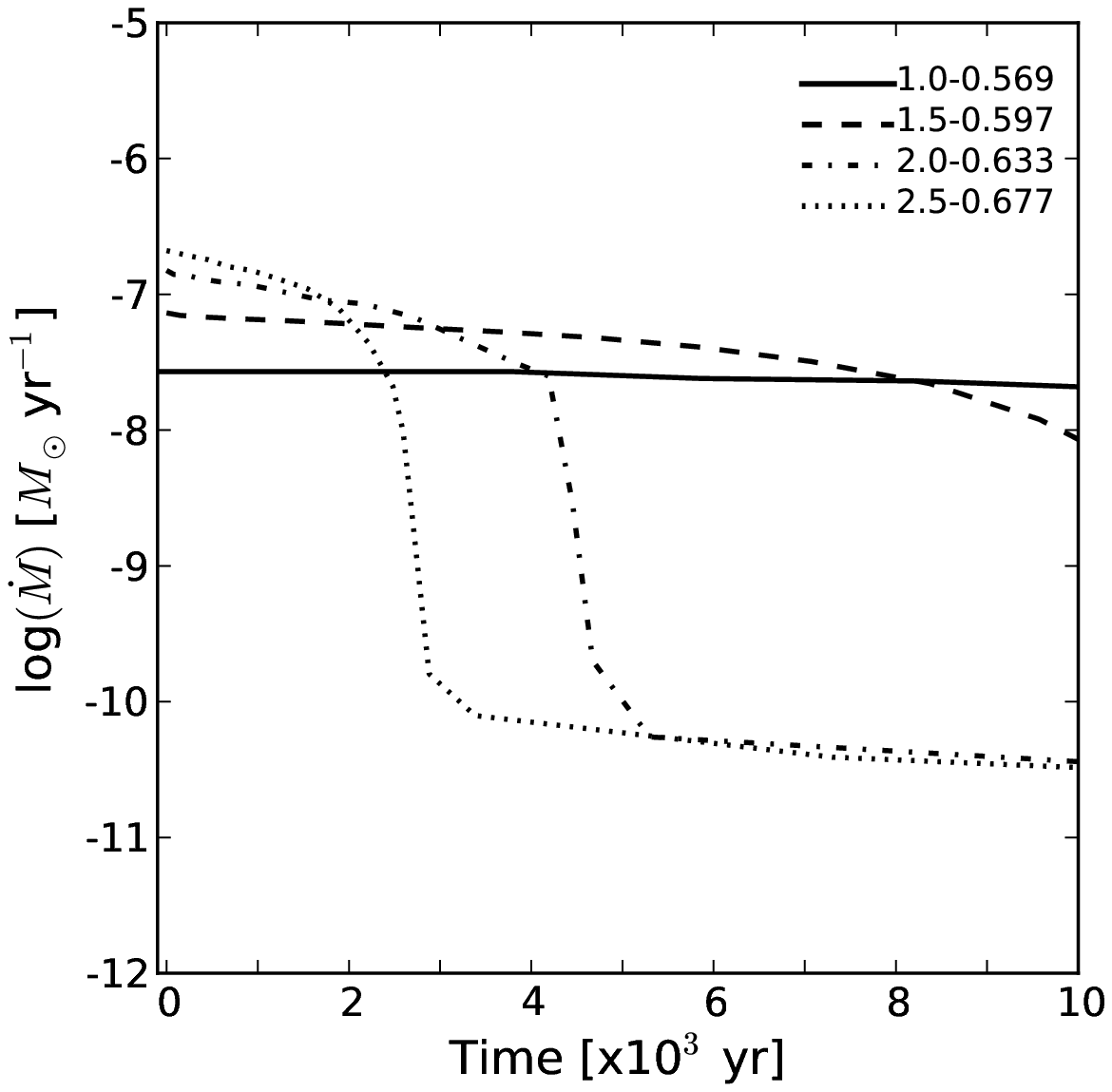}~
\includegraphics[width=.45\linewidth]{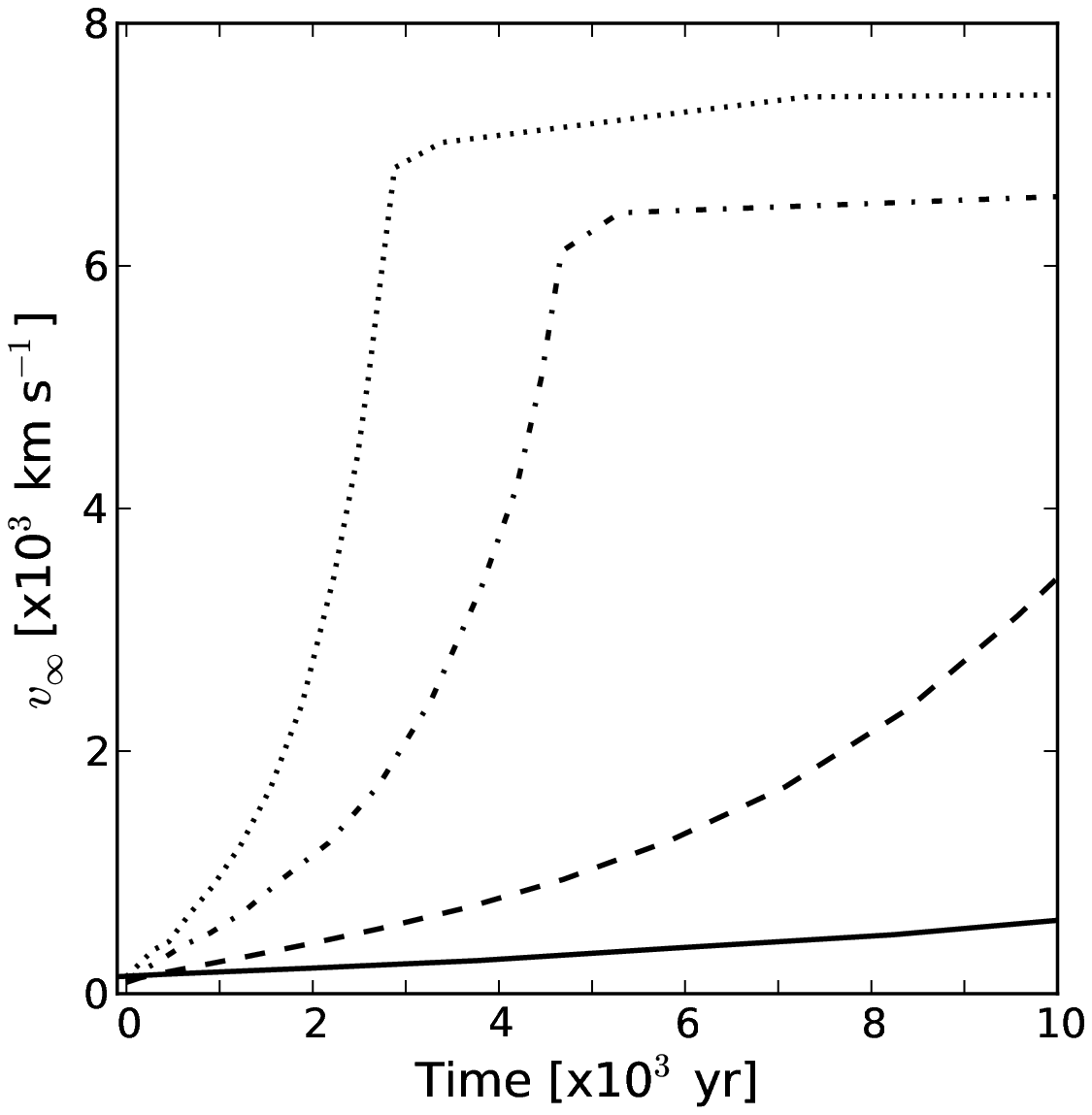}\\
\includegraphics[width=.45\linewidth]{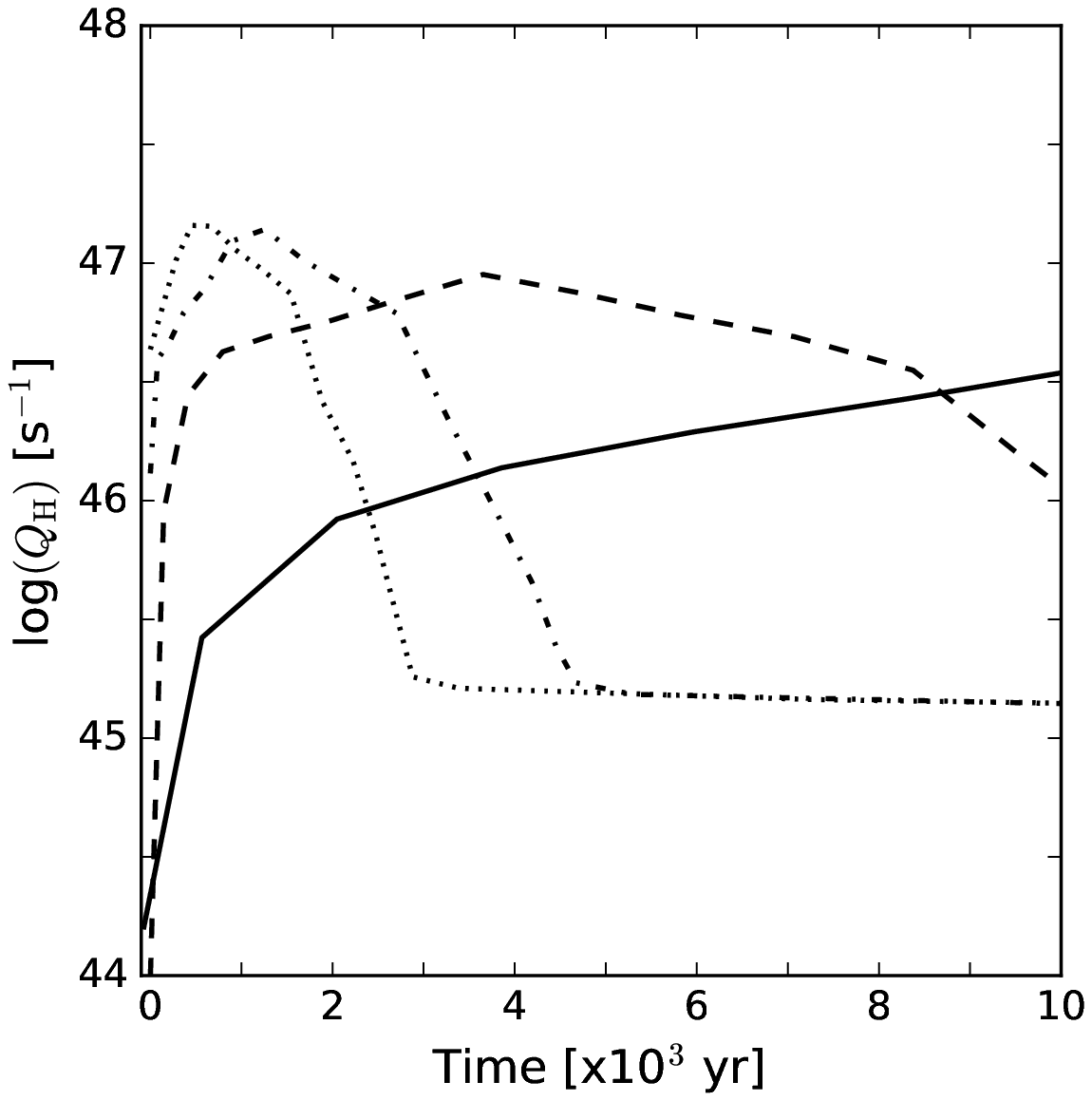}~
\includegraphics[width=.45\linewidth]{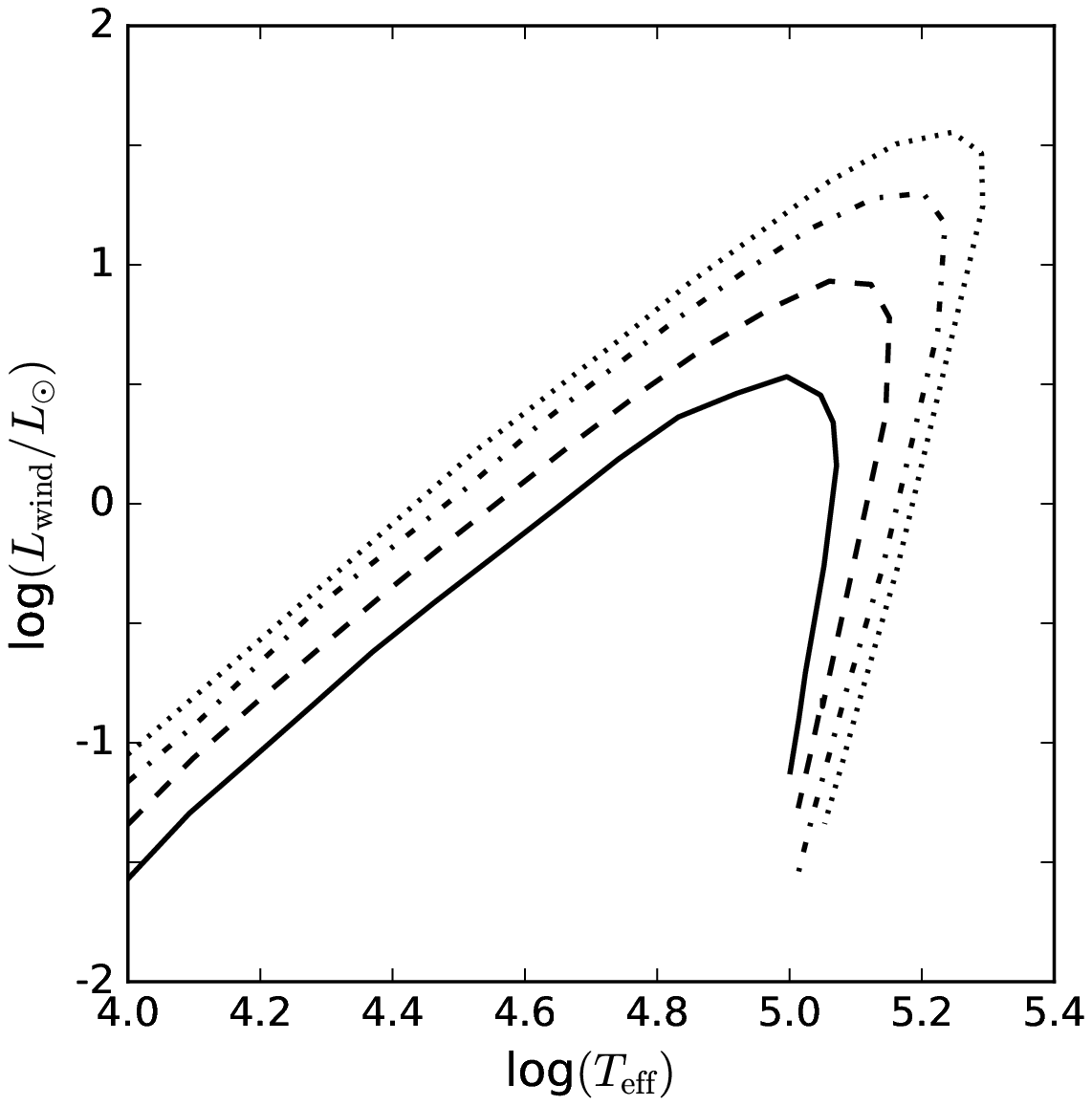}
\caption{(Upper left) Mass-loss rate, (upper right) stellar wind
  velocity, and (bottom left) ionizing photon flux as a function of
  time, and (bottom right) stellar wind luminosity as a function of
  effective temperature for the different stellar models used in this
  work. The initial-final mass of each model is marked on the upper
  left panel.}
  \label{fig:wind_parameters}
\end{figure*} 

\section{One dimensional results}
\label{sec:appb}
\begin{figure}
\includegraphics[width=\linewidth]{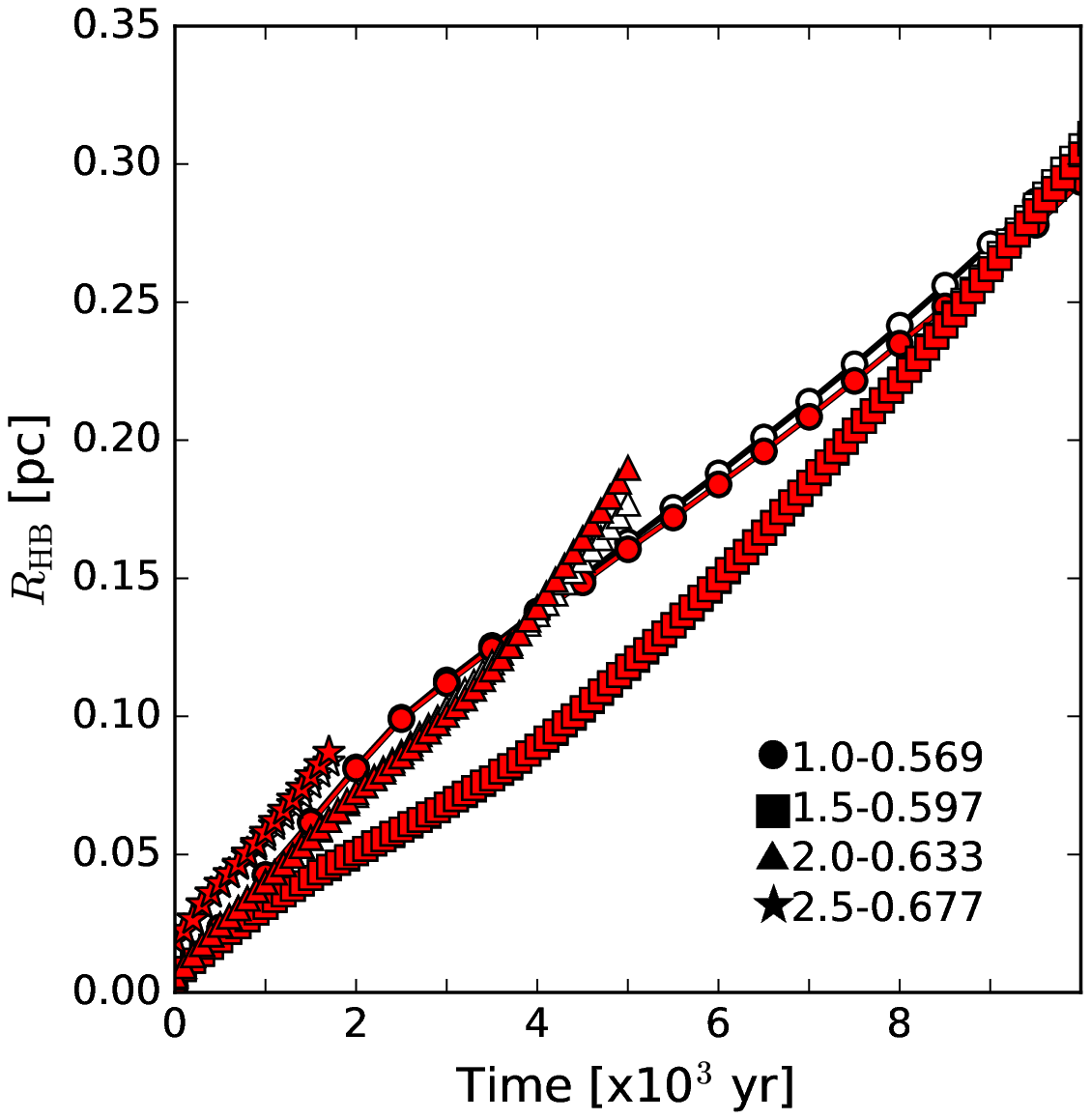}
\caption{Mean radius of the hot bubble as a function of time for the
  three principal stellar models for the 1D simulations. Filled
  symbols are for the models that include thermal condutcion, while
  open symbols are for the models that do not.}
  \label{fig:1D_hotbubble}
\end{figure}

In this appendix we present the results for models with and without
thermal conduction as obtained from the 1D version of our
radiation-hydrodynamic code. These simulations were performed at the
same numerical resolution as the 2D models.

First, we show the expansion of the hot bubble as a function of
time. For the 1D models this is simply the position of the contact
discontinuity as a function of time and is shown in
Figure~\ref{fig:1D_hotbubble}. The variety of expansion behaviours is
due to the different stellar wind histories and initial density
distributions of the different models. As shown by \citet{Steffen2008}
the cases with and without thermal conduction evolve virtually at the
same rate.

We focus particularly on the 1.5-0.597 models, as was done for the 2D
models in the main paper.  Figure~\ref{fig:1D_DEM}-top panel presents
examples of the DEM at $[2, 4,$ and $8]\times 10^{3}$~yr. This figure
shows that for the case without thermal conduction the DEM is
restricted to discontinuous, narrow temperature ranges: the lower
temperature gas comes from a few dense cells at the contact
discontinuity whose relative contribution would be lower if a higher
grid resolution were used. The higher temperature DEM comes from the
shocked fast stellar wind that fills most of the volume of the hot
bubble. Only the latest time results show the DEM distributed over a
wider temperature range.

On the other hand, the 1D cases with thermal conduction (see
Fig.~\ref{fig:1D_DEM}) show a more continuous distribution similar to
those presented in Fig.~\ref{fig:DEM_time} (although for gas with
$\log T < 6$ the temperature bin resolution has an effect). This is
because gas at this temperature belongs to the conduction layer; in
the 2D case the instabilities increase the surface area of the contact
discontinuity, and hence the quantity of gas in the conduction layer
as compared to the 1D case.

Figure~\ref{fig:1D_avtemp} shows the average temperature of both sets
of 1D simulations calculated using Eqs.~\ref{eq:DEM}
and~\ref{eq:avtemp}. The average temperature of the 1D simulations
without conduction starts at $\log T_\mathrm{A} = 6$ but ends up much
higher, with $\log T_\mathrm{A} \sim 7$, far higher than any
temperature reported from observations. The average temperature of the
1D simulations with conduction hovers around $\log T_\mathrm{A} =
6.3$. Even in this case the average temperature is higher than the
corresponding 2D results.

Fig.~\ref{fig:1D_spec} shows the corresponding spectra, computed using
{\sc chianti}, resulting from the DEMs of Fig.~\ref{fig:1D_DEM}.  The
shape of the spectra from the 1D simulations without conduction is
very flat in the 0.3--2.0~keV energy range due to the much higher gas
temperatures. The fluxes at low energy for the 1D cases with
conduction are an order of magnitude higher than those without
conduction. The fluxes of the 1D simulations are more varied than the
2D cases because there is not the continuum in the DEM distribution
that we obtain in the 2D models.

\begin{figure}
\includegraphics[height=12cm]{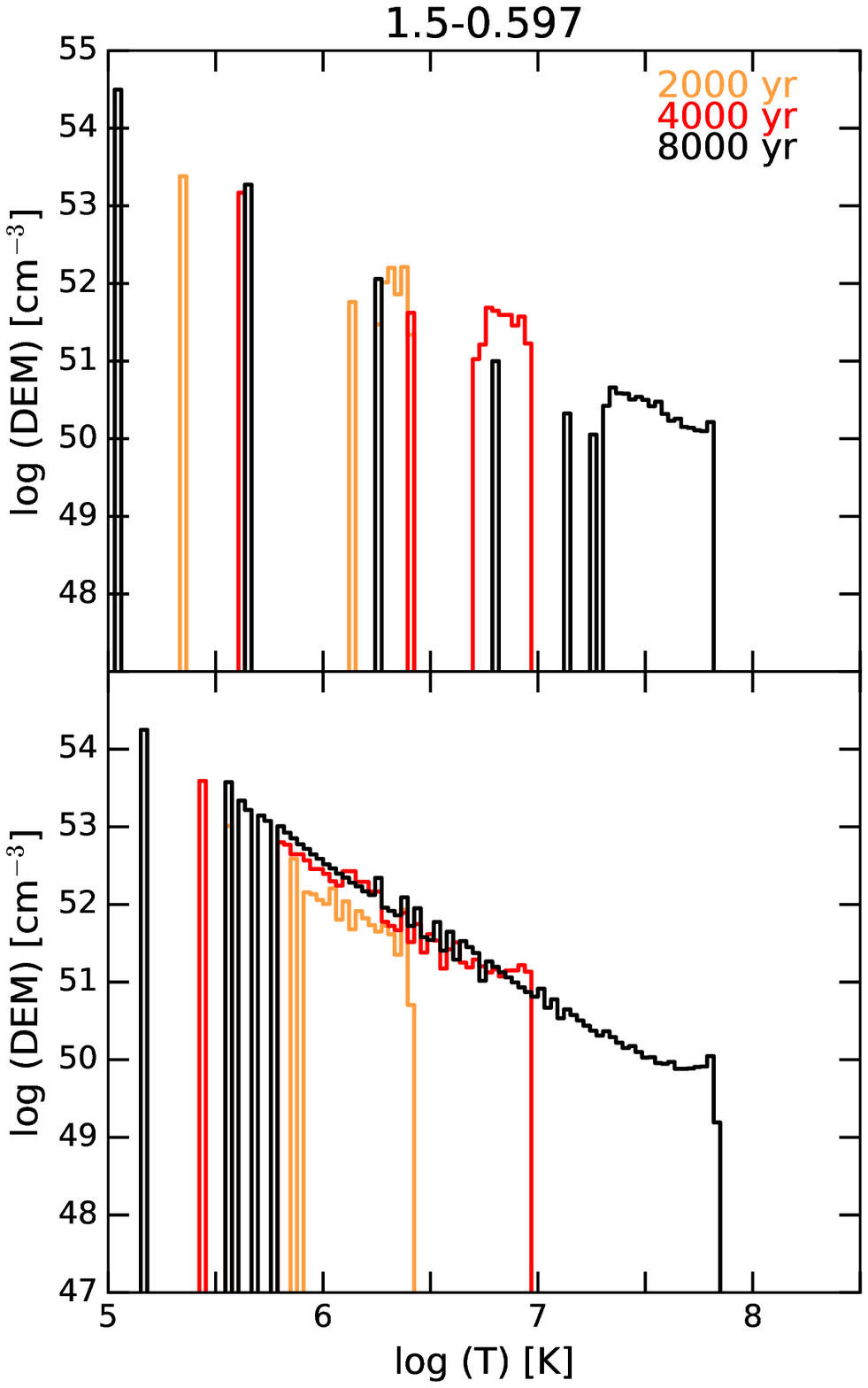}~
\caption{Differential emission measure (DEM) for the 1.5-0.597
  model as computed from 1D numerical results without (top panels) and
  with (bottom panels) thermal conduction. The colours represent
  different times, as indicated in the upper panel.}
  \label{fig:1D_DEM}
\end{figure}

\begin{figure*}
\includegraphics[width=0.9\textwidth]{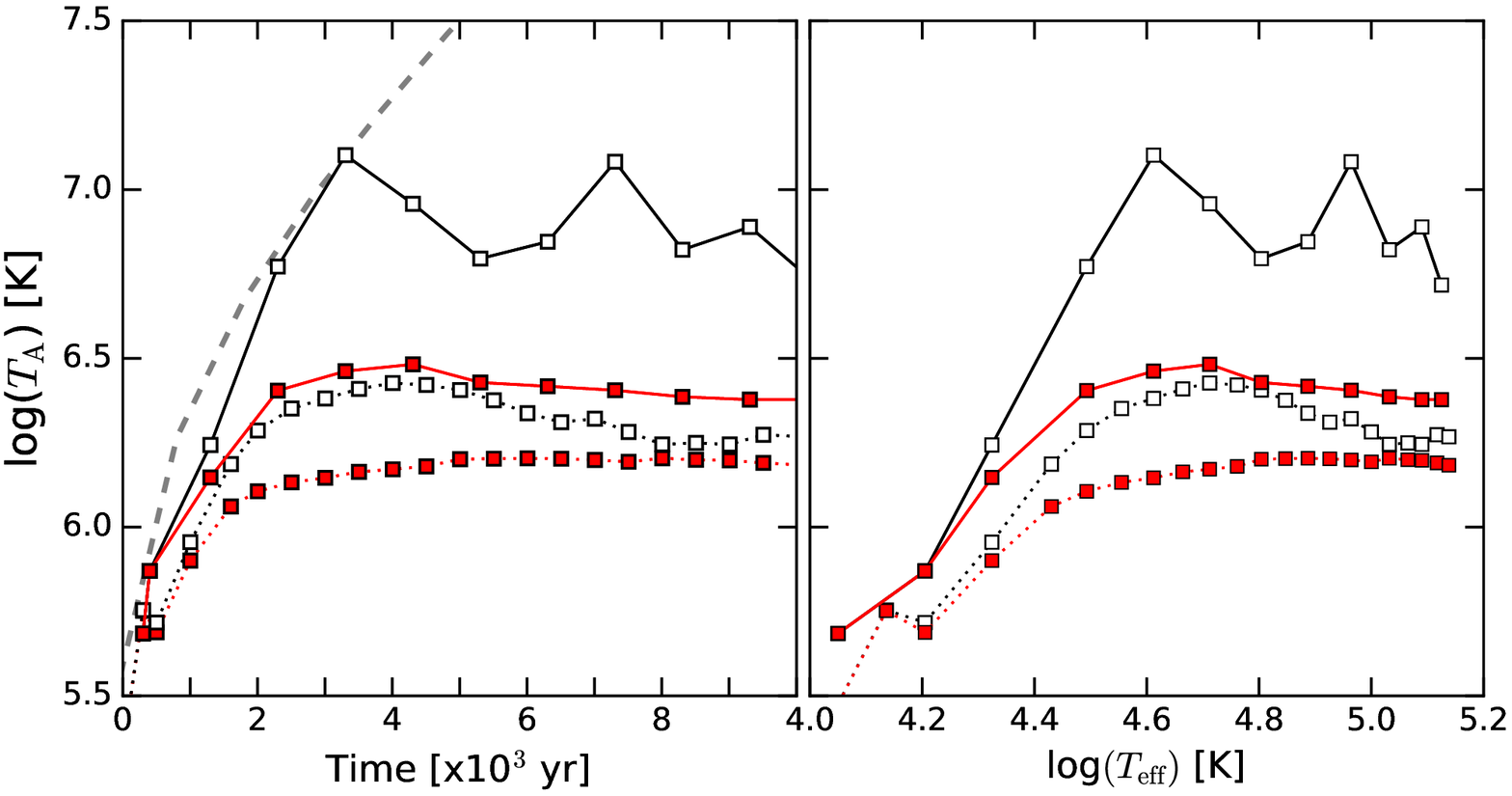}
\caption{Average temperature for the 1.5-0.597 model from 1D numerical
  simulations. Left panel: average temperature as a function of
  time. Right panel: average temperature as a function of stellar
  effective temperature. The continuous lines with filled symbols are
  for the case with thermal conduction, the continuous lines with open
  symbols are for the case without thermal conduction. The dotted
  lines show the corresponding 2D results for comparison. The dashed
  gray line shows the evolution of the post-shock temperature computed
  as $T=3 \mu m_\mathrm{H} v_\mathrm{\infty}^{2} / 16 k_\mathrm{B}$.}
\label{fig:1D_avtemp}
\end{figure*}

\begin{figure}
\includegraphics[height=12cm]{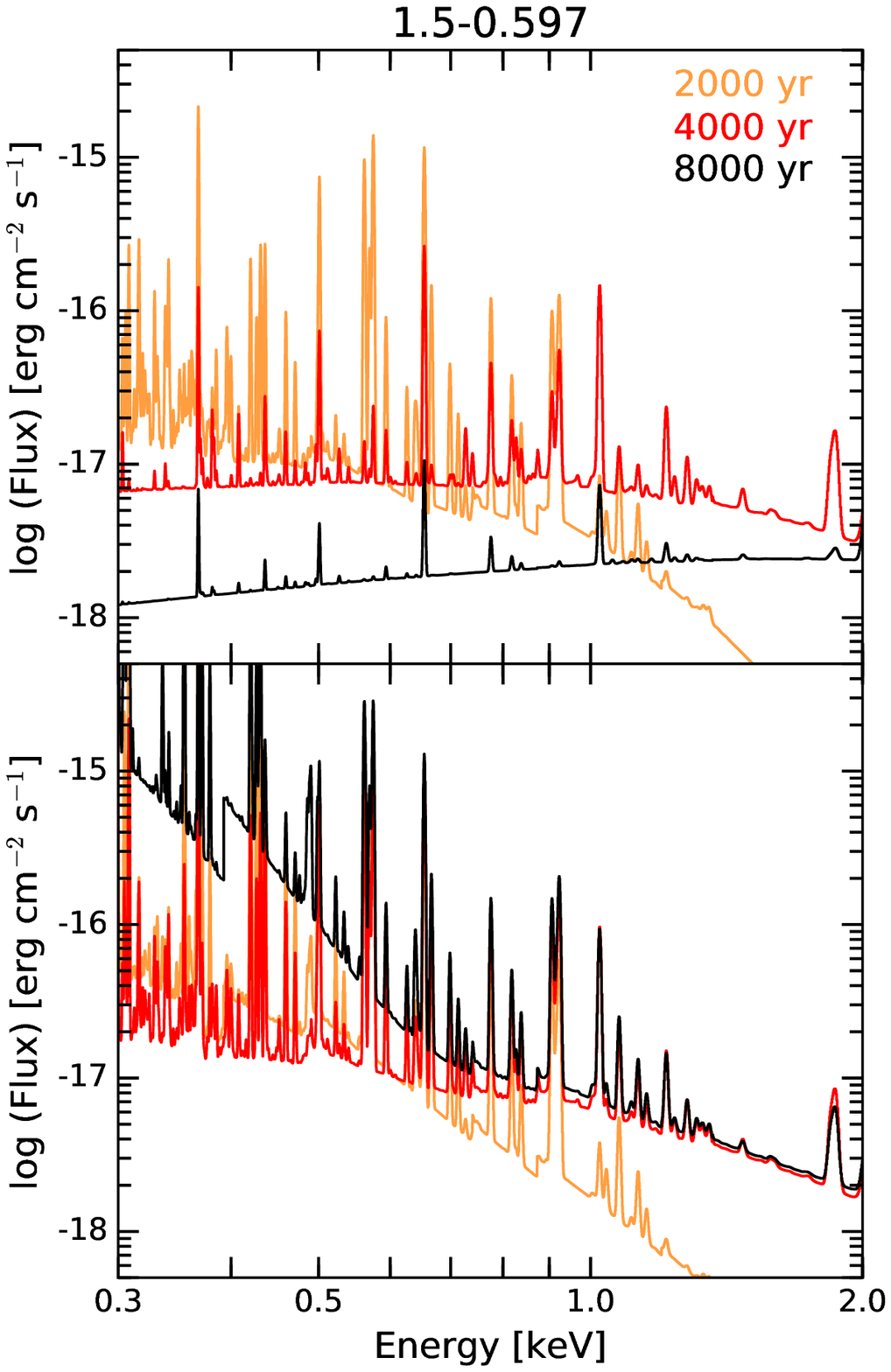}
\caption{Spectra generated from the DEMs presented in
  Fig.~\protect\ref{fig:1D_DEM} for the 1D numerical results without
  (top panels) and with (bottom panels) thermal conduction. The
  colours represent the same times as indicated in
  Fig.~\protect\ref{fig:1D_DEM}.}
  \label{fig:1D_spec}
\end{figure}

\end{document}